\documentclass[prx, 11pt, reprint, amsmath,amssymb,showpacs,floatfix,longbibliography, aps, onecolumn, superscriptaddress, nofootinbib]{revtex4-2}
\usepackage[utf8]{inputenc}
\usepackage{CJKutf8}

\usepackage{natbib}

\usepackage[left=1in,right=1in,top=1in,bottom=1in]{geometry}
\usepackage{setspace} 
\usepackage{color}
\usepackage[dvipsnames]{xcolor}
\usepackage{hyperref}
\usepackage{enumitem}
\usepackage{cancel}
\hypersetup{colorlinks=true,citecolor=blue,linkcolor=blue, urlcolor=blue}
\hypersetup{linktocpage}
\usepackage{graphicx}
\usepackage{subcaption}
\usepackage{bm} 
\usepackage[dvipsnames]{xcolor}
\usepackage{mathtools} 

\usepackage{environ} 
\NewEnviron{eqs}{%
\begin{equation}\begin{split}
    \BODY
\end{split}\end{equation}
}

\usepackage[most]{tcolorbox}

\definecolor{examplecolor}{RGB}{0,0,157}
\newcounter{tcbexample}
\newenvironment{tcbexample}[1][]{%
  \refstepcounter{tcbexample}%
  \begin{tcolorbox}[
    enhanced,
    breakable,
    pad at break*=4mm,
    attach boxed title to top left={yshift=-3mm, yshifttext=-1mm},
    colback=examplecolor!2!white,
    colframe=examplecolor!95!black,
    colbacktitle=examplecolor!95!black,
    fonttitle=\bfseries,
    boxed title style={size=small,colframe=examplecolor!65!black},
    parbox=false,
    title={Example~\thetcbexample\if\relax\detokenize{#1}\relax\else: #1\fi}
  ]%
}{%
  \end{tcolorbox}%
}

\newtcolorbox{graybox}{
    enhanced,
    breakable,
    pad at break*=4mm,
    colback=black!5!white,
    boxrule=0pt,
    sharp corners,
    parbox=false,
}

\makeatletter
\newcommand{\pagefootnotemark}{%
  \refstepcounter{footnote}%
  \protected@xdef\@thefnmark{\thefootnote}%
  \@footnotemark
}
\newcommand{\pagefootnotetext}[1]{%
  \protected@xdef\@thefnmark{\thefootnote}%
  \@footnotetext{#1}%
}
\makeatother

\usepackage{tabularray} 
\UseTblrLibrary{booktabs} 

\usepackage{physics}
\usepackage{dsfont}

\usepackage{centernot}

\usepackage[normalem]{ulem}

\usepackage{subcaption}

\usepackage{caption}
\usepackage{ragged2e} 
\makeatletter 
\def\l@subsubsection#1#2{}
\makeatother

\makeatletter 
\newcommand{\hideappsubsections}{%
  \def\l@subsection##1##2{}%
}
\makeatother

\definecolor{myBlue}{HTML}{0e35b3}

\newcommand{\ZZ}{{\mathbb Z}}
\newcommand{\RR}{{\mathbb R}}

\newcommand{\LL}{\mathsf{L}}
\newcommand{\cH}{{\mathcal H}}

\newcommand{\cG}{\mathcal G}

\newcommand{\Z}{{\mathbb Z}}
\def\U{\mathrm{U}}

\newcommand{\Bigggvert}{%
    \mathopen{\vcenter{\hbox{\scalebox{1}[1.35]{$\Bigg|$}}}}}
\newcommand{\Bigggrangle}{%
    \mathclose{\vcenter{\hbox{\scalebox{1}[1.35]{$\Bigg\rangle$}}}}} 

\definecolor{tyler}{rgb}{1,.2,0}

\definecolor{bluem}{rgb}{0.0, 0.5, 0.69}

\newtheorem{definition}{Definition}
\newtheorem{mainresult}{Main Result}

\newcommand{\cg}{\mathbb{C}[G]}
\newcommand{\cgw}{\mathbb{C}_\omega[G]}
\newcommand{\Endom}{\mathrm{End}}
\newcommand{\Gammomega}{\Gamma}

\newcommand{\glog}[1]{\ket{#1}_\mathsf{L}}
\newcommand{\multomega}{\times_\omega}
\newcommand{\invomega}[1]{\bar{#1}^{\,\omega}}

\begin{document}

\title{Engineering logical gates with irrep surface codes}

\author{Yabo Li (\begin{CJK*}{UTF8}{gbsn}李雅博\end{CJK*})}
\email{liyb.poiuy@gmail.com}
\affiliation{Department of Physics and Astronomy, Purdue University, West Lafayette, IN, 47907}
\affiliation{Center for Quantum Phenomena, Department of Physics, New York University, 726 Broadway, New York, New York 10003, USA}
\author{Zijian Song (\begin{CJK*}{UTF8}{gbsn}宋子健\end{CJK*})}
\affiliation{C. N. Yang Institute for Theoretical Physics, State University of New York at Stony Brook, Stony Brook NY 11794, USA}
\author{Vieri Mattei}
\affiliation{Department of Physics and Astronomy, Purdue University, West Lafayette, IN, 47907}
\author{Naren Manjunath}
\affiliation{Perimeter Institute for Theoretical Physics, Waterloo, Ontario N2L 2Y5, Canada}
\author{Apoorv Tiwari}
\affiliation{Center for Quantum Mathematics \& 
Danish Institute for Advanced Study (Danish IAS), Southern Denmark University, Campusvej 55, 5230 Odense, Denmark}
\author{{Tyler D.~Ellison}}
\affiliation{Department of Physics and Astronomy, Purdue University, West Lafayette, IN, 47907}
\affiliation{Purdue Quantum Science and Engineering Institute, Purdue University, West Lafayette, IN, 47907}

\begin{abstract}
        Fault-tolerant quantum computation can incur a substantial resource cost when a desired logical operation must be constructed from a restricted set of logical primitives. To make such operations directly available, we introduce irrep surface codes (ISCs), two-dimensional topological quantum error-correcting codes engineered to transversally implement a prescribed set of logical gates. The target logical gates only need to satisfy two conditions: (i) they generate a finite group, which, up to global phases, is forced by the Eastin-Knill theorem, and (ii) they are irreducible, meaning that any operator commuting with the set of gates must be proportional to the identity. This accommodates arbitrary finite groups of diagonal gates and arbitrary reversible classical gates, since each can be embedded into a larger irreducible finite group of gates. It also enables logical gates outside every finite level of the Clifford hierarchy. To incorporate these operations into surface-code computations, we further develop lattice-surgery protocols to transfer the encoded information between surface codes and ISCs obeying one of two sufficient conditions. Together, these results establish a systematic method for tailoring topological codes to desired logical operations and integrating them into conventional computations. 
\end{abstract}

\maketitle

\setcounter{tocdepth}{1}
\tableofcontents

\section{Introduction}

Quantum error-correcting codes are necessary for protecting quantum information from decoherence. Performing a computation, however, requires manipulating the encoded information while maintaining the fault tolerance. Consequently, the choice of code directly determines how logical operations are implemented and their required resources. For Pauli stabilizer codes, such as the surface code~\cite{bravyi1998surfacecode}, there are well-established means of implementing logical Clifford gates~\cite{Steane1996CSS, Moussa2016folded, Fowler2012practical, Bombin2010twist, Yoder2017twist, Horsman2012latticesurgery, Litinski2019gameof, cross2025improvedqldpcsurgerylogical, he2025extractorsqldpcarchitecturesefficient, Williamson2026gauging}. These gates alone are insufficient for universal computation and must be supplemented with a non-Clifford resource, such as a logical $T$ gate~\cite{boykin1999CliffodplusT, nebe2000invariantscliffordgroups, Bravyi2005magic}.

Providing this non-Clifford resource efficiently is a central problem in fault-tolerant quantum computation. Approaches include magic-state distillation and cultivation~\cite{Bravyi2005magic, Chamberland2020lowoverhead,Itogawa2025zerolevel,gidney2024magicstatecultivationgrowing}, employing transversal non-Clifford gates in higher-dimensional Pauli stabilizer codes~\cite{Bombin2015GCC, Bombin2016dimensionaljumping, Brown2016GCC, Beverland2021comparison, Brown20202DnonClifford, daguerre2025codeswitching}, and leveraging non-Clifford operations enabled by non-Pauli topological stabilizer codes~\cite{huang2025D4, davydova2025D4, sajith2025noncliffordS3, bauer2025planarfaulttolerantcircuitsnonclifford, kobayashi2025cliffordhierarchystabilizercodes, hsin2025automorphismgaugetheorieshigher, warman2026transversalcliffordhierarchygatesnonabelian, manjunath2026GSC, li_anyon_2026}. Evaluating and optimizing these approaches remains an important step in reducing the overhead of universal quantum computation.

Even with an implementation of a logical non-Clifford gate, synthesizing a desired operation can require substantial resources. For example, under optimal gate-by-gate synthesis, the number of $T$ gates required for a typical single-qubit gate to a precision $\epsilon$ is, to leading order, $3\log_2(1/\epsilon)$~\cite{morisaki2025optimalancillafreecliffordtsynthesis, fowler2010constructingarbitrarysteanecode}. This suggests that synthesizing an individual gate to a precision $10^{-6}$, relevant to large-scale quantum chemistry simulations~\cite{Luo2025chemsimulation, Nguyen2026chemsimulation}, would require approximately $60$ $T$ gates. More sophisticated strategies can reduce the average cost per operation across a large computation~\cite{zhang2026quantumcircuitcompilationconstant}, but the target operations are still realized indirectly from a restricted set of logical primitives. The resource cost of synthesis motivates constructing codes that not only implement a non-Clifford gate but implement a target logical gate directly.

Non-Pauli topological codes provide a promising route to engineering codes for target logical gates. Recent constructions realize roots of $T$ through constant-depth circuits~\cite{warman2026transversalcliffordhierarchygatesnonabelian} and arbitrary reversible classical gates through transversal operations in group surface codes~\cite{manjunath2026GSC}. These constructions address valuable classes of operations, but they raise the broader question: which groups of gates can be implemented transversally in scalable quantum error-correcting codes?

\begin{table}[t]
    \centering
    \renewcommand{\arraystretch}{1.5}
    \setlength{\tabcolsep}{5pt}

    \resizebox{0.99\textwidth}{!}{%
    \begin{tabular}{|c|c|c||c|c||c|c||c|}
    \hline
        \multicolumn{3}{|c||}{Group and dimensions} &
        \multicolumn{2}{|c||}{Transversal gates -- representations} &
        \multicolumn{2}{|c||}{Transversal gates -- intertwiners} &
        Reference \\ \hline

        $G$ & $|G|$ & $d_{\Gamma}$ &
        Generating set & Max level &
        Additional generators & Max level & \\

        \noalign{\hrule height 1.5pt}

        $\Z_2 \times \Z_2$ & 4 & 2 &
        $X,~Z$ & 1 &
        $H,S$ & 2 & Ex.~\ref{example1} \\
        \hline

        $\rule[-2.25em]{0pt}{4.5em}D_{2^m}$ & $2^{m+1}$ & 2 &
        $X,~
        \begin{pmatrix}
            1 & 0 \\
            0 & e^{i \frac{2\pi}{2^m}}
        \end{pmatrix}$
        & $m$ &
        $\begin{pmatrix}
            1 & 0 \\
            0 & e^{i \frac{2\pi}{2^{m+1}}}
        \end{pmatrix}$
        & $m+1$ & Ex.~\ref{example4} \\
        \hline

        $\rule[-2.25em]{0pt}{4.5em}D_{2m+1}$ & $4m+2$ & 2 &
        $X,~
        \begin{pmatrix}
            1 & 0 \\
            0 & e^{i \frac{2\pi}{2m+1}}
        \end{pmatrix}$
        & $\infty$ &
        $Z$ & $\infty$ & Ex.~\ref{example4} \\
        \hline

        $\rule[-2.25em]{0pt}{4.5em}A_5$ & 60 & 2 &
        $X$,~$Z$,~
        $\frac{1}{2}
        \begin{pmatrix}
            \varphi-i\varphi^{-1} & i\\
            i & \varphi+i\varphi^{-1}
        \end{pmatrix}$
        & $\infty$ & None
        & $\infty$ & Ex.~\ref{example3} \\
        \hline

        $\rule[-2.75em]{0pt}{5.5em}A_5$ & 60 & 4 &
        $X_1,X_2,~\frac{1}{4}
    \begin{pmatrix}
    -\varphi & -\varphi & -\varphi & 2-3\varphi\\
    \varphi-2 & \varphi-2 & \varphi+2 & -\varphi\\
    \varphi+2 & \varphi-2 & \varphi-2 & -\varphi\\
    \varphi-2 & \varphi+2 & \varphi-2 & -\varphi
    \end{pmatrix}$
        & $\infty$ & $\text{CX}_{12},~\text{SWAP}_{12}$
        & $\infty$ & App.~\ref{app: A5-2qubits} \\
        \hline

        $D_4^2 \rtimes \Z_2$ & 128 & 4 &
        $X_1,~X_2,~\text{CH}_{12}$
        & 3 & $S_1,~\text{CX}_{12}$
        & 3 & App.~\ref{app: CH}
        \\
        \hline

        $\Z_2^7 \rtimes \Z_2^3$ &
        $2^{10}$ &
        $8$ &
        $X_1,~X_2,~X_3,~\text{CCZ}_{123}$ 
        & 3 & $\{S_i\},~\{\text{CX}_{ij}\}$
        & 3 & App.~\ref{app: arbitrary diagonal gates}
        \\
        \hline

        $\mathcal{D}_X \rtimes \Z_2^m$ &
        $2^m \times |\mathcal{D}_X|$ &
        $2^m$ &
        $X_1,~ X_2,~ \dots,~ X_m,~\mathcal{D}$ (diagonal)
        & $\geq 1$ & Depends on $\mathcal{D}$
        & $\geq 1$ & App.~\ref{app: arbitrary diagonal gates} \\
        \hline

    \end{tabular}%
    }

    \caption{
        Examples of transversal logical gates in ISCs. $G$ is the quotient group in Eq.~\eqref{eq: quotient group} obtained by modding out by global phases. $|G|$ denotes the number of elements in $G$; it equals the local Hilbert space dimension on each edge of the ISC. $d_{\Gamma}$ is the dimension of the projective irrep $\Gamma$; it equals the dimension of the corner qudit in the ISC, as well as the dimension of the code space. The gates in $\cG$ form a projective irrep of $G$ with the representation operators in the fourth column (Result~\ref{res: ISCs}). The sixth column lists additional generators of transversal gates from projective intertwiners (Result~\ref{res: additional transversal}). 
        The max level denotes the highest level of the qubit Clifford hierarchy accessible in each case, with the symbol $\infty$ meaning that at least one gate is outside any finite level. Note that every representation operator yields a projective intertwiner, so the max level in the second to last column is strictly larger than max level of the representation operators. Other notation used: $\text{CH}$ is the controlled Hadamard gate. $\varphi = \frac{1 + \sqrt{5}}{2}$ is the golden ratio. $\mathcal{D}$ is some desired finite group of diagonal gates on $m$ qubits, and $\mathcal{D}_X$ is the full group of diagonal gates obtained by conjugating elements of $\mathcal{D}$ with $m$-qubit Pauli $X$ operators.
    }
    \label{tab: summary}
\end{table}

In this work, we substantially broaden the class of transversal gates admitted by two-dimensional topological quantum error-correcting codes. In particular, we engineer topological codes that transversally implement any irreducible finite group of gates, where irreducibility is defined as follows:
\begin{definition}[Irreducible group of gates] \label{def: irreducible group of gates}
A group of gates $\cG$ on a finite-dimensional Hilbert space $\cH$ is irreducible if the only operators on $\cH$ that commute with every gate in $\cG$ are proportional to the identity.
\end{definition}
In other words, block diagonalizing the group of gates gives a single block. Notably, this includes arbitrary finite groups of diagonal gates or reversible classical gates, after adding a sufficient number of generalized Pauli operators. Therefore, we directly extend the logical gate sets of Refs.~\cite{warman2026transversalcliffordhierarchygatesnonabelian, manjunath2026GSC}.\footnote{Note that although ISCs include the logical gates of group surface codes, the resource overheads can be substantially different.}

More specifically, we introduce irrep surface codes (ISCs), which are, in general, non-Pauli stabilizer codes with an underlying non-Abelian topological order. Importantly, the ISC based on the irreducible finite group of gates $\cG$ admits a transversal implementation of the gates in $\cG$. This is summarized by our first main result: 
\begin{mainresult}[Irrep surface codes] \label{res: ISCs}
For any irreducible finite group of gates $\cG$ acting on $\cH$, we construct a two-dimensional topological code encoding $\cH$ and implementing every gate in $\cG$ transversally. 
\end{mainresult}
An ISC can also admit transversal gates beyond those in the irreducible finite group of gates $\cG$. Specifically, any unitary that preserves $\cG$ under conjugation up to global phases has a transversal implementation:
\begin{mainresult}[Additional transversal gates] \label{res: additional transversal}
Let $\cG$ be an irreducible finite group of gates on $\cH$. Suppose a unitary $W$ on $\cH$ satisfies $W\Gamma_1W^\dagger=e^{i\theta}\Gamma_2$ for every $\Gamma_1\in\cG$ and some $\Gamma_2\in\cG$, $\theta\in\mathbb{R}$. Then the ISC based on $\cG$ admits a transversal implementation of $W$.
\end{mainresult}
Table~\ref{tab: summary} presents examples of the transversal gate sets enabled by ISCs.

To incorporate these gates into a universal computation, we further provide a means to transfer encoded information between an ISC and surface codes. This allows the transversal gates of an ISC to be combined with logical Clifford gates available in the surface codes:
\begin{mainresult}[Transferring information] \label{res: transfer}
    We give a lattice-surgery protocol for transferring information between a surface code and an ISC whenever the irreducible finite group of gates $\cG$ satisfies one of the following two conditions:
    \begin{enumerate} [label=(\roman*)]
        \item $\cG$ contains the generalized Pauli $X$ and $Z$ operators, up to phases, for the logical qudit being transferred.
        \item $\cG$ is generated by diagonal gates and the generalized Pauli $X$ operators for each subsystem of $\cH$.
    \end{enumerate}
\end{mainresult}
Together, these results provide a route to engineering quantum error-correcting codes tailored to the gates required by an algorithm and further incorporating them into a surface-code computation. 

The organization of the manuscript is as follows. In Section~\ref{sec: irrep surface codes}, we introduce the defining data of ISCs and use it to specify the physical degrees of freedom and the code space. We then establish the code distance and construct the transversal logical gates of Results~\ref{res: ISCs} and \ref{res: additional transversal}. In Section~\ref{sec: transferring information}, we leverage lattice surgery to prepare logical states and transfer encoded information between ISCs and surface codes. We conclude by commenting on future directions in Section~\ref{sec: discussion}. 
The appendices provide complementary descriptions of ISCs and their logical operations. Appendix~\ref{app: irrep review} provides a review of the twisted group algebras underlying ISCs. Appendix~\ref{app: isc from gsc} derives ISCs from projective group surface codes, making the connection to the construction in Ref.~\cite{manjunath2026GSC} explicit. Appendix~\ref{app: minimal isc} gives a disentangling circuit that reduces an ISC to a minimal model. Appendix~\ref{app: line-like logical} develops the deformations of logical operators used to establish the code distance. Appendix~\ref{app: more examples} contains details on the examples in Table~\ref{tab: summary}.

\section{Irrep surface codes} \label{sec: irrep surface codes}

The goal of this section is to construct an ISC admitting a transversal implementation of a given irreducible finite group of gates $\cG$, as defined in Def.~\ref{def: irreducible group of gates}. The group $\cG$ contains more information than is needed to define an ISC. In particular, it includes global phases that do not produce physically meaningful transversal gates. Therefore, our first step is to remove the global phases from $\cG$. This naturally leads us to an irreducible projective representation $\Gamma$ of a finite group $G$, which is the defining data of an ISC. We define irreducible projective representations of a finite group, referred to as projective irreps, in Section~\ref{sec: input data}.
In the subsequent two sections, Sections~\ref{sec: physical Hilbert space} and \ref{sec: code space}, we use this data to define an ISC and compute the dimension of the code space. We then demonstrate that an ISC derived from $\cG$ admits a transversal implementation of the gates in $\cG$ in Section~\ref{sec: logical operators}. We further show that there is a transversal gate associated to any unitary operator that, through conjugation, maps $\cG$ back to itself up to global phases.   

\subsection{Input data} \label{sec: input data}

\begin{figure}[tb]
        \centering
        \includegraphics[width=0.9\textwidth]{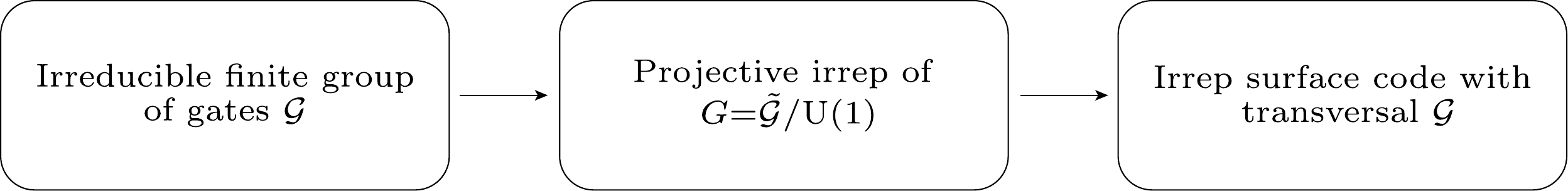}
    \caption{From $\cG$ to an ISC. Starting with an irreducible finite group $\cG$, we define a projective irrep of the quotient group $G$. We then use the projective irrep of $G$ to build an ISC. This ISC admits the gates of $\cG$ as transversal logical operators, up to a global phase.}
    \label{fig: ISC flow chart}
\end{figure}

To motivate the introduction of projective irreps, we consider an irreducible finite group of gates $\cG$. The gates in $\cG$ generally include global phases, which have no effect on a computation. We therefore remove the redundant information associated to the global phases. To accomplish this, we first take a step in the opposite direction and enlarge $\cG$ to include arbitrary global phases. This gives us the $\U(1)$ completion of $\cG$:
\begin{align} \label{eq: U1 completion}
    \tilde{\cG} = \{e^{i\theta}\Gamma \, : \, \theta \in \mathbb{R}, \, \Gamma \in \cG\},
\end{align}
ensuring that $\U(1)$ is a subgroup.
We can then quotient out by the phases to obtain the finite group $G$:
\begin{align} \label{eq: quotient group}
    G = \tilde{\cG}/\U(1).
\end{align}
The elements of $G$ are equivalence classes of gates, where two gates belong in the same equivalence class if they only differ by a global phase. The product of two elements in $G$ is determined from the multiplication of gates. Letting $[\,\Gamma_1\,]$ and $[\,\Gamma_2\,]$ be equivalence classes labeled by the gates $\Gamma_1$ and $\Gamma_2$, we have:
\begin{align}\label{eq: equivalence classes}
    [\,\Gamma_1\,][\,\Gamma_2\,]=[\,\Gamma_1\Gamma_2\,].
\end{align}
At this point, $G$ is an abstract group whose elements label equivalence classes of gates, rather than unitary operators. To construct an ISC, we need to associate a concrete gate to each element of $G$. This can be done by choosing a representative gate from each equivalence class.\footnote{Formally, this choice is determined by a map from $G$ to $\tilde{\cG}$ that respects the equivalence classes. This is known as a section.} We let $\Gamma^g$ denote the representative gate for the equivalence class associated to $g \in G$. The representative gates need only reproduce the group multiplication in $G$ up to a global phase. The gates $\Gamma^g\Gamma^h$ and $\Gamma^{gh}$ belong to the same equivalence class, so they differ only by a phase. This allows us to write:
\begin{align} \label{eq: projective rep1}
    \Gamma^g \Gamma^h = \omega(g,h)\Gamma^{gh},
\end{align}
where $\omega(g,h)$ is valued in $\U(1)$. 

The map $\Gamma$ obtained above gives precisely a projective representation of $G$, as defined in the next section. Moreover, because the original group of gates $\cG$ is irreducible, this projective representation is also irreducible. We are thus naturally led from $\cG$ to a projective irrep $\Gamma$ of the finite group $G$. The construction of an ISC depends only on the projective irrep, rather than the group of gates $\cG$. Therefore, we now define projective irreps formally and introduce the properties needed for the ISC construction.

\subsubsection{Projective irreps}

To unpack the definition of a projective irrep, we first recall that an ordinary representation of $G$ assigns each element of $G$ to a unitary operator, such that the unitary operators reproduce the group multiplication of $G$. For the construction of an ISC, we only require that the group multiplication of $G$ is reproduced up to a global phase. This gives us a \textit{projective} representation of $G$. More concretely, a projective representation $\Gamma$ maps the elements $g,h \in G$ to unitary operators $\Gamma^g$ and $\Gamma^h$ acting on a Hilbert space $V_\Gamma$ and satisfying:
\begin{align} \label{eq: projective rep}
    \Gamma^g\Gamma^h=\omega(g,h)\Gamma^{gh},
\end{align}
where $\omega(g,h)$ belongs to $\U(1)$. We denote the dimension of the representation as:
\begin{align}
    d_\Gamma = \mathrm{dim}\,V_\Gamma,
\end{align}
such that each $\Gamma^g$ is a $d_\Gamma \times d_\Gamma$ unitary operator.

A projective representation is further called irreducible if the only operators that commute with $\Gamma^g$ for all $g \in G$ are those proportional to the identity. This implies that the Hilbert space $V_\Gamma$ cannot be decomposed into smaller nontrivial subspaces that are preserved by every $\Gamma^g$. The representation therefore acts on $V_\Gamma$ as a single irreducible block, rather than as a direct sum of independent representations, described in more detail in Appendix~\ref{app: irrep review}. We refer to such representations as projective irreps. 

It is convenient to also introduce a notion of equivalence on projective representations. We say $\Gamma$ and $\hat{\Gamma}$ are equivalent, if there exists a unitary operator $W$ that maps between their representation operators up to a phase:
\begin{align} \label{eq: projective equivalence}
    W\Gamma^g W^\dagger =\beta(g) \hat{\Gamma}^g,
\end{align}
where $\beta(g)$ is some $g$-dependent phase.\footnote{Starting from an irreducible finite group of gates $\cG$, the phase $\beta(g)$ accounts for the ambiguity in choosing the representative gate for each equivalence class. The choices of representative gates yield equivalent projective irreps.} In other words, they are the same representation up to a change of basis on $V_\Gamma$ and a redefinition of the global phases. The unitary operator $W$ is referred to as a \textit{projective intertwiner} of $\Gamma$ and $\hat{\Gamma}$. In Section~\ref{sec: logical operators}, we show that certain projective intertwiners give rise to transversal logical operators of the ISC.
We note that although ISCs based on equivalent projective irreps can differ in their microscopic details, they share the same essential properties, including the dimension of the code space and the transversal logical gates.

Before turning to examples of projective irreps, there is one property of the phase factor $\omega(g,h)$ in Eq.~\eqref{eq: projective rep} that is worth making explicit for the construction of ISCs. The associativity of the representation operators requires:
\begin{align}
    (\Gamma^g\Gamma^h)\Gamma^k=\Gamma^g(\Gamma^h\Gamma^k),
\end{align}
which implies that the phases produced must satisfy:
\begin{align} \label{eq: cocycle condition}
    \omega(g,h)\omega(gh,k)=\omega(g,hk)\omega(h,k).
\end{align}
This is known as the cocycle condition, and we refer to any function $\omega: G \times G \to \U(1)$ obeying this condition as a cocycle. 

We conclude this section by giving two examples of the group $\cG$ and their associated projective irreps.

\begin{tcbexample}[{$\cG=\langle X, Z\rangle$}]\label{example1}
\vspace{.2cm}

As our first example, let us take the group of gates $\cG$ to be generated by Pauli $X$ and $Z$. The group $\cG$ is certainly finite, as the elements are explicitly $\{\pm I,\pm X,\pm iY,\pm Z\}$, where $I$ is the $2 \times 2$ identity matrix. It is also irreducible, because the only operators commuting with $X$ and $Z$ are of the form $e^{i \theta} I$, i.e., proportional to the identity. 


To compute the quotient group $G$, we first identify the $\text{U}(1)$ completion $\tilde{\cG}$ in Eq.~\eqref{eq: U1 completion} as the group with generators $\langle e^{i \theta} I, X, Z : \theta \in \RR\rangle$. Modding out by global phases, we find that $G \cong  \Z_2 \times \Z_2$,\pagefootnotemark~where the equivalence classes satisfy, for example: 
\begin{align}
        [\,X\,][\,Z\,] = [\,Y\,],
\end{align}
with the phases absorbed into the equivalence classes.

    We let $x$ and $z$ denote the equivalence classes $[\,X\,]$ and $[\,Z\,]$, respectively. A projective irrep of $G$ is then specified by choosing the matrix representatives
    \begin{align}
        \Gamma^1=I, \quad \Gamma^x = X, \quad \Gamma^z = Z, \quad \Gamma^{xz}= X Z = - iY.
    \end{align}
    The projective irrep is two-dimensional, so $d_{\Gamma} = 2$. An arbitrary element of $G$ can be decomposed as $a = x^{a_x} z^{a_z}$, where $a_x,a_z \in \{0,1\}$. For the above choice of $\Gamma$, the anticommutation relation between $X$ and $Z$ implies:
\begin{equation}\label{eq: projective equivalence XZ}
        X^{a_x} Z^{a_z} X^{b_x} Z^{b_z} = (-1)^{b_x a_z} X^{a_x + b_x} Z^{a_z + b_z}~,
\end{equation}
which is Eq.~\eqref{eq: projective rep} with 
\begin{equation}
    \omega(a,b) = (-1)^{b_x a_z}~.
\end{equation}
The cocycle condition Eq.~\eqref{eq: cocycle condition} can be readily verified for this choice of $\omega$. 

We could have also satisfied Eq.~\eqref{eq: equivalence classes} by choosing the representation $\hat{\Gamma}^x = X$, $\hat{\Gamma}^z = Z$ and $\hat{\Gamma}^{xz} = +iY$. This is a {different} choice of representative for the $xz$ equivalence class referred to below Eq.~\eqref{eq: equivalence classes}, and results in a different but equally valid cocycle 
\begin{equation}
    \hat{\omega}(a,b) = (-1)^{a_x b_z}.
\end{equation}
In fact, we have
\begin{equation}
    Y \Gamma^a Y = -\hat{\Gamma}^a, \quad a \neq 1,
\end{equation}
implying that $\Gamma$ and $\hat{\Gamma}$ are equivalent projective irreps based on the criterion in Eq.~\eqref{eq: projective equivalence}; taking $W = Y$ and setting $\beta(1) = 1$ and $\beta(x) = \beta(z) = \beta(xz) = -1$. 

\end{tcbexample}
\pagefootnotetext{We could also have, of course, found $G$ by directly quotienting out the $\Z_2$ subgroup of $\cG$ generated by $-I$.}

\begin{tcbexample}[{$\cG=\langle X, P_m \rangle$}]\label{example2}
\vspace{.2cm}

    Next, we consider the group of gates $\cG$ generated by $X$ and the phase gate $P_m$, defined as:
    \begin{align} \label{eq: Pm def}
         P_m = \begin{pmatrix}
        1 & 0 \\ 
        0 & e^{i \frac{2\pi}{m}}
        \end{pmatrix},
    \end{align}
    where $m>1$ is a fixed positive integer. To see that $\cG$ is finite, we note the commutation relations of $X$ and $P_m$:
    \begin{align} \label{eq: XP commutation}
        P_m X = e^{i \frac{2\pi}{m}} X P_m^{-1}.
    \end{align}
    Therefore, by commuting factors of $P_m$ past $X$, we can put any element of $\cG$ in the form:
    \begin{align}
        e^{i \frac{2\pi }{m}a} X^b P_m^{c}, \quad a \in \Z_m, \,\,b \in \Z_2,\,\,c\in \Z_m,
    \end{align}
    showing that $\cG$ must have finitely many elements. To see that $\cG$ is irreducible, we first notice that operators that commute with $P_m$ must be diagonal. Of the diagonal operators, the only ones that also commute with $X$ are those that are proportional to the identity, so $\cG$ is irreducible.
    
    The quotient group $G$, formed by modding out by global phases, is the dihedral group $D_m$. This has the abstract presentation:
        \begin{equation} \label{eq: sr generators}
            D_m = \langle s,r| s^2 = r^m =  1, \, sr sr= 1  \rangle.
        \end{equation}
    To see that this is the quotient group, we take the generators $s$ and $r$ to be represented as:
        \begin{align}
            \Gamma^s=X, \quad \Gamma^r = P_m.
        \end{align}
    These satisfy the relations of $D_m$ up to a phase:
    \begin{align}
        \left(\Gamma^s\right)^2=\left(\Gamma^r\right)^m=1, \quad \Gamma^s\Gamma^r\Gamma^s\Gamma^r=e^{i\frac{2\pi}{m}}.
    \end{align}
        
    The representations of the generators only determine the projective irrep up to global phases. We need to further fix the phase of the representations for every $g \in D_m$. Letting $g = s^{g_s} r^{g_r}$ denote a general group element, for $g_s \in \{0,1\}$ and $g_r \in \{0,\ldots,m-1\}$, we take $g$ to be represented by:
        \begin{equation} \label{eq: Dm representation operators}
            \Gamma^g = X^{g_s} P_{m}^{g_r}.
        \end{equation}
        
    To determine the corresponding cocycle, we compute $\Gamma^g\Gamma^h$, which gives us:
        \begin{equation}
            X^{g_s} P_m^{g_r} X^{h_s} P_m^{h_r} = e^{i \frac{2\pi}{m} g_r h_s} X^{g_s + h_s} P_m^{(-1)^{h_s} g_r + h_r}.
        \end{equation}
        This implies that the cocycle is:
        \begin{equation}
            \omega(g,h) = e^{i \frac{2\pi}{m} g_r h_s}.
        \end{equation} 
        We conclude by commenting on specific cases. 
        \begin{itemize}
            \item When $m=2$, we have $P_2 = Z$, so we recover Example~\ref{example1}.
            \item The case $m=3$ corresponds to $G = S_3$, the permutation group on 3 elements. In this case, and in fact whenever $m$ is odd, the cocycle $\omega$ is trivial. We can remove the $\U(1)$ phase factor in Eq.~\eqref{eq: XP commutation} by replacing $P_m$ with $\lambda P_m$, where $\lambda^2 = e^{-i \frac{2\pi}{m}}$. Crucially, since $m$ is odd, we can choose $\lambda = e^{i \frac{2\pi}{m} \frac{m-1}{2}}$ which is still an $m$-th root of unity, so this redefinition does not change the group structure of $\cG$. 
            \item The case $m=4$ corresponds to $G = D_4$ and $P_4$ is the single-qubit phase gate $S$. The redefinition discussed above does not preserve the group relations within $\cG$ for even $m$, since $\frac{m-1}{2}$ is not an integer.   
        \end{itemize}
        
\end{tcbexample}

\subsection{Physical Hilbert space} \label{sec: physical Hilbert space}

We are now prepared to define an ISC from a projective irrep $\Gamma$ of a finite group $G$. We let $\omega$ denote the corresponding cocycle, as in Eq.~\eqref{eq: projective rep}. We take the dimension of the projective irrep to be $d_\Gamma$. The projective irrep may be derived from an irreducible finite group of gates $\cG$, as described in the previous section. Before getting started, we point out that ISCs are quantum double models for the group $G$, with specific boundary conditions determined by the projective irrep. The language of quantum double models is not essential to follow the text, but we refer to Refs.~\cite{Kitaev2003computationbyanyons,beigi2011boundaries,cong2016boundaries,cui2018topological,lo2025universal} for background.

\begin{figure}[tb]
        \centering
        \includegraphics[width=0.4\textwidth]{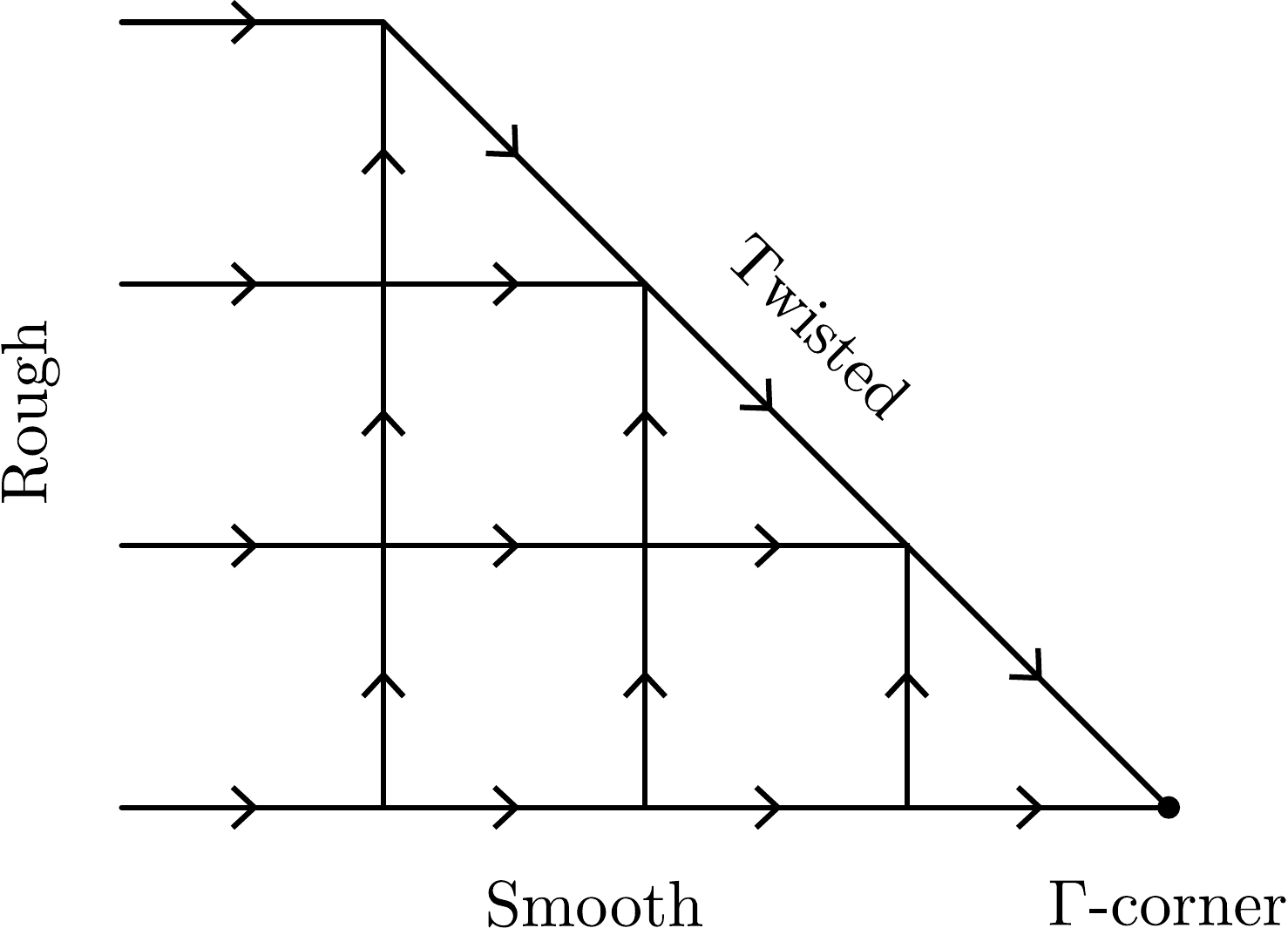}
    \caption{Lattice for the ISC. The ISC is defined on a triangular patch with three different boundary conditions: rough, smooth, and twisted. These are distinguished by their associated stabilizers. The edges are assigned an orientation, which is used in the definition of the stabilizers.}
    \label{fig: setup}
\end{figure}

The ISC is defined on a triangular patch with three types of boundaries, labeled by rough, smooth, and twisted. The distinction between these boundaries will become clear once the stabilizers are introduced in the next section. We also assign an orientation to each edge to arrive at the setup shown in Fig.~\ref{fig: setup}.

The physical Hilbert space is built by placing a $|G|$-dimensional qudit at every edge and a $d_\Gamma$-dimensional qudit in the bottom-right corner.\footnote{We point out that the $|G|$- and $d_\Gamma$-dimensional qudits can always be taken to be a $|G|$- or $d_\Gamma$-dimensional subspaces of a suitable number of qubits.} We refer to this as the $\Gamma$-corner, or simply ``the corner'', when it is clear from context. Since the dimension of a qudit on an edge is $|G|$, it admits a basis labeled by the elements of $G$. We index a basis for the $d_\Gamma$-dimensional space in the $\Gamma$-corner by $i$. A basis for the full Hilbert space can then be labeled as $\ket{\bm{g_e},i}$, where $\bm{g_e}$ denotes a choice of group element for each edge, and $i$ labels the basis state at the corner. We refer to this as the group basis, assuming there is a natural choice of basis for the $d_\Gamma$ qudit.

In the next section, we define the code space by specifying a generating set of stabilizers. For this purpose, it is convenient to introduce the following operators. We first define the operators $L^g_e$ and $R^g_e$, whose action on the edge qudit at $e$ can be understood through group multiplication. More specifically, for an arbitrary $g,h \in G$ and edge $e$, the operators $L^g_e$ and $R^g_e$ act on a group-labeled state $\ket{h}$ as:
\begin{align}
    L^g_e |h\rangle = |gh\rangle, \qquad R^g_e|h\rangle = |h\bar{g}\rangle,
\end{align}
where $\bar{g}$ is the inverse of $g$.
These operators form a representation of $G$ at the edge $e$, since they satisfy:
\begin{align}
    L_e^gL_e^h=L_e^{gh}, \qquad R_e^gR_e^h=R_e^{gh}.
\end{align}
They are also mutually commuting, for any $g,h \in G$, we have:
\begin{align}
    L_e^gR_e^h=R_e^hL_e^g.
\end{align}
It is useful to further introduce twisted versions of these operators, denoted $\tilde{L}_e^g$ and $\tilde{R}_e^g$. The twisted left and right group multiplication operators are defined by:
\begin{align}
    \tilde{L}^g_e |h\rangle = \omega(g,h)|gh\rangle, \qquad \tilde{R}^g_e|h\rangle = \omega(h\bar{g},g)^{-1}|h\bar{g}\rangle,
\end{align}
where the phase $\omega(g,h)$ is determined by the projective irrep, as in Eq.~\eqref{eq: projective rep}. 
Using the condition in Eq.~\eqref{eq: cocycle condition}, these operators satisfy:
\begin{align} \label{eq: projective representation L and R}
    \tilde{L}^g_e\tilde{L}^h_e=\omega(g,h)\tilde{L}^{gh}_e, \qquad \tilde{R}^g_e\tilde{R}^h_e=\omega(g,h)^{-1}\tilde{R}^{gh}_e.
\end{align}
It can further be checked using Eq.~\eqref{eq: cocycle condition} that the twisted left and right group multiplication operators are mutually commuting, i.e., for arbitrary $g,h \in G$, we find:
\begin{align}
    \tilde{L}^g\tilde{R}^h=\tilde{R}^h\tilde{L}^g.
\end{align}
For a mathematical interpretation of these operators, we refer to Appendix~\ref{app: irrep review}.

\subsection{Code space} \label{sec: code space}

We define the code space by specifying a generating set of stabilizers. There are two types of stabilizers, associated to the vertices and the plaquettes. Apart from the vertex stabilizers at the $\Gamma$-corner, which we treat separately, the vertex stabilizers are products of the group multiplication operators around the vertex. For an arbitrary $g\in G$, they are graphically represented as follows, depending on the location of $v$ in the lattice:
\begin{align} \label{eq: vertex stabilizers}
    A_v^g =
    \underbrace{\vphantom{\rule[-4em]{1pt}{4em}}\vcenter{\hbox{\includegraphics[scale=.25]{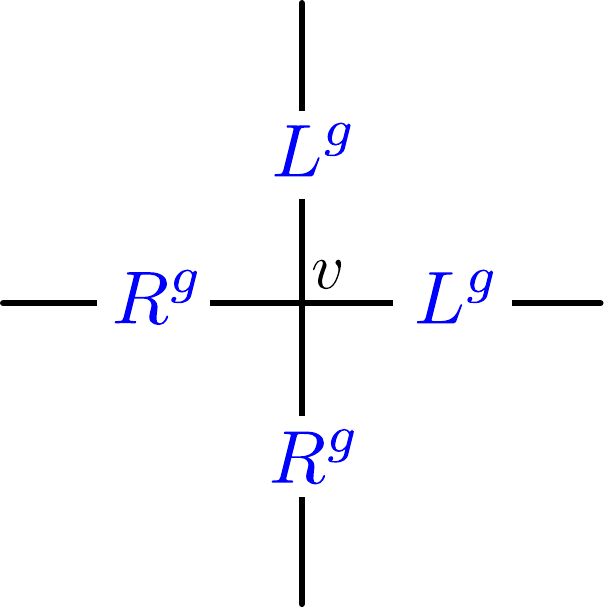}}}}_{\text{Bulk and rough boundary}}, \qquad 
    \underbrace{\vphantom{\rule[-4em]{1pt}{4em}}\vcenter{\hbox{\includegraphics[scale=.25]{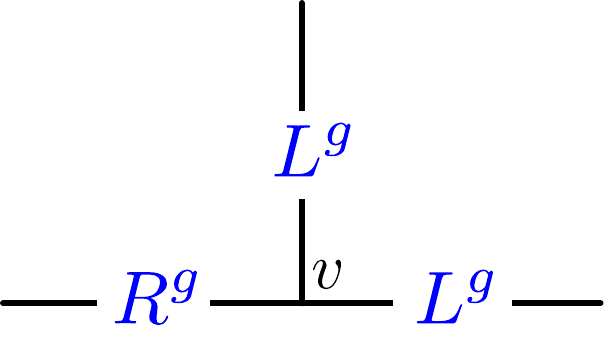}}}}_{\text{Smooth boundary}}, \qquad 
    \underbrace{\vphantom{\rule[-4em]{1pt}{4em}}\vcenter{\hbox{\includegraphics[scale=.25]{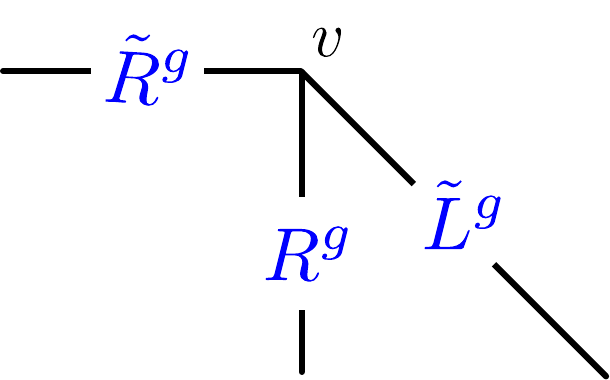}}}, \quad \vcenter{\hbox{\includegraphics[scale=.25]{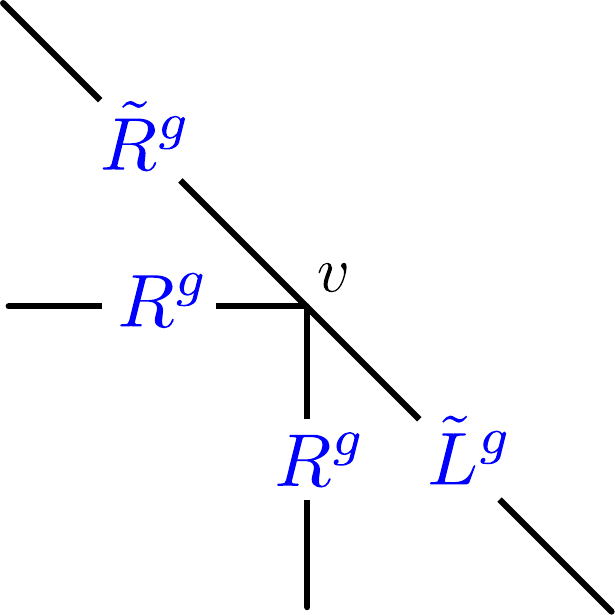}}}}_{\text{Twisted boundary}}.
\end{align}
Here, we use the convention that $L^g$ and $\tilde{L}^g$ act on the edges pointing away from the vertex, while $R^g$ and $\tilde{R}^g$ act on the edges pointing toward the vertex. Note that the twisted group multiplication operators only act on the twisted boundary.

The vertex stabilizers in the corner are modified to account for the qudit in the $\Gamma$-corner. For an arbitrary $g \in G$, we define the vertex stabilizer in the corner as:
\begin{align}
    A_c^g =\underbrace{\vphantom{\rule[-2em]{1pt}{2em}}\vcenter{\hbox{\includegraphics[scale=.25]{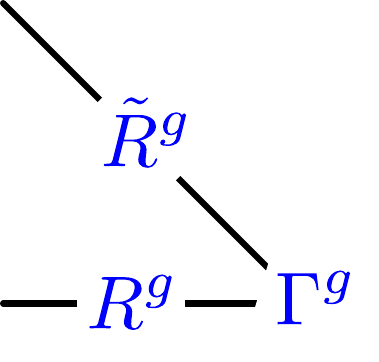}}}}_{\text{Corner}}.
\end{align}
Here, $\Gamma^g$ is the representation of $g$ determined by the input projective irrep. It is important to include $\Gamma^g$ in the definition of the corner stabilizer so that it satisfies the relation below, in Eq.~\eqref{eq: vertex G representation}. In Appendix~\ref{app: isc from gsc}, we provide intuition for the corner stabilizer by giving a heuristic derivation of the ISC from a group surface code, as defined in Ref.~\cite{manjunath2026GSC}. 

We next consider the relations of the vertex stabilizers. Due to the fact the left and right multiplication commute, the vertex stabilizers at the vertices $v$ and $w$, with $v\neq w$, are mutually commuting, i.e., for all $g,h \in G$, we have:
\begin{align}
    A_v^gA_w^h=A_w^hA_v^g.
\end{align}
The vertex stabilizers $A_v^g$ and $A_v^h$ at the same vertex, however, need not commute. In general, the vertex stabilizers form a representation of the group $G$ at each vertex:
\begin{align} \label{eq: vertex G representation}
    A_v^gA_v^h=A_v^{gh}.
\end{align}
Importantly, this holds for the vertices on the twisted boundary and at the $\Gamma$-corner. The phases appearing from Eqs.~\eqref{eq: projective rep} and \eqref{eq: projective representation L and R} cancel to ensure that the vertex stabilizers indeed form a representation of $G$.

The plaquette stabilizers, on the other hand, are diagonal in the group basis and are defined by: 
\begin{equation}
\begin{gathered}
    \underbrace{\vphantom{\rule[-2.5em]{1pt}{2.5em}}B_p \Bigggvert \,\vcenter{\hbox{\includegraphics[scale=.25]{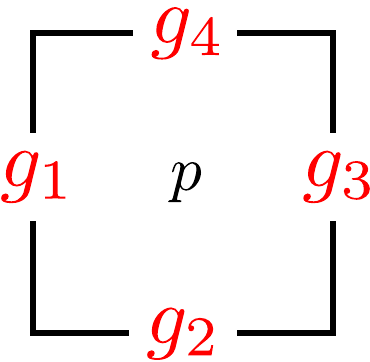}}} \,\Bigggrangle = \delta_{\bar{g}_1g_2g_3\bar{g}_4,1}\Bigggvert \,\vcenter{\hbox{\includegraphics[scale=.25]{Figures/Bpbulk.pdf}}}\, \Bigggrangle}_{\text{Bulk and smooth boundary}}, \qquad 
    \underbrace{\vphantom{\rule[-2.5em]{1pt}{2.5em}}B_p \Bigggvert \,\,\vcenter{\hbox{\includegraphics[scale=.25]{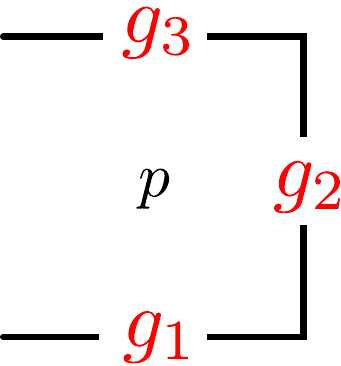}}} \,\Bigggrangle= \delta_{g_1g_2\bar{g}_3,1}\Bigggvert \,\,\vcenter{\hbox{\includegraphics[scale=.25]{Figures/Bprough.pdf}}}\, \Bigggrangle}_{\text{Rough boundary}}, \\[1em]
    \underbrace{\vphantom{\rule[-2.5em]{1pt}{2.5em}}B_p \Bigggvert \,\vcenter{\hbox{\includegraphics[scale=.25]{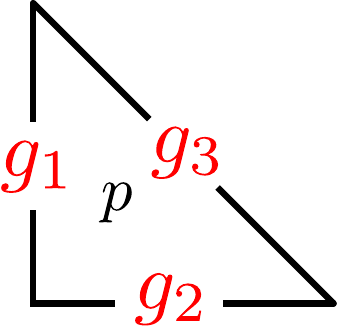}}}\, \Bigggrangle= \delta_{\bar{g}_1g_2\bar{g}_3,1}\Bigggvert \,\vcenter{\hbox{\includegraphics[scale=.25]{Figures/Bptwisted.pdf}}}\, \Bigggrangle}_{\text{Twisted boundary}},
\end{gathered}
\end{equation}
where the figures only show a portion of the configuration of group elements in the group basis. The delta functions are determined by taking a counterclockwise product of the group elements around a plaquette, with the caveat that, if the orientation of the edge is clockwise, then we multiply by the inverse of the group element. We note that the starting point for the product of group elements does not matter, by virtue of the fact that the product is set to $1$. 

The plaquette stabilizers commute with each other, since they are diagonal in the group basis. It can also be checked that they commute with the vertex stabilizers. For arbitrary plaquettes $p$ and $q$, vertex $v$, and group element $g \in G$, we therefore have:
\begin{align}
    B_pB_q=B_qB_p, \qquad B_pA_v^g = A_v^g B_p.
\end{align}
With this, we have defined a generating set of stabilizers, and we take the code space to be their mutual $+1$ eigenspace:
\begin{align}
    \cH_C = \{ \ket{\psi} \, : \, A_v^g\ket{\psi} = B_p \ket{\psi} = \ket{\psi}, \, \forall v,p,g \}.
\end{align}
As argued below, the dimension of this space is $d_\Gamma$. We show this by constructing an explicit basis for the code space. 

\subsubsection{Code space dimension}
To get started, we consider the set of group basis states $\{\ket{\bm{1_e}, i}\}_{i=1}^{d_\Gamma}$. These are the group basis states with the identity $1 \in G$ at every edge and a basis state of the $\Gamma$-corner qudit indexed by $i$. They are certainly in the $+1$ eigenspace of the plaquette stabilizers, but they are not eigenstates of the vertex stabilizers. Therefore, to form code states, we project them to the mutual $+1$ eigenspace of the vertex stabilizers. Letting $\Pi_A$ be the projector onto this subspace, we then consider the set of states $\{\Pi_A\ket{\bm{1_e}, i}\}_{i=1}^{d_\Gamma}$ and argue that they, in fact, form a basis for the logical space.

To establish that the states $\{\Pi_A\ket{\bm{1_e}, i}\}_{i=1}^{d_\Gamma}$ form a basis, we need to show that they are mutually orthogonal and span the code space. To make this explicit, we begin by expanding the projector $\Pi_A$. First, we define the projector onto the $+1$ eigenspace of the vertex stabilizers at a single vertex $v$:
\begin{align}
    A_v^{G}=\frac{1}{|G|}\sum_{g \in G}A_v^g.
\end{align}
By changing the summation, this operator satisfies, for all $g \in G$:
\begin{align} \label{eq: absorbing gauge transformations}
    A_v^GA_v^G=A_v^G, \qquad A_v^g A_v^G=A^G_vA_v^g=A_v^G.
\end{align}
With this, the projector $\Pi_A$ can be written as:

\begin{align} \label{eq: PiA decomposition}
    \Pi_A= \prod_vA_v^G=  \frac{1}{|G|^{N_v}}\prod_v\sum_{g_v}A_v^{g_v},
\end{align}
where $N_v$ is the number of vertices.

We are now able to show that the states in $\{\Pi_A\ket{\bm{1_e}, i}\}_{i=1}^{d_\Gamma}$ are orthogonal. Consider the states indexed by $i$ and $j$ with $i\neq j$. Their inner product is then:
\begin{align}
    \langle \bm{1_e}, j|\Pi_A\Pi_A|\bm{1_e},i\rangle =\frac{1}{|G|^{N_v}} \bigg \langle \bm{1_e}, j\bigg |\prod_v\sum_{g_v}A_v^{g_v}\bigg |\bm{1_e},i\bigg \rangle,
\end{align}
where we have used that $\Pi_A$ is a projector and inserted the expression in Eq.~\eqref{eq: PiA decomposition} for $\Pi_A$. The only choice of group elements ${g_v}$ that maps the configuration of trivial group elements $\bm{1_e}$ back to itself is $g_v=1$ for every vertex $v$. Therefore, the inner product reduces to:
\begin{align}
    \langle \bm{1_e}, j|\Pi_A\Pi_A|\bm{1_e},i\rangle = \frac{1}{|G|^{N_v}}\langle \bm{1_e}, j|\bm{1_e},i\rangle = 0,
\end{align}
where in the last equality we have used the orthogonality of the states labeled by $i$ and $j$.

What remains to show is that the set of states $\{\Pi_A\ket{\bm{1_e}, i}\}_{i=1}^{d_\Gamma}$ spans the code space. To accomplish this, we consider the group basis states that are $+1$ eigenstates of the plaquette stabilizers. Following Ref.~\cite{manjunath2026GSC}, we refer to these as flux-free states.\footnote{This name is motivated by an interpretation of the plaquette stabilizers in terms of lattice gauge theory, where they enforce zero flux through a plaquette.} They span the mutual $+1$ eigenspace of the plaquette stabilizers, which includes the code space. Therefore, all that we need to show is that an arbitrary flux-free state $\ket{\bm{g_e},j}$, when projected to the $+1$ eigenspace of the vertex stabilizers by $\Pi_A$, is in the span of the states $\{\Pi_A\ket{\bm{1_e}, i}\}_{i=1}^{d_\Gamma}$.

\begin{figure}[tb]
        \centering
        \includegraphics[width=.7\textwidth]{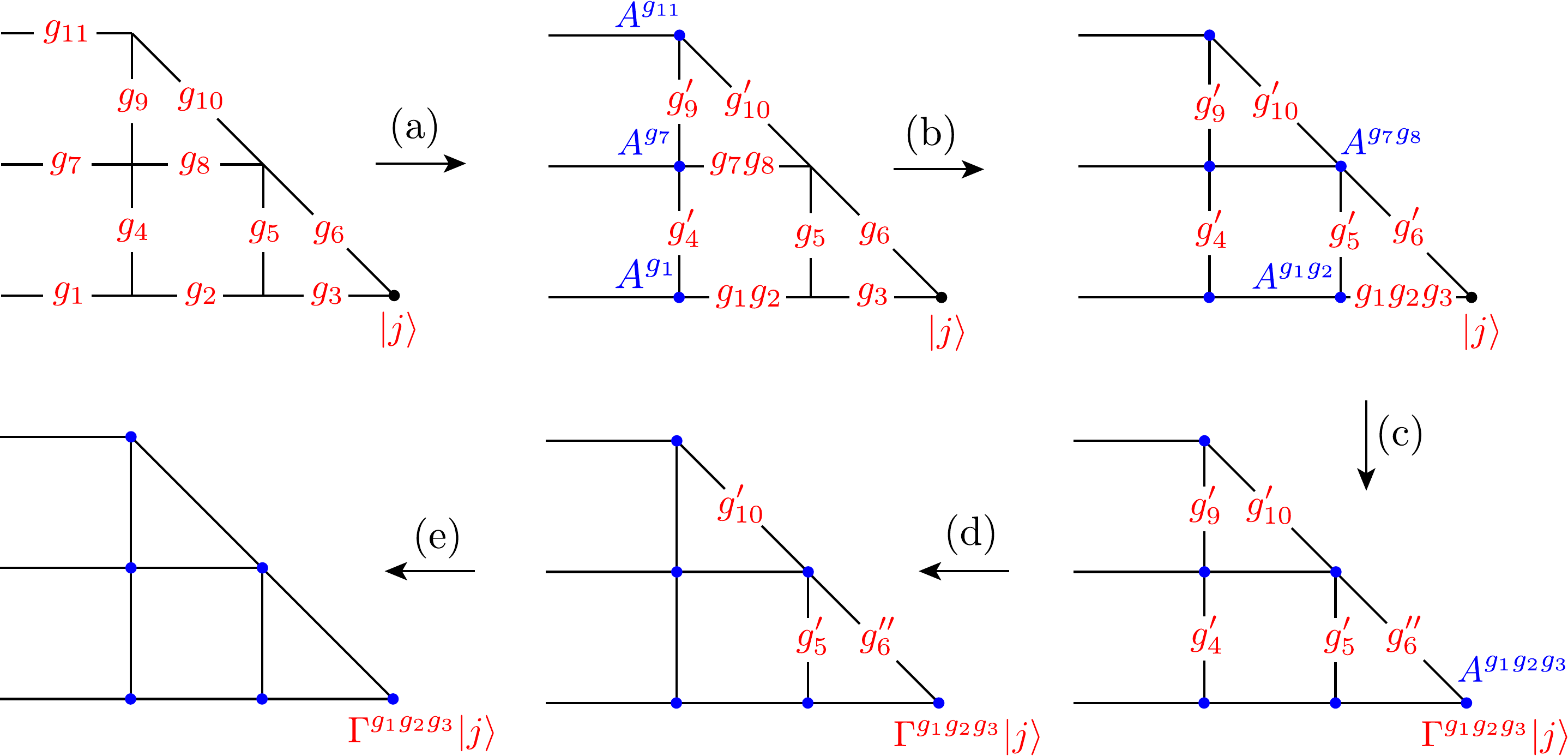}
    \caption{Mapping a flux-free state to the corner gauge. (a) Starting with an arbitrary flux-free state, we apply vertex stabilizers to map the leftmost edges to the identity. This produces group elements labeled with primes on the connected edges. (b) We apply vertex stabilizers on the next column of vertices to map the horizontal edges on their left to the identity element. (c) After removing the group elements from the horizontal edges, we apply the corner stabilizer. This enacts a representation operator on the $\Gamma$-corner. (d) Next, we use that the configuration is a flux-free state to determine that the primed group elements must be the identity. (e) We continue to use the flux-free condition to conclude that all of the edges have been put into the state $\ket{\bm{1_e}}$. This leaves us in the corner gauge.}
    \label{fig: cornergauge}
\end{figure}

We first observe that there is a product of vertex stabilizers that maps $\ket{\bm{g_e},j}$ to a state with configuration $\bm{1_e}$ on the edges; explicitly:
\begin{align} \label{eq: going to the corner gauge}
    \prod_vA_v^{k_v}\ket{\bm{g_e},j}=\ket{\bm{1_e},\varphi},
\end{align}
for some state $|\varphi\rangle$ on the $d_\Gamma$-corner qubits. We show an algorithm for determining the vertex-dependent product of vertex stabilizers in Fig.~\ref{fig: cornergauge}. Effectively, the group variables are swept into the $\Gamma$-corner, where a representation operator may act on the corner qudit. 

Explicitly projecting the flux-free state $\ket{\bm{g_e},j}$ to the $+1$ eigenspace of the vertex stabilizers, we find:
\begin{eqs}
    \Pi_A \ket{\bm{g_e},j} 
    = \Pi_A\prod_vA_v^{\bar{k}_v}\ket{\bm{1_e},\varphi} =\Pi_A \ket{\bm{1_e},\varphi}.
\end{eqs}
Here, we have used the transformation in Eq.~\eqref{eq: going to the corner gauge} to obtain the first equality and the relation in Eq.~\eqref{eq: absorbing gauge transformations} for the second equality. The state $\ket{\varphi}$ can then be decomposed using the basis of the corner qubits, for some coefficients $c_i$:
\begin{align}
    \Pi_A \ket{\bm{1_e},\varphi}=\sum_{i}c_i \Pi_A \ket{\bm{1_e},i}.
\end{align}
Therefore, we see that $\Pi_A \ket{\bm{g_e},j}$ is in the span of the states $\{\Pi_A\ket{\bm{1_e}, i}\}_{i=1}^{d_\Gamma}$:
\begin{align}
    \Pi_A \ket{\bm{g_e},j}=\sum_{i}c_i \Pi_A \ket{\bm{1_e},i}.
\end{align}
We conclude that the states in $\{\Pi_A\ket{\bm{1_e}, i}\}_{i=1}^{d_\Gamma}$ form a basis for the code space, and thus the code space has dimension $d_\Gamma$. To simplify the notation for the logical basis states, we define the logical state $\ket{i}_\LL$ as:
\begin{align}
    \ket{i}_\LL= \Pi_A\ket{\bm{1_e}, i}.
\end{align}
We give an independent derivation of the dimension of the code space in Appendix~\ref{app: minimal isc}. There we use circuits to sequentially disentangle the ISC into a minimal ISC. We can then immediately see that the dimension of the code space is $d_\Gamma$.

When an operation preserves the $+1$ eigenspace of the vertex stabilizers, as for the logical operators in the next section, we find it convenient to use the group basis states $\{\ket{\bm{1_e}, i}\}_{i=1}^{d_\Gamma}$ to represent the logical states. Using language from lattice gauge theory, we refer to this as the corner gauge. We caution that the corner gauge makes it appear as though the logical information is encoded in the corner qubits and susceptible to local errors. We prove in the next section that this is not the case by establishing the Knill-Laflamme condition for local errors.

\subsection{Logical operators} \label{sec: logical operators}

In this section, we describe two types of transversal logical operators for the ISC. The first set of logical operators implement the representation operators $\Gamma^g$ on the logical level. Therefore, if the projective irrep is derived from an irreducible group of gates $\cG$, these operators allow us to perform the gates in $\cG$, up to a global phase. We then describe deformations of the logical operators, which allow us to prove that the code distance of the ISC is exactly $d$, the linear size of the system. The second set of transversal logical operators correspond to automorphisms of $G$ that map the projective irrep to one that is equivalent, in the sense of Eq.~\eqref{eq: projective equivalence}. 

\subsubsection{Representation operators} \label{sec: logical representation operators}

Let us begin with the implementation of the representation operators $\Gamma^g$. We denote the corresponding logical operator for $g\in G$ as $\Gamma^g_\LL$. It is represented graphically as:
\begin{align}\label{eq: logical representation operator}
    \Gamma_\LL^g = \quad \vcenter{\hbox{\includegraphics[scale=.25]{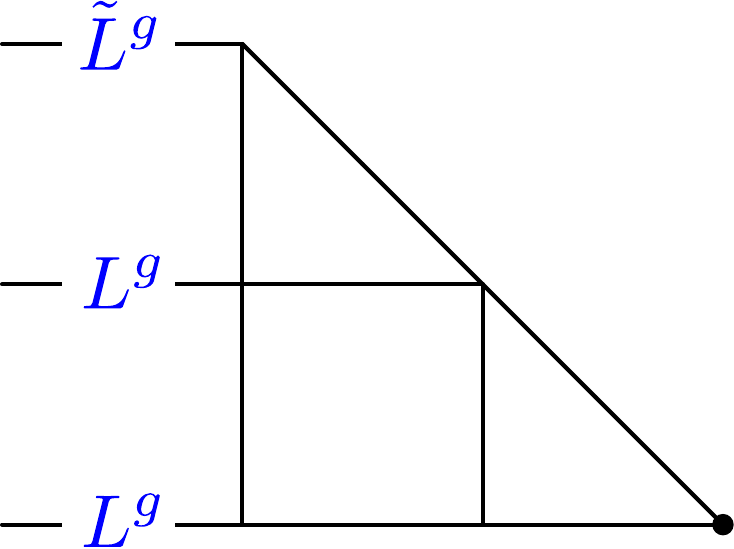}}}\!\!\!\!\!\!.
\end{align}
In words, we apply left multiplication operators on the rough edges, with the twisted left multiplication applied to the top edge. The first thing to note is that it commutes with all of the stabilizers. It commutes with the vertex stabilizers, due to the commutativity of left and right multiplication. It also commutes with the plaquette stabilizers, as can be seen by direct computation. 

To determine the effect of $\Gamma_\LL^g$ on a logical state $\ket{j}_\LL$, we make use of the corner gauge. This is valid, since $\Gamma_\LL^g$ preserves the $+1$ eigenspace of the vertex stabilizers. Acting with $\Gamma_\LL^g$ on the corner gauge state $\ket{\bm{1_e},j}$ gives us:
\begin{align}
    \Gamma_\LL^g:\quad \vcenter{\hbox{\includegraphics[scale=.25]{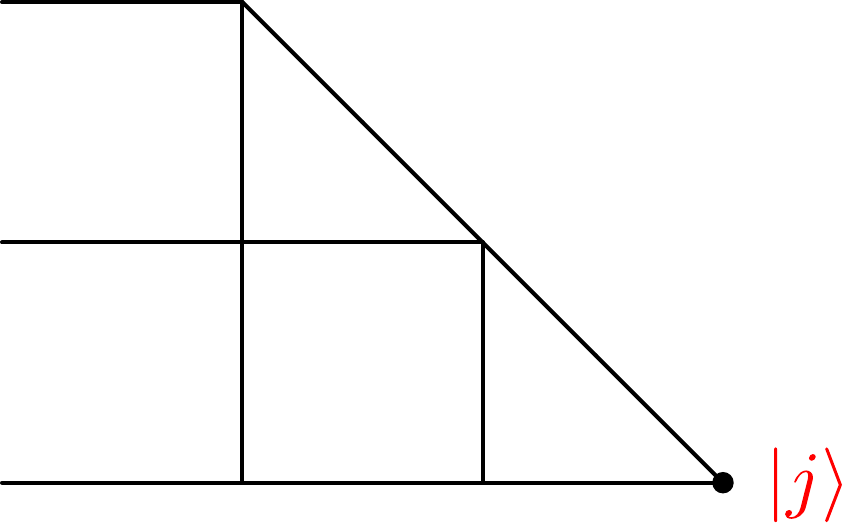}}} \mapsto \quad \vcenter{\hbox{\includegraphics[scale=.25]{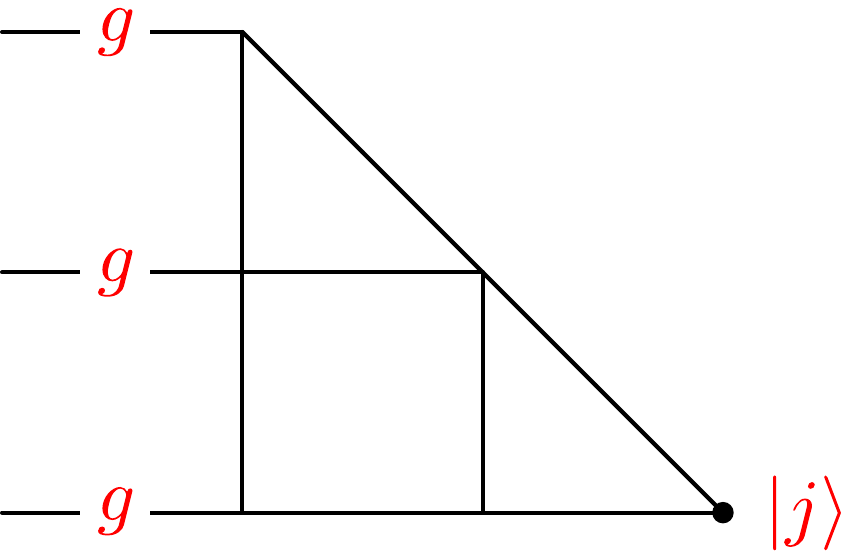}}}\!\!\!\!\!\!\!\!\!\!\!\!\!\!\!\!.
\end{align}
Applying vertex stabilizers to shuffle the group elements to the right, we obtain:
\begin{align}
    \vcenter{\hbox{\includegraphics[scale=.25]{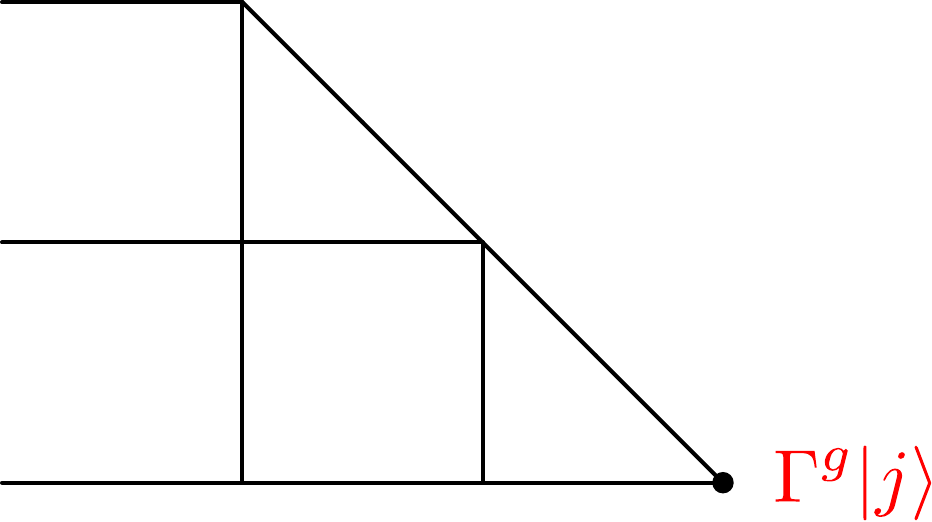}}}\!\!\!\!\!\!\!\!\!\!\!\!\!\!\!\!,
\end{align}
where the $\Gamma^g$ is applied to the state $\ket{j}$ in the corner in the last step by the corner stabilizer $A_c^g$.
Therefore, we see that $\Gamma_\LL^g$ indeed acts as $\Gamma^g$ on the logical states and maps $\ket{j}_\LL$ as:
\begin{align}
    \boxed{\Gamma_\LL^g \ket{j}_\LL = \sum_i \Gamma^g_{ij}|i\rangle_\LL.}
\end{align}
In Appendix~\ref{app: line-like logical}, we construct alternative representations for a logical $\Gamma^g$, by moving $\Gamma_\LL^g$ into the bulk. We find that, in the code space, $\Gamma_\LL^g$ can be deformed to an operator depicted schematically as:
\begin{align}\label{eq: deformed logicals}
    \vcenter{\hbox{\includegraphics[scale=.15]{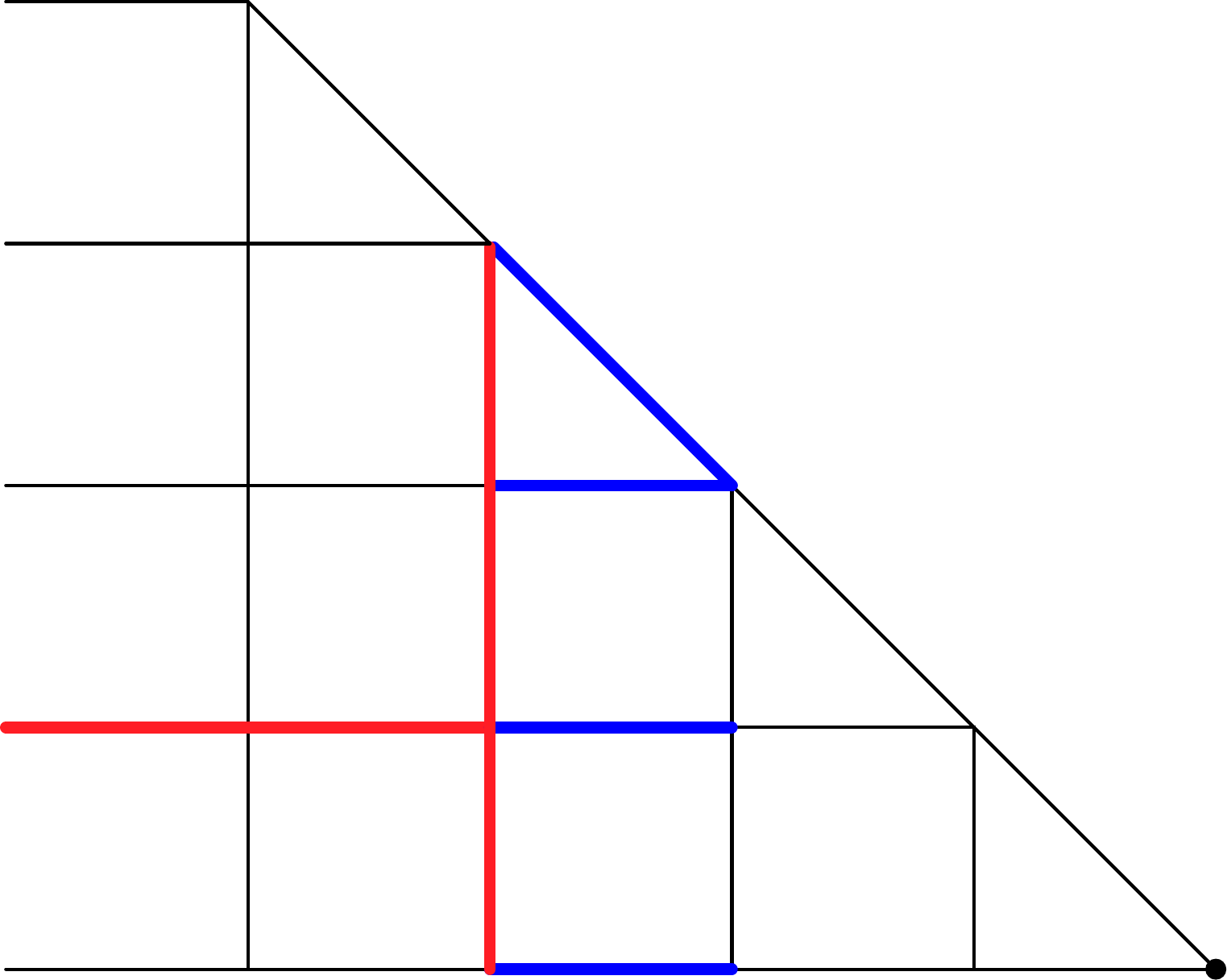}}}\!\!\!\!\!\!\!\!\!\!\!\!\!\!\!\!.
\end{align}
Here, the blue edges represent group multiplication operators applied in a column. The multiplications are controlled on certain products of group elements supported on the red edges. Such logical operators can be defined for any choice of column and row of edges. 

The deformability of the logical operators allows us to establish that the code distance is exactly $d$, where $d$ is the linear size of the triangular patch. To make this explicit, we note that linear combinations of the logical representation operators generate the full matrix algebra on the logical space, as reviewed in Appendix~\ref{app: irrep review}. This puts strong constraints on the allowed supports of logical operators. For example, suppose an operator is supported on the complement of the rough boundary edges. Then it must commute with $\Gamma^g_\LL$ in Eq.~\eqref{eq: logical representation operator} for every $g \in G$. This means that, if it were a logical operator, it would have to commute with the full matrix algebra of logical operators. According to Schur's lemma, this is only possible if its action is proportional to the identity on the logical space. A similar conclusion holds for any of the deformations of $\Gamma^g_\LL$, as pictured in Eq.~\eqref{eq: deformed logicals}.

We now suppose that an error $E$ is supported on $d-1$ edges. Since the triangular patch has dimensions $d \times d$, there must be a column and row of edges that avoids the support of $E$. According to Appendix~\ref{app: line-like logical}, we can deform $\Gamma^g_\LL$ to find a representation whose support is non-overlapping with $E$. Therefore, the action of $E$ on the logical subspace must be proportional to the identity, showing that it is a detectable error or a stabilizer. This tells us that the code distance is lower bounded by $d$. On the other hand, since the logical operator $\Gamma_\LL^g$ has a weight of $d$, we conclude that the code distance of the ISC is exactly $d$.

We point out that, for non-Abelian groups, the deformations of $\Gamma^g_\LL$ are generally not transversal. This is due to the product of group elements along the red edges in Eq.~\eqref{eq: deformed logicals}. However, if $h \in G$ belongs to the center of $G$, meaning that it commutes with every element of $G$, then the deformation of $\Gamma^h_\LL$ to the smooth boundary is transversal. We discuss this in more detail in Appendix~\ref{app: line-like logical}. 

We conclude the discussion of logical representation operators with an example. We emphasize that $\Gamma_\LL^g$ allows for the implementation of logical gates that are not only non-Clifford but beyond any finite level of the Clifford hierarchy, as in the example below.

\begin{tcbexample}[$G=A_5$]\label{example3}
\vspace{.2cm}
The alternating group $A_5$ is the group of even permutations of 5 objects, with the standard presentation:
\begin{equation}
    A_5=
\langle a,b \mid a^5=b^2=(ab)^3=1\rangle.
\end{equation}
In the usual cycle notation of permutations,
\begin{equation}
    a=(1\,2\,3\,4\,5),
\quad
b=(1\,2)(3\,4).
\end{equation}
The group $A_5$ admits a two-dimensional projective irrep, with the representation generators\pagefootnotemark
\begin{equation}
    \Gamma^{a} = \frac{1}{2}
    \begin{pmatrix}
    \varphi-i\varphi^{-1} & i\\
    i & \varphi+i\varphi^{-1}
    \end{pmatrix}, \quad
    \Gamma^{b} = X.
\end{equation}
Here, $\varphi$ is the golden ratio $\varphi=\frac{1+\sqrt{5}}{2}$. One can check that these matrices obey the relations
\begin{equation}
    (\Gamma^a)^5=-I, \quad (\Gamma^b)^2
    =I, \quad (\Gamma^a \Gamma^b)^3=iI,
\end{equation}
establishing that they reproduce the group multiplication of $A_5$ up to global phases. Note that Pauli $Z$ is generated by $\Gamma^a$ and $X$: we have $Z = ((\Gamma^a)^2 X)^2 (\Gamma^a)^{-1} X \Gamma^a$.

The ISC defined by this projective irrep has a Hilbert space of dimension $|A_5|=60$ at each edge and a single qubit at the $\Gamma$-corner. The logical representation operators in Eq.~\eqref{eq: logical representation operator} implement the gates generated by $\Gamma^a$ and $\Gamma^b$. The gate $\Gamma^a$, in particular, is beyond any finite level of the Clifford hierarchy. This can be seen by conjugating $\Gamma^a$ by $SX$, where $S$ is the phase gate:
\begin{align}
    S=
    \begin{pmatrix}
        1&0\\
        0&i
    \end{pmatrix}.
\end{align}
This produces the gate $\Phi$ in Ref.~\cite{teixeira2023transversal}, which is proven to be outside of any finite level of the Clifford hierarchy. We also comment that Ref.~\cite{teixeira2023transversal} proposed a family of codes with distance $d=3$ admitting transversal implementations of $\Gamma^a$. It would be interesting to study the relation between their codes and ISCs. 

\end{tcbexample}
\pagefootnotetext{Note that this only specifies the projective irrep up to phases for the other elements of $A_5$.}

\subsubsection{Projective intertwiners} \label{sec: logical projective intertwiners}

The second class of transversal gates arises from projective intertwiners, defined in Eq.~\eqref{eq: projective equivalence}. More specifically, we consider projective intertwiners between $\Gamma$ and $\Gamma \circ \phi$, where $\phi$ is some automorphism of $G$. These are unitary operators $W$ satisfying:
\begin{align} \label{eq: allowed automorphisms}
    W\Gamma^g W^\dagger  = \beta(g)\Gamma^{\phi(g)},
\end{align}
for all $g \in G$ and some $g$-dependent phase $\beta(g)$. As shown below, this condition is necessary to ensure that the transversal gate preserves the set of vertex stabilizers on the twisted boundary and at the $\Gamma$-corner. It is also important to note that $W$ is not uniquely determined by the automorphism $\phi$. Moreover, as illustrated in the examples below, there can be nontrivial logical operators that correspond to the trivial automorphism. 

\begin{graybox}
\noindent \textbf{Remark:} The unitary $W$ has a particularly natural interpretation when the projective irrep is derived from an irreducible group of gates $\cG$. In this case, Eq.~\eqref{eq: allowed automorphisms} is equivalent to requiring that $W$ preserves $\tilde{\cG}$ under conjugation, where $\tilde{\cG}$ is the $\U(1)$ completion of $\cG$ defined in Eq.~\eqref{eq: U1 completion}:
\begin{align}
    W\tilde{\cG}W^\dagger = \tilde{\cG}.
\end{align}
The condition in Eq.~\eqref{eq: allowed automorphisms} immediately implies that $W$ preserves $\tilde{\cG}$. Conversely, any unitary $W$ that preserves $\tilde{\cG}$ under conjugation induces an automorphism of the quotient group $G$, since conjugation by $W$ leaves global phases unchanged. By construction, $W$ and the automorphism satisfy Eq.~\eqref{eq: allowed automorphisms}.
\end{graybox}

We now define a transversal logical operator $W_\LL$ implementing the operator $W$ on the logical space. We first define single-edge unitaries based on the automorphism $\phi$ and the phases $\beta(g)$. We let $U_e^\phi$ be the unitary operator that enacts the automorphism on the group basis state at edge $e$:
\begin{align}
    U_e^\phi |g\rangle = |\phi(g)\rangle.
\end{align}
We also define an operator $V_e^{\beta}$ that multiplies by the $g$-dependent phase $\beta(g)$:
\begin{align}
    V^{\beta}_e |g\rangle = \beta(g) |g\rangle.
\end{align}
Finally, we define the operator $W_c$ on the $\Gamma$-corner, which acts as $W$:
\begin{align}
    W_c |j\rangle = \sum_i W_{ij}|i\rangle.
\end{align}
This allows us to define the logical operator $W_\LL$ as:
\begin{align} \label{eq: W logical operator}
    W_\LL = \Bigg [ \prod_{e} U^\phi_e \Bigg ] \Bigg [ \prod_{e \in {E}_\text{twist}}  V^\beta_e \Bigg ] W_c,
\end{align}
where ${E}_\text{twist}$ denotes the set of edges along the twisted boundary.

To argue that $W_\LL$ is a logical operator, we show that it preserves the stabilizer group under conjugation. First, the group of vertex stabilizers is preserved, since they are mapped according to:
\begin{align} \label{eq: transformations of Avg}
    W_\LL A_v^g W_\LL^\dagger = A_v^{\phi(g)}.
\end{align}
This can be seen by noting the following transformations at the edges for the group multiplication operators:
\begin{align}
    U_e^\phi L^g_e \left( U^\phi_e \right )^\dagger = L^{\phi(g)}, \qquad U_e^\phi R^g_e \left( U^\phi_e \right )^\dagger = R^{\phi(g)}.
\end{align}
A similar condition holds for the twisted group multiplication operators using $U^\phi_eV^\beta_e$:\footnote{These transformations can be derived by making the following observation. The product $W\Gamma^g W^\dagger W \Gamma^h W^\dagger$ can be evaluated in two ways: by performing the multiplication then using the relation in Eq.~\eqref{eq: allowed automorphisms} or vice-versa. This results in the condition: 
\begin{align}\label{eq: beta_omega_phi}
    \omega(\phi(g),\phi(h)) = \frac{\beta(gh)}{\beta(g)\beta(h)}\omega(g,h).
\end{align}}
\begin{align}
    \left (U_e^\phi V^\beta_e \right ) \tilde{L}^g_e \left( U_e^\phi V^\beta_e \right )^\dagger = \beta(g) \tilde{L}^{\phi(g)}, \qquad \left(  U_e^\phi V^\beta_e \right ) \tilde{R}^g_e \left( U_e^\phi V^\beta_e \right )^\dagger = \beta(g)^{-1}\tilde{R}^{\phi(g)}.
\end{align}
The transformation of $\Gamma^g_c$ under $W_c$ follows directly from Eq.~\eqref{eq: allowed automorphisms}. Putting it all together, we obtain the mapping in Eq.~\eqref{eq: transformations of Avg}. The commutativity of $W_\LL$ with the plaquette stabilizers follows from the fact that $\phi$ is a homomorphism, and $V^\beta_e$ is diagonal in the group basis. Therefore, $W_\LL$ preserves the stabilizer group under conjugation.

To determine the action of $W_\LL$ on the logical space, we consider its action on a state $\ket{\bm{1_e}, j}$ in the corner gauge. This gives us:
\begin{align}
    W_\LL: \quad \vcenter{\hbox{\includegraphics[scale=.25]{Figures/logicalGamma1.pdf}}} \mapsto \quad \vcenter{\hbox{\includegraphics[scale=.25]{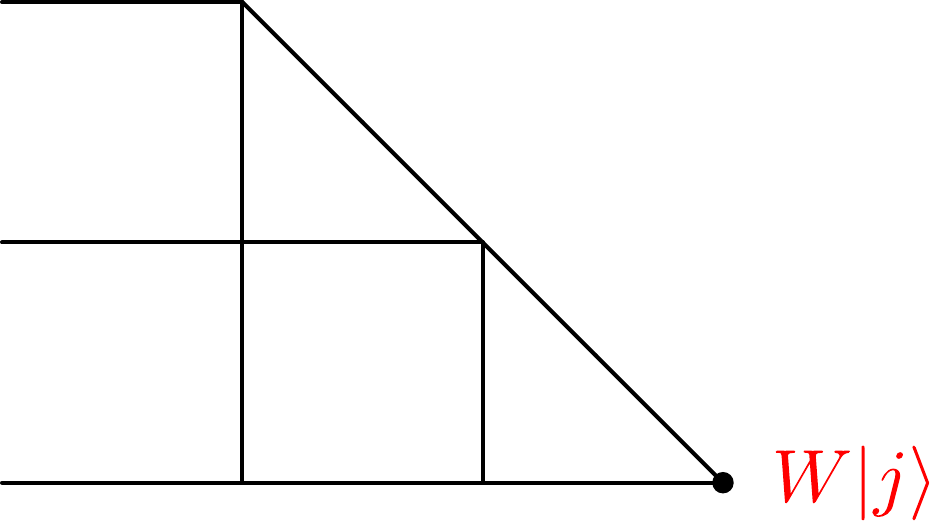}}}\!\!\!\!\!\!\!\!\!\!\!\!\!\!\!\!.
\end{align}
Therefore, $W_\LL$ acts on the code state $\ket{j}_\LL$ as:
\begin{align}
    \boxed{W_\LL \ket{j}_\LL = \sum_i W_{ij}|i\rangle_\LL.}
\end{align}

\begin{graybox}
\noindent \textbf{Remark:} Each representation operator $\Gamma^g$ defines a projective intertwiner. Indeed, conjugating by $\Gamma^g$, we find:
\begin{align}
    \Gamma^g \Gamma^h (\Gamma^g)^\dagger = \omega(g,h) \omega(gh\bar{g},g)^{-1}\Gamma^{gh\bar{g}}.
\end{align}
This implies that the phase $\beta(h)$ is:
\begin{align}
    \beta(h) = \omega(g,h) \omega(gh\bar{g},g)^{-1},
\end{align}
and the automorphism is an inner automorphism implemented by conjugation with $g$. For this special case, the logical operator $W_\LL$ in Eq.~\eqref{eq: W logical operator} can be rewritten as:
\begin{align}
    W_\LL = \Bigg [ \prod_{e} U^\phi_e \Bigg ] \Bigg [ \prod_{e \in {E}_\text{twist}}  V^\beta_e \Bigg ] W_c = \Gamma^g_\LL \prod_v A_v^g,
\end{align}
where the phases from $V^\beta_e$ are reproduced on the right-hand side by the twisted group multiplication operators. In the code space, $W_\LL$ is equivalent to $\Gamma_\LL^g$.
\end{graybox}

\begin{graybox}
\noindent \textbf{Remark:} When there is a projective intertwiner for a trivial automorphism, the associated logical operator $W_\LL$ can be represented solely on the twisted boundary:
 \begin{equation} 
    W_\LL = \Bigg [ \prod_{e \in {E}_\text{twist}} V^\beta_{e} \Bigg ] W_c.
\end{equation}
 We show in Appendix~\ref{app: line-like logical} that this can be further deformed (in the code space) to a logical operator supported solely on the smooth boundary:
 \begin{align} \label{eq: W on smooth boundary}
     W_\LL = \Bigg [ \prod_{e \in {E}_\text{smooth}} V^\beta_{e} \Bigg ] W_c,
 \end{align}
 where $E_\text{smooth}$ are the edges on the smooth boundary. We also note that, since $\phi$ is trivial, the logical operators $\Gamma^g_\LL$ and $W_\LL$ commute up to a phase determined by $\beta(g)$:
\begin{align}
    W_\LL \Gamma^g_\LL W_\LL^\dagger = \beta(g) \Gamma^g_\LL.
\end{align} 
The representation of $W_\LL$ on the smooth boundary is particularly important in Section~\ref{sec: transferring information} where it is used for smooth-smooth lattice surgery.
\end{graybox}

\begin{graybox}
    \noindent \textbf{Remark:} If the goal is to implement a logical gate $W$, we can choose a projective irrep that admits $W$ as a projective intertwiner. For example, suppose that $W$ is supported on $k$ qudits, with possibly varying dimensions. Starting with the group generated by the Pauli operators on the qudits, namely $\langle X_i,\, Z_i: i \in \{1,\ldots, k\}\rangle$, we define:
    \begin{align}
        \cG = \langle W^\ell X_iW^{-\ell},\, W^\ell Z_iW^{-\ell} : i \in \{1,\ldots, k\}, \,\,\ell \in \ZZ\rangle.
    \end{align}
    This group of gates is irreducible, because it contains the Pauli $X$ and $Z$ operators. If it is finite, then we can carry out the prescription in Section~\ref{sec: input data} to derive a projective irrep. By construction, the projective irrep will admit $W$ as a projective intertwiner. Note that, since we have included the Pauli $X$ and $Z$ operators, we can also transfer information to surface codes by Result~\ref{res: transfer}.
\end{graybox}

We finish this section with illustrative examples of $W_\LL$.

\begin{tcbexample}[$G=\Z_2 \times \Z_2$]\label{example3.5}
\vspace{.2cm}

Using the notation from Example~\ref{example1}, we denote the generators of $\Z_2 \times \Z_2$ as $x$ and $z$. We also consider the projective irrep in Example~\ref{example1} with representation operators:
\begin{align}
    \Gamma^1=I, \quad \Gamma^x=X, \quad \Gamma^z=Z, \quad \Gamma^{xz}=XZ.
\end{align}
We first note that there are projective intertwiners associated to the representation operators themselves. Indeed, taking $W=X$ as an example, we find:
    \begin{align}
        X\Gamma^xX=\Gamma^x, \quad X\Gamma^zX=-\Gamma^z, \quad X\Gamma^{xz}X=-\Gamma^{xz}.
    \end{align}
    Looking back at Eq.~\eqref{eq: allowed automorphisms}, we see that this projective intertwiner corresponds to a trivial automorphism and a nontrivial phase function $\beta$. The function $\beta$ satisfies:
    \begin{align}
        \beta(x^{a_x}z^{a_z})=(-1)^{a_z}.
    \end{align}
    Therefore, a logical $X$ can be implemented transversally; although, this was already the case, since it is a representation operator.

    For a more interesting example, we consider the single-site Clifford operators $H$ and $S$. Taking $H$ as an example, we see that it acts on the representation operators as
    \begin{align}
        H\Gamma^xH=\Gamma^z, \quad H\Gamma^zH=\Gamma^x, \quad H\Gamma^{xz}H=-\Gamma^{xz}.
    \end{align}
    Thus, $\phi$ is the nontrivial automorphism of $\Z_2 \times \Z_2$ that interchanges $x$ and $z$ and the function $\beta$ satisfies:
    \begin{align}
        \beta(x^{a_x}z^{a_z})=(-1)^{a_xa_z}.
    \end{align}
    This implies that the Hadamard gate can be implemented transversally with Eq.~\eqref{eq: W logical operator}. 
    
    Notably, this example reproduces the transversal Clifford gates of the color code, and indeed has the same underlying $\Z_2 \times \Z_2$ topological order. In this case, the $X$ and $Z$ operators can be represented on each of the three boundaries of the ISC. This agrees with the color boundaries of the color code used, for example, in Ref.~\cite{Thomsen2024colorcodesurgery}. 
\end{tcbexample}

\begin{tcbexample}[$G=D_m$]\label{example4}
\vspace{.2cm}
        
     We again consider the projective irrep of $G=D_m$ in Example~\ref{example2}. Using the notation from Example~\ref{example2}, the representation operators for the generators $s$ and $r$ are $\Gamma^s = X$ and $\Gamma^r = P_{m}$, where $P_m$ is the phase gate defined in Eq.~\eqref{eq: Pm def}. The ISC hosts a Hilbert space of dimension $|D_m|=2m$ at each edge and a qubit in the corner. There is a single logical qubit, since the dimension of the projective irrep is $d_\Gamma = 2$. 
     Here, we focus on the projective intertwiners and show the following results. 
     \begin{enumerate} 
         \item When $m = 2k+1$ is odd, there is a projective intertwiner corresponding to the Pauli $Z$ gate, which is not a representation operator.
         \item When $m = 2^k$ is a power of two, the intertwiners are generated by the phase gate $P_{2^{k+1}}$. 
     \end{enumerate}
    Further details on the projective intertwiners can be found in Appendix~\ref{app: more examples}. Note that the case $m=2$ was considered in Example~\ref{example3.5}.\\

    \noindent \underline{$m=2k + 1$:} 
     For odd $m$, there are two types of projective intertwiners. The first are given by the representation operators $\Gamma^g$, which can be implemented directly on the rough boundary. The second type of projective intertwiner are those for which the automorphism is trivial but $\beta(g)$ is a nontrivial phase. 
     For example, we can implement a transversal gate corresponding to $W = Z$, which acts as:
     \begin{align}
         Z\Gamma^sZ=-\Gamma^s, \quad Z\Gamma^rZ=\Gamma^r.
     \end{align}
     In this case, we have the phase function $\beta$ given by:
     \begin{align}
         \beta(s^a r^b) = (-1)^a.
     \end{align} 
     It is worth noting that, for odd $m$, $Z$ cannot be realized as $\Gamma^g$ for any $g \in D_m$. Nevertheless, the ISC has a logical $Z$ operator. This logical operator can be fully supported on the smooth boundary, since the automorphism is trivial (see Appendix~\ref{app: line-like logical}). This is especially useful for transferring information, as noted in Section~\ref{sec: transferring information}. We argue in Appendix~\ref{app: more examples} that the representation operators and $Z$, in fact, generate all of the projective intertwiners.\\

    \noindent \underline{$m = 2^k$:} In this case, there are again two types of projective intertwiners. In addition to the representation operators, there is a projective intertwiner given by $W=P_{2^{k+1}}$. This acts on the representation operators as:
    \begin{align}
       P_{2^{k+1}} \Gamma^s P_{2^{k+1}}^\dagger =, \quad P_{2^{k+1}}\Gamma^r P_{2^{k+1}}^\dagger=\Gamma^r.
    \end{align}
    This projective intertwiner belongs to the $(k+1)$-th level of the Clifford hierarchy~\cite{Cui2017diagonal}.  In particular, when $m = 4$, the unitary $W$ is the $T$ gate. This shows that an ISC for $D_4$ admits a transversal logical $T$ gate, which complements prior work discussing non-Clifford gates in related codes based on the $D_4$ topological order~\cite{davydova2025D4,hsin2025automorphismgaugetheorieshigher,manjunath2026GSC}.
\end{tcbexample}

\section{Transferring information} \label{sec: transferring information}

In this section, we describe how information can be transferred between an ordinary surface code and an ISC using lattice surgery~\cite{Horsman2012latticesurgery,cowtan2022algebraic,cowtan2025homologyhopfalgebrasquantum,huang2026hybridlatticesurgerynonclifford}. We note that generalized notions of lattice surgery~\cite{cross2025improvedqldpcsurgerylogical, he2025extractorsqldpcarchitecturesefficient, Williamson2026gauging} can be used to transfer information between ISCs and a wider range of CSS codes, but we restrict to the surface code for concreteness. For simplicity, we describe the construction for a single surface code and an ISC encoding a single logical qubit. The generalization to multiple qubits and qudits is straightforward.

Starting with an irreducible finite group of gates $\cG$, we consider two cases, as stated in Result~\ref{res: transfer}.
\begin{enumerate} [label=(\roman*)]
    \item[] \textbf{Case 1:} $\cG$ includes both Pauli $X$ and $Z$. 
\end{enumerate}
This ensures that the ISC has logical operators $\Gamma^x_\LL$ and $\Gamma^z_\LL$, supported on the rough boundary, that transversally implement the logical Pauli $X$ and $Z$ operators, respectively. Throughout this section, we use $x$ and $z$ to denote the elements of $G$ represented by $X$ and $Z$ Pauli operators.
\begin{enumerate} [label=(\roman*),start=2]
    \item[] \textbf{Case 2:} $\cG$ is generated by Pauli $X$ and a finite group of diagonal gates.
\end{enumerate}
If Pauli $Z$ does not belong to $\cG$, then there is no transversal implementation of the logical Pauli $Z$ operator supported on the rough boundary. However, this form of $\cG$ ensures that there is a logical $Z$ operator with a transversal representation $W_\LL$ supported on the smooth boundary, as described in Eq.~\eqref{eq: W on smooth boundary} and elaborated on in Appendix~\ref{app: line-like logical}.

The protocol for transferring information is then built from three primitives: a rough-rough lattice surgery, smooth-smooth lattice surgery, and the readout of a logical $X$ and $Z$ operator of a surface code. Here, by rough-rough, we mean that the rough boundary of the surface code is merged with the rough boundary for the ISC, likewise for smooth-smooth. We combine these elements in Section~\ref{sec: teleportation} to transfer information.

\subsection{Rough-rough surgery} \label{sec: roughrough surgery}

\begin{figure*}[t]
        \centering
        \includegraphics[scale=.15]{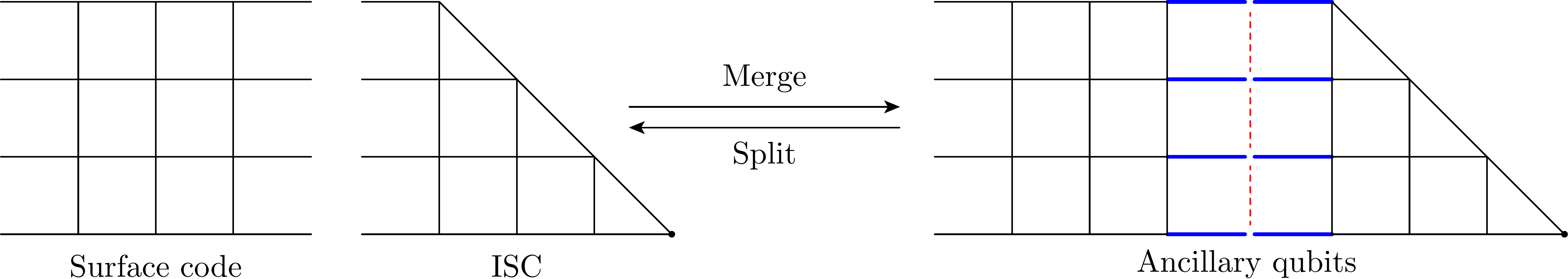}
            \caption{For the rough-rough surgery, we merge an ISC and a surface code patch from the left, along with a column of ancillary qubits on the seam (denoted by the dashed red lines) by measuring the seam stabilizers. During this merging, a measurement of the joint logical operator (denoted by the thickened blue edges) is performed. The merged code can be split by making a $Z$ measurement on each ancillary qubit on the seam, and local actions conditioned on the measurement outcomes.} 
    \label{fig: rough surgery}
\end{figure*}

We first consider lattice surgery between the rough boundaries of the surface code and the ISC. We orient the codes as shown in Fig.~\ref{fig: rough surgery}, with the surface code to the left of the ISC. We show that through merging and splitting of the codes, we can perform a measurement of $X_\LL \otimes \Gamma^x_\LL$ or $X_\LL \otimes \Gamma^z_\LL$, where $X_\LL$ denotes a logical Pauli $X$ operator of the surface code.

We start by discussing how to merge the two codes. We initialize a column of ancillary qubits in state $\ket{0}$ along the seam. We then define vertex stabilizers for the merged code on the seam:
\begin{align}
\vcenter{\hbox{\includegraphics[scale=.25]{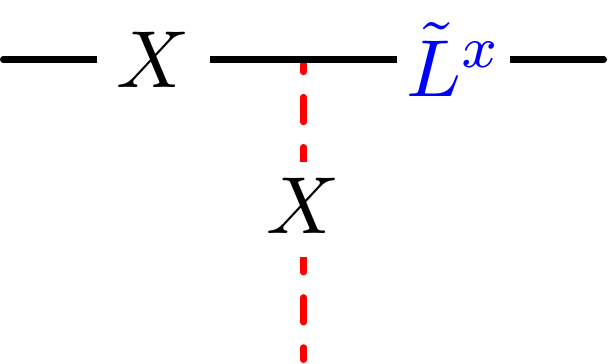}}},\qquad
\vcenter{\hbox{\includegraphics[scale=.25]{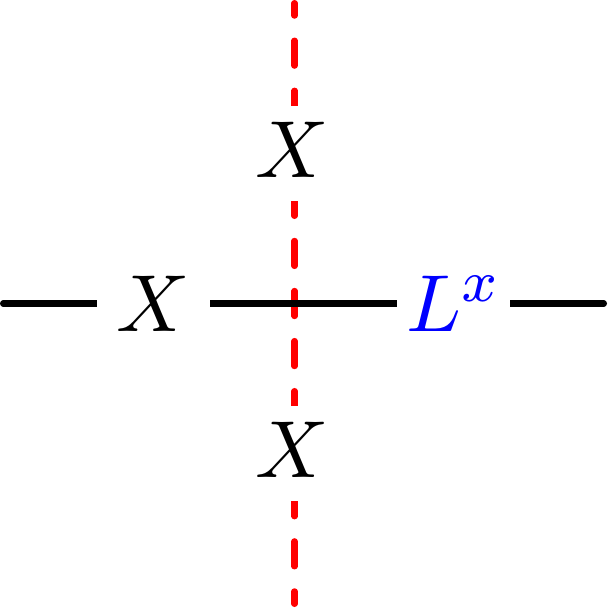}}},\qquad
\vcenter{\hbox{\includegraphics[scale=.25]{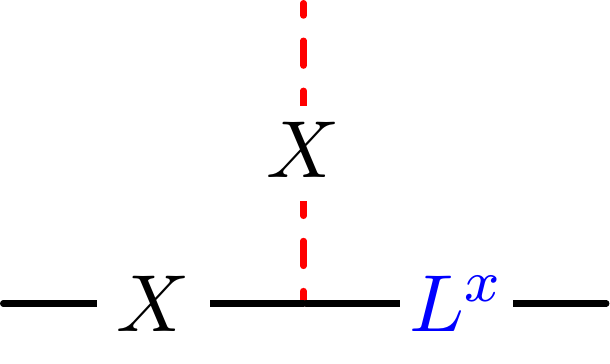}}}.
\label{eq:seamxxstabilizers}
\end{align}
Each vertex stabilizer acts on one ISC edge qudit, one surface-code qubit, and at most two ancillary qubits, shown in a red dashed line. The plaquette stabilizers of the merged code enforce the flux-free conditions, with an embedding of each seam qubit into a $\mathbb{Z}_2$ subgroup of $G$ generated by $x$. As we introduce ancillary qubits along the seam in state $|0\rangle$, the flux-free conditions along the seam are automatically guaranteed by the original open plaquette stabilizers of the ISC. 

The codes are merged by measuring the vertex stabilizers in Eq.~\eqref{eq:seamxxstabilizers} and recording the corresponding measurement outcomes $\{s_i=\pm 1\}$.\footnote{Here, we assume that the stabilizers of the merged code are measured $O(d)$ times and error correction is performed to ensure that the measurement outcomes are fault tolerant.} The product of the vertex stabilizers along the seam is equal to the product of logical operators $X_\LL \otimes \Gamma_\LL^{x}$. These measurements therefore effectively perform a joint logical measurement of $X_\LL \otimes \Gamma_\LL^{x}$.
Operationally, they project onto a subspace of the product of the code spaces of the ISC and the surface code via the projector 
\begin{equation}
\label{eq:projector PXX}
    \begin{aligned}
        P_{XX}=\frac{1+\left(\prod_i s_i\right)X_\LL \otimes \Gamma_\LL^{x}}{2},
    \end{aligned}
\end{equation}
where the measurement outcome is given by $\prod_i s_i$. This merging process with a joint measurement of $X_\LL \otimes \Gamma_\LL^{x}$ can be represented using the following circuit:
\begin{eqs}
    \vcenter{\hbox{\includegraphics[scale=.25]{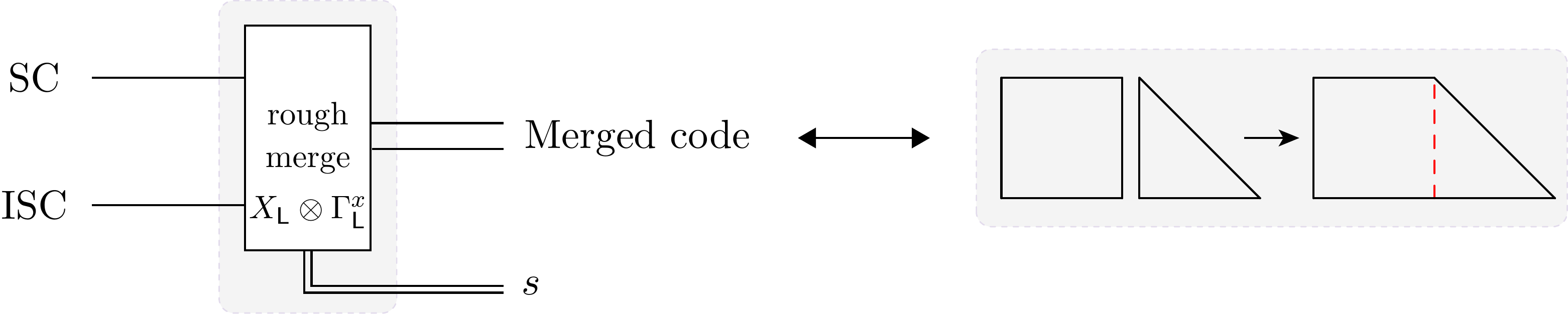}}}.
    \label{eq: rough merge}
\end{eqs}

Next, we describe how to split the merged code back into a surface code and an ISC, which we refer to as a rough split. This is achieved by measuring the ancillary qubits in the $Z$ basis. Subsequently, the ancillary qubits measured in the $\ket{1}$ state are flipped to the $\ket{0}$ state by applying the vertex stabilizers in Eq.~\eqref{eq:seamxxstabilizers}. We show below an example of the feedforward along the seam needed to map the ancilla to the $\ket{0}$ state:
\begin{align}
\vcenter{\hbox{\includegraphics[scale=.25]{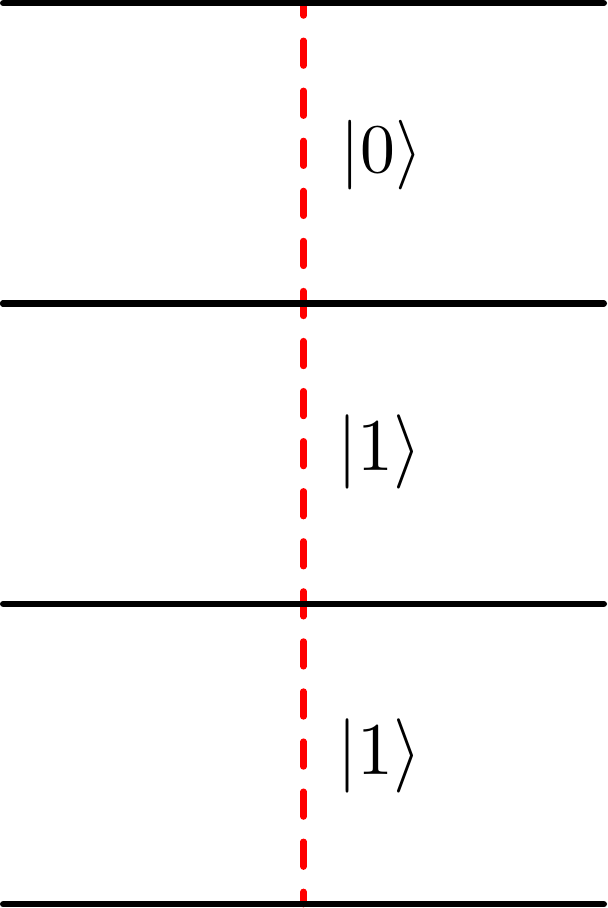}}}\quad \to \quad \vcenter{\hbox{\includegraphics[scale=.25]{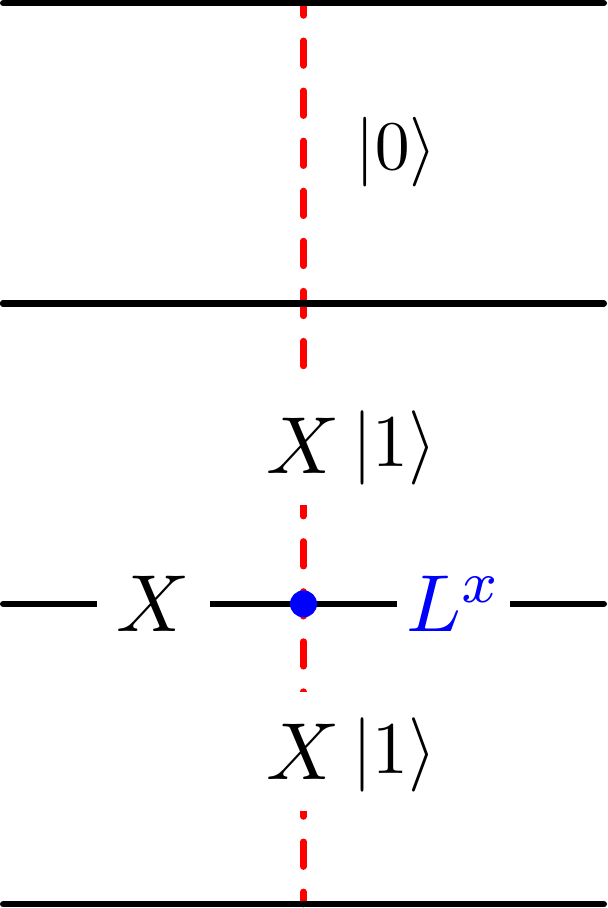}}}.
\end{align}
Since the logical operator $X_\LL \otimes \Gamma_\LL^{x}$ was just measured by the merge, the system is in a definite eigenstate of $X_\LL \otimes \Gamma_\LL^{x}$. The action of the vertex stabilizers of the merged code therefore does not create any logical faults.

We present the rough split as the following circuit:
\begin{eqs}
    \vcenter{\hbox{\includegraphics[scale=.25]{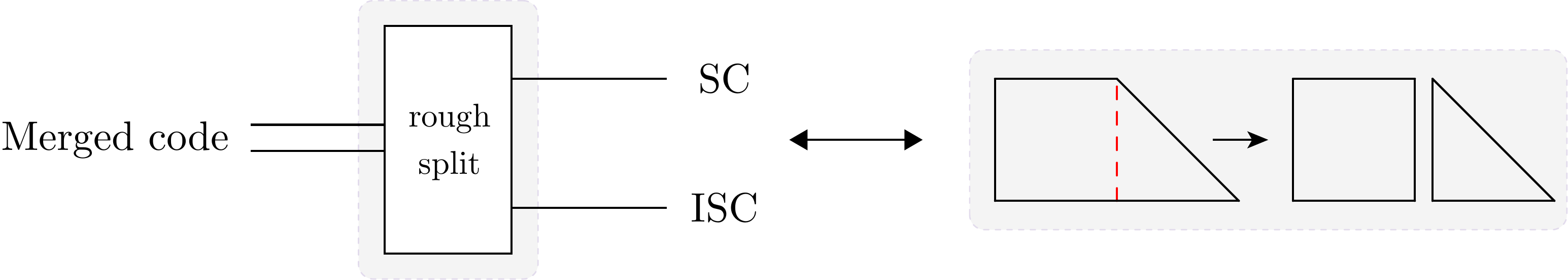}}}\,.
    \label{eq: rough split}
\end{eqs}

The composition of a rough merge followed by a rough split is nothing other than a non-destructive joint logical measurement. This is equivalent to applying the projection $P_{XX}$ in Eq.~\eqref{eq:projector PXX}. We denote the joint measurement circuit as:
\begin{eqs}
   \vcenter{\hbox{ \includegraphics[scale=.25]{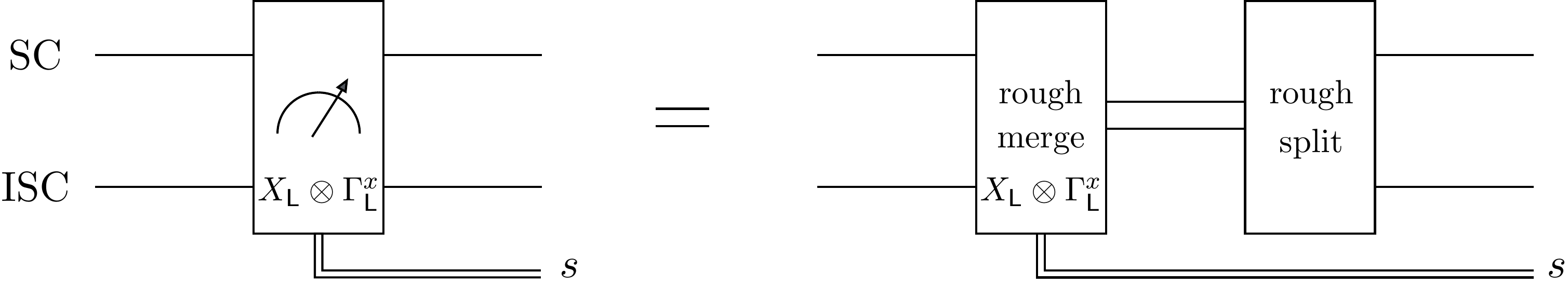}}}\,.
    \label{eq: rough joint measurement}
\end{eqs}

The above procedures can be straightforwardly adapted for other logical operators supported on the rough boundary of the ISC. In particular, replacing $x$ by $z$ in the seam vertex stabilizers yields a joint measurement of $X_\LL \otimes \Gamma_\LL^{z}$.

\subsection{Smooth-smooth surgery} \label{sec: smoothsmooth surgery}
We next consider lattice surgery between the smooth boundaries of a surface code and the ISC. In general, as discussed in Section~\ref{sec: logical operators} and Appendix~\ref{app: line-like logical}, there are two types of logical operators that can be represented transversally on the smooth boundary of the ISC: $W_\LL$ from projective intertwiners with trivial automorphism, and $\Gamma_\LL^z$ for $z$ in the center of $G$. In this subsection, we provide the details of a smooth-smooth surgery for the first type and comment on the second type. The logical operator associated with projective intertwiners with trivial automorphism is defined in Eq.~\eqref{eq: W on smooth boundary}. To avoid confusion, we note that, in this section, we use $W_\LL$ only for the special case of a line-like operator supported on the smooth boundary, instead of those associated with general projective intertwiners, as discussed in Section~\ref{sec: logical projective intertwiners}. More specifically, we focus on the case where $W_\LL$ corresponds to the logical $Z$ in the ISC, as seen in Example~\ref{example4}.

\begin{figure*}[t]
        \centering
        \includegraphics[scale=.15]{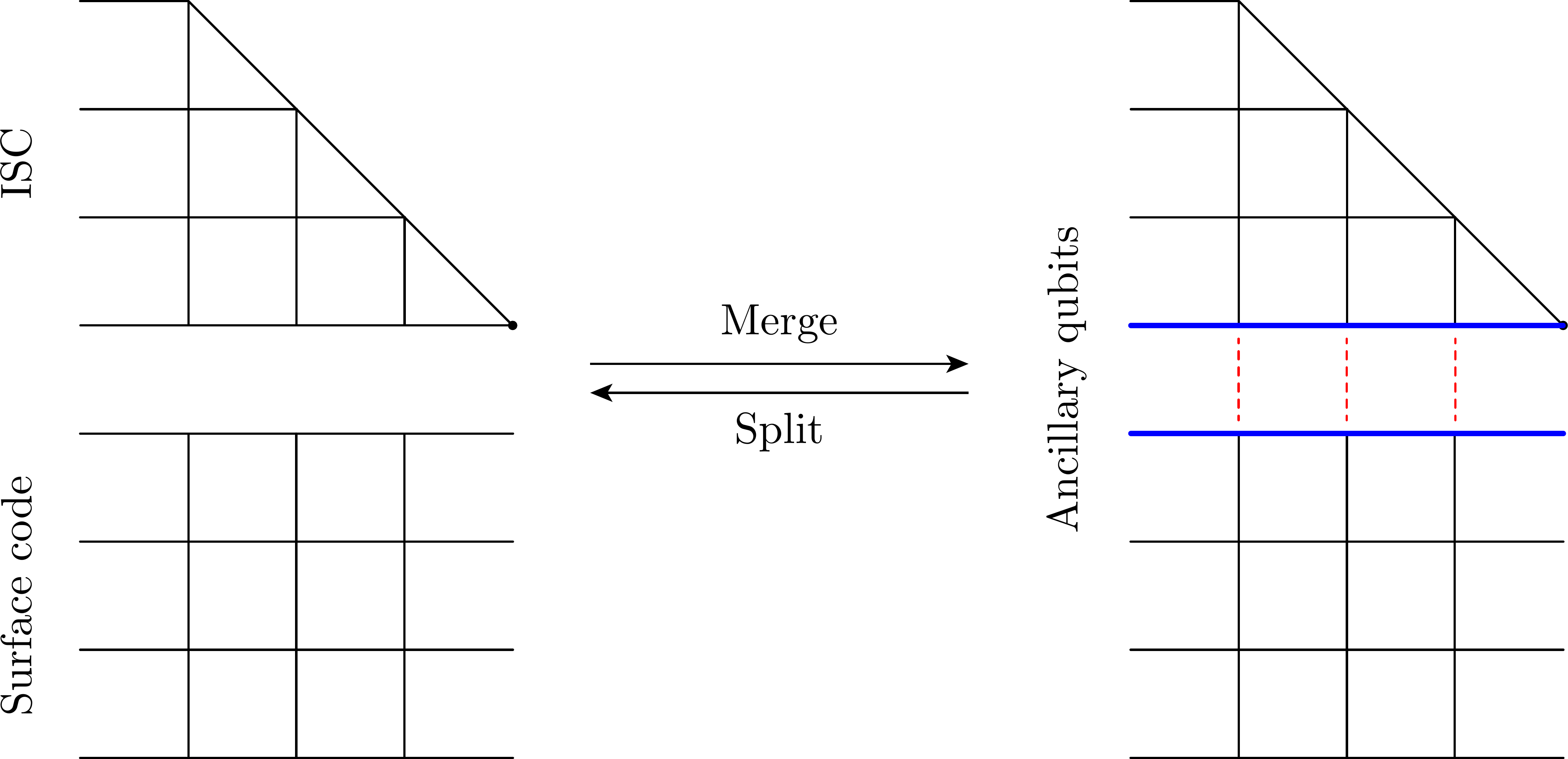}
            \caption{For the smooth-smooth surgery, we merge an ISC and a surface code patch from the bottom, along with a row of ancillary qubits on the seam (denoted by the dashed red lines) by measuring the seam stabilizers. During this merging, a measurement of the joint logical operator (denoted by the thickened blue edges) is performed. The merged code can be split by making an $X$ measurement on each ancillary qubit on the seam, and local actions conditioned on the measurement outcomes.} 
    \label{fig: smooth surgery}
\end{figure*}

Consider the orientation of the codes shown in Fig.~\ref{fig: smooth surgery}, where an ordinary surface code is below an ISC. Similar to the rough boundaries, we merge the two codes along the smooth boundary with a row of ancillary qubits initialized in the state $\ket{+}$. We define plaquette stabilizers of the merged code along the seam as:
\begin{align}
\vcenter{\hbox{\includegraphics[scale=.25]{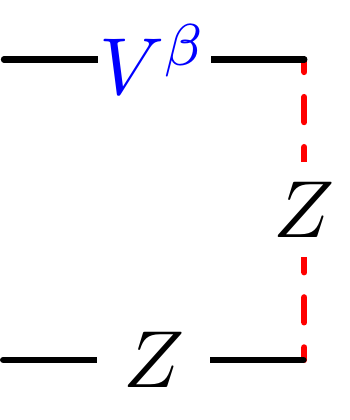}}},\qquad
\vcenter{\hbox{\includegraphics[scale=.25]{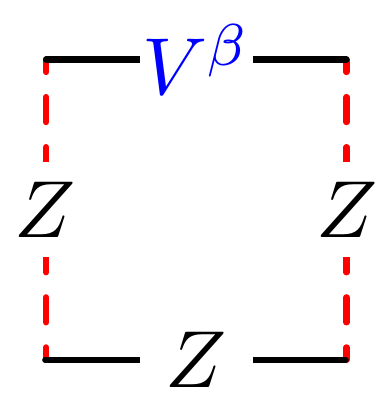}}},\qquad
\vcenter{\hbox{\includegraphics[scale=.25]{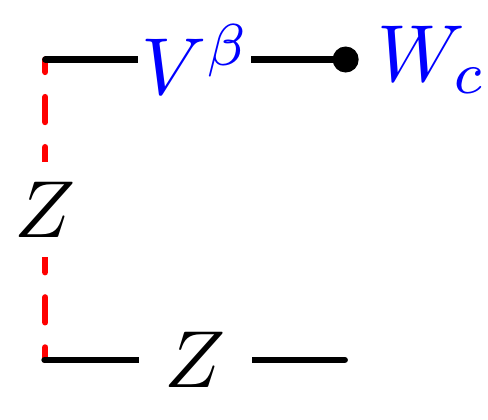}}}.
\label{eq:seamchistabilizers}
\end{align}
Since $W_\LL$ corresponds to the logical $Z$ operator, $V^\beta$ acts on the $G$ qudit as a one-dimensional representation of order $2$. In other words, $\beta(g)^2=1$ for any element $g\in G$. The vertex stabilizers of the merged code along the upper-side of the seam are defined as follows:
\begin{align}
\vcenter{\hbox{\includegraphics[scale=.25]{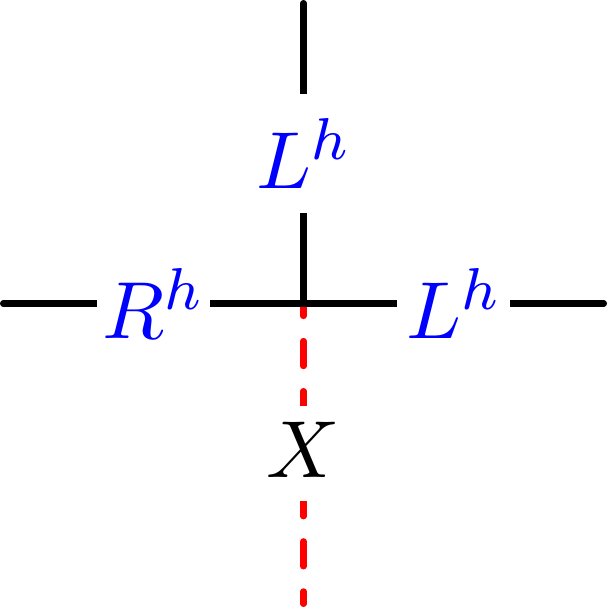}}},\qquad
\vcenter{\hbox{\includegraphics[scale=.25]{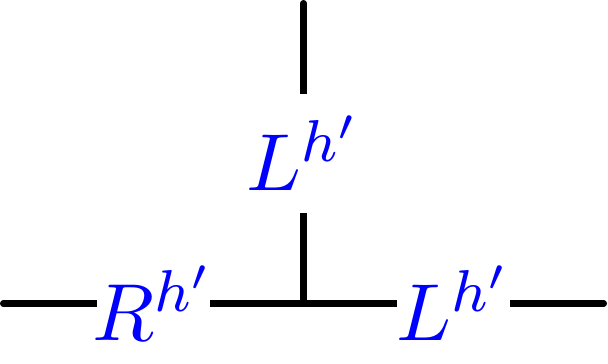}}},
\end{align}
where $h,h' \in G$ are such that $\beta(h) = -1$ and $\beta(h') = 1$. The vertex stabilizers along the lower-side of the seam are the usual vertex stabilizers of the surface code. As the ancillary qubits are introduced in the state $|+\rangle$, the vertex stabilizers are automatically satisfied, due to the truncated vertex stabilizers of the codes before merging. 

To perform the smooth-smooth merge, we measure the plaquette stabilizers along the seam and record the corresponding measurement outcomes $\{s_i=\pm 1\}$.\footnote{Again, the measurements are repeated and error correction is performed to ensure fault tolerance.} Taking the product of the plaquettes along the seam, we obtain the product of logical operators $Z_\LL\otimes W_\LL$, where $Z_\LL$ is the logical Pauli $Z$ of the surface code. The measurements therefore effectively perform a joint logical measurement of $Z_\LL \otimes W_\LL$,
\begin{equation}
\label{eq:projector PZZ}
    \begin{aligned}
        P_{ZZ}=\frac{1+\left(\prod_i s_i\right)Z_\LL \otimes W_\LL}{2},
    \end{aligned}
\end{equation}
where the measurement outcome is given by $\prod_i s_i$. We present this merging process with a joint measurement of $Z_\LL \otimes W_\LL$ using the following circuit:
\begin{eqs}
    \vcenter{\hbox{\includegraphics[scale=.25]{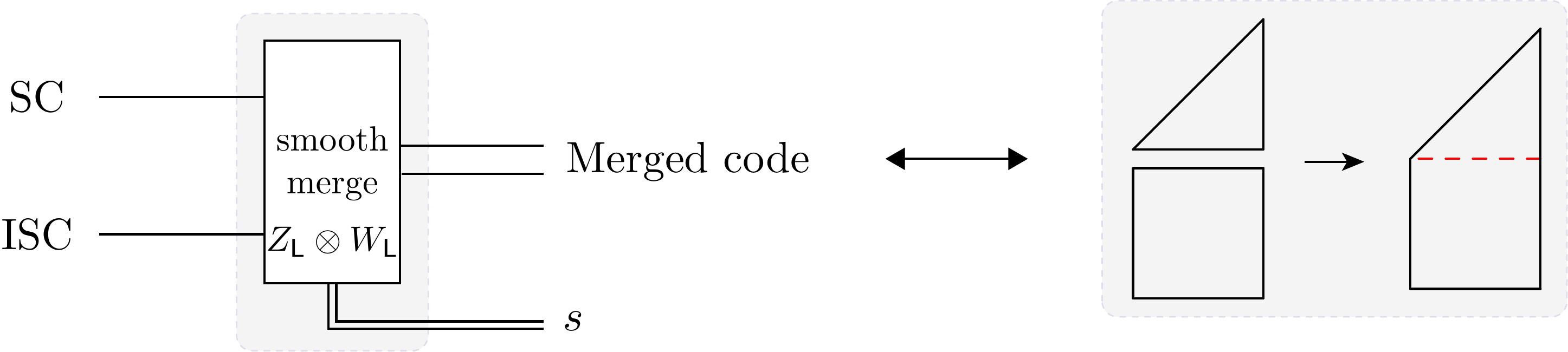}}}\,.
    \label{eq: smooth merge}
\end{eqs}

The merged code can then be split into the surface code and the ISC, as shown in Fig.~\ref{fig: smooth surgery}. Analogous to the rough split, the smooth split is achieved by measuring all ancillary qubits in the $X$ basis, followed by applying the plaquette stabilizers in Eq.~\eqref{eq:seamchistabilizers} to flip the measurements of $|-\rangle$ to the state $\ket{+}$.

We present the rough split as the following circuit:
\begin{eqs}
    \vcenter{\hbox{\includegraphics[scale=.25]{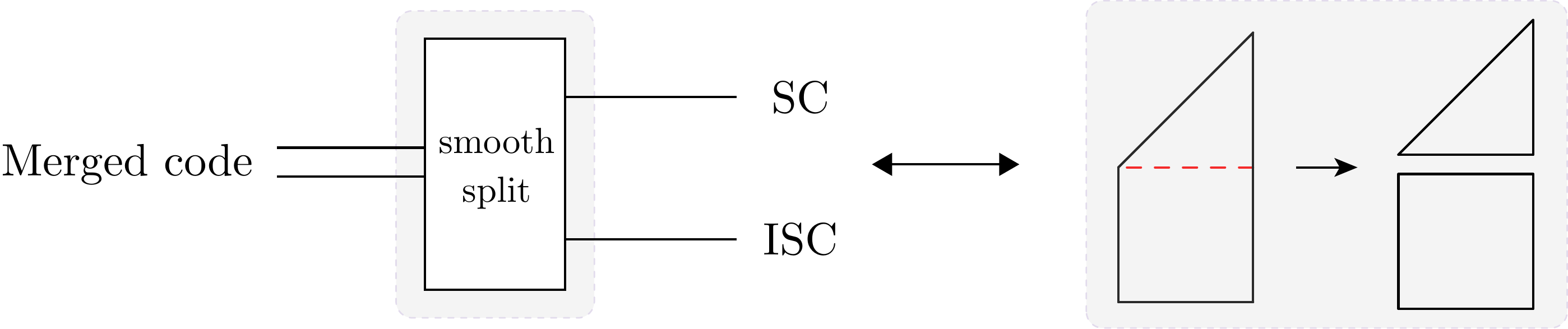}}}\,.
    \label{eq: smooth split}
\end{eqs}

The composition of smooth merge followed by a smooth split produces a non-destructive joint logical measurement or $Z_\LL\otimes W_\LL$, equivalent to applying $P_{ZZ}$ in Eq.~\eqref{eq:projector PZZ}. We denote the joint measurement circuit as:
\begin{eqs}
    \vcenter{\hbox{\includegraphics[scale=.25]{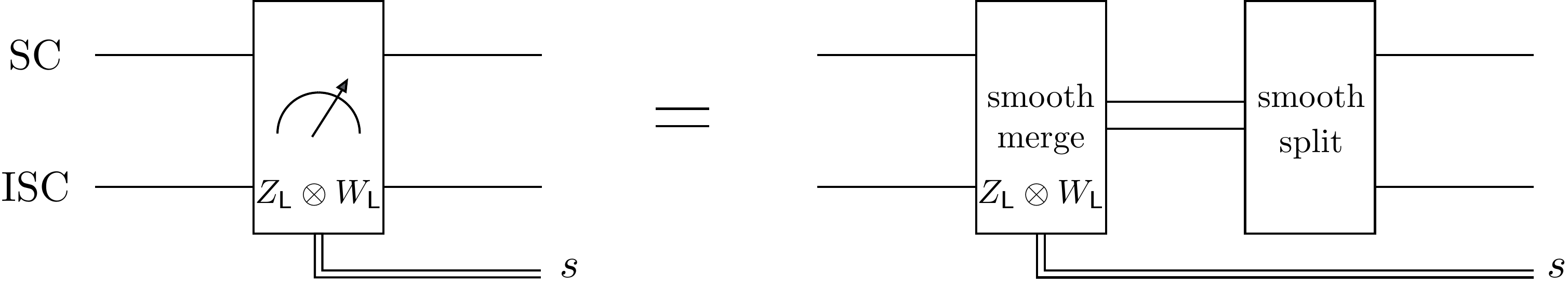}}}\,.
    \label{eq: smooth joint measurement}
\end{eqs}
All of the above procedures can be used in the same way for other logical operators supported on the smooth boundary of ISC. In particular, when the element $z$ is in the center of $G$, the logical operator $\Gamma^z_\LL$ can be represented transversally on the smooth boundary. We can therefore perform smooth-smooth surgery to implement a joint measurement of $Z_\LL \otimes \Gamma_\LL^{z}$, by replacing the $V^\beta$ and $W_c$ in the vertex stabilizers and conditioned actions above by $V^{\theta_z}$ and $\Gamma^z$. See Appendix~\ref{app: line-like logical} for more details on the representation of $\Gamma_\LL^z$ on the smooth boundary and the phase gate $V^{\theta_z}$.

\subsection{Surface code measurements} \label{sec: logical measurement}

The final ingredients needed for the teleportation protocol are logical $X$ and $Z$ measurements of surface code. These can be performed using standard methods~\cite{Fowler2012practical, Horsman2012latticesurgery}.  For the surface code,  representative logical operators take the form $Z_\LL=\prod_{e\in\gamma} Z_e$, where $\gamma$ is a path connecting the two rough boundaries, and $X_\LL=\prod_{e\in\bar{\gamma}} X_e$, where $\bar{\gamma}$ is a path in the dual lattice connecting the two smooth boundaries. Destructive measurements of $Z_\LL$ and $X_\LL$ can be performed by measuring all of the qubits in the $Z$ or $X$ basis, respectively. After performing error correction, path $\gamma$ or $\bar{\gamma}$ can be used to determine the logical measurement outcome. We present these logical measurements as the following circuits:
\begin{eqs}
    \vcenter{\hbox{\includegraphics[scale=.25]{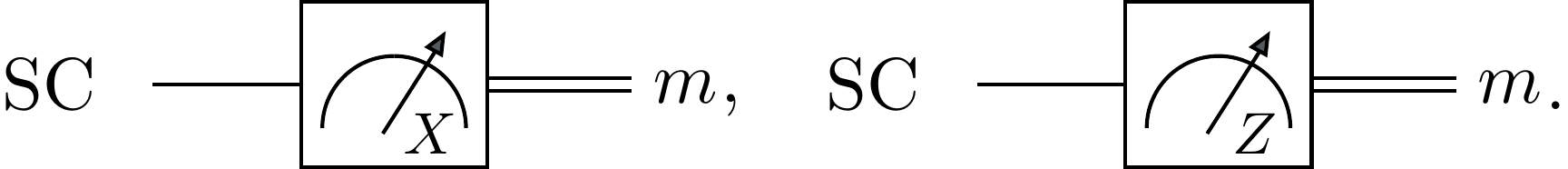}}}
    \label{eq: logical measurement}
\end{eqs}
In the figure above, $m=0,1$ correspond to the logical measurement outcomes $+1,-1$, respectively.

\subsection{Teleportation} \label{sec: teleportation}

Having described the protocols for merging, splitting, and surface code logical measurements, we now use them to describe gadgets for ISC logical state preparation and teleporting information into and out of the ISC.
The teleportation requires only the following set of primitive operations: 
\begin{itemize}
    \item Joint logical measurements of $X_\LL\otimes \Gamma^x_\LL$ between the surface code and ISC (Sec.~\ref{sec: roughrough surgery}).
    \item Joint logical measurement of $X_{\LL}\otimes \Gamma^{z}_\LL$ or $Z_\LL\otimes W_{\LL}$ between the surface code and ISC (Sec.~\ref{sec: smoothsmooth surgery}).
    \item Preparation of the ISC in an arbitrary state.
    \item Preparation of ancillary surface code patches in the logical states $|0\rangle_{\LL}$ or $|+\rangle_{\LL}$.
    \item Application of the transversal logical operators $\Gamma^{x}_{\LL}$, and either $\Gamma^{z}_{\LL}$ or $W_\LL$ on the ISC, conditioned on measurement outcomes. 
\end{itemize}
To simplify the discussion below, we assume that the ISC admits a logical Pauli $Z$ through the representation operator $\Gamma^{z}_{\LL}$. This can be replaced by $W_\LL$ enacting a Pauli $Z$ operator, if it is admitted by the ISC. We note that since all the conditional operations are implemented as transversal logical operators, each feed-forward step has a constant depth. For simplicity, we also assume that the ISC encodes one logical qubit and discuss the generalization at the end of the section.

\subsubsection{Initialization in $|0\rangle_{\LL}$ and $|+\rangle_{\LL}$ through surgery}

We now describe the procedure for preparing the ISC logical state in $|0\rangle_\LL$ or $|+\rangle_\LL$ by using rough-rough lattice surgery. We assume that the ISC can be initialized in some (possibly incoherent) code state $|\Phi\rangle_\LL$. \\

    \noindent {\bf Preparing $\ket{0}_\LL$:} We prepare an ancillary surface code patch in the logical state $\ket{+}_\LL$.
    Let the ISC be in an arbitrary code state $|\Phi\rangle_\LL$.
    Using rough-rough surgery, we measure the joint logical $X_{\LL}\otimes \Gamma_{\LL}^{z}$.
    Let the measurement outcome be $(-1)^s$ with $s\in \{0,1\}$.
    Up to normalization, the post-measurement state is
    \begin{eqs}
        \frac{1+(-1)^{s}X_{\LL}\otimes \Gamma_{\LL}^{z}}{2} \Big( |+\rangle_{\LL}\otimes |\Phi\rangle_{\LL}\Big) = |+\rangle_{\LL}\otimes \frac{1+(-1)^{s} \Gamma_{\LL}^{z}}{2}\ket{\Phi}_{\LL}\,.
    \end{eqs}
    where we have used $X_{\LL}\ket{+}_{\LL}=\ket{+}_{\LL}$.
    Hence, the measurement projects the ISC onto $\ket{s}_{\LL}$ and leaves the surface code unentangled with the ISC. We can discard the surface code patch or use it subsequently for other purposes. Finally, if $s=1$ we apply $\Gamma^{x}_{\LL}$, which maps the logical state from $\ket{1}_{\LL}$ to $\ket{0}_{\LL}$.
    The procedure to prepare the ISC logical state $\ket{0}_{\LL}$ can be summarized by the following circuit:
    \begin{eqs}\label{eq:0 state prep}
    \vcenter{\hbox{\includegraphics[scale=.25]{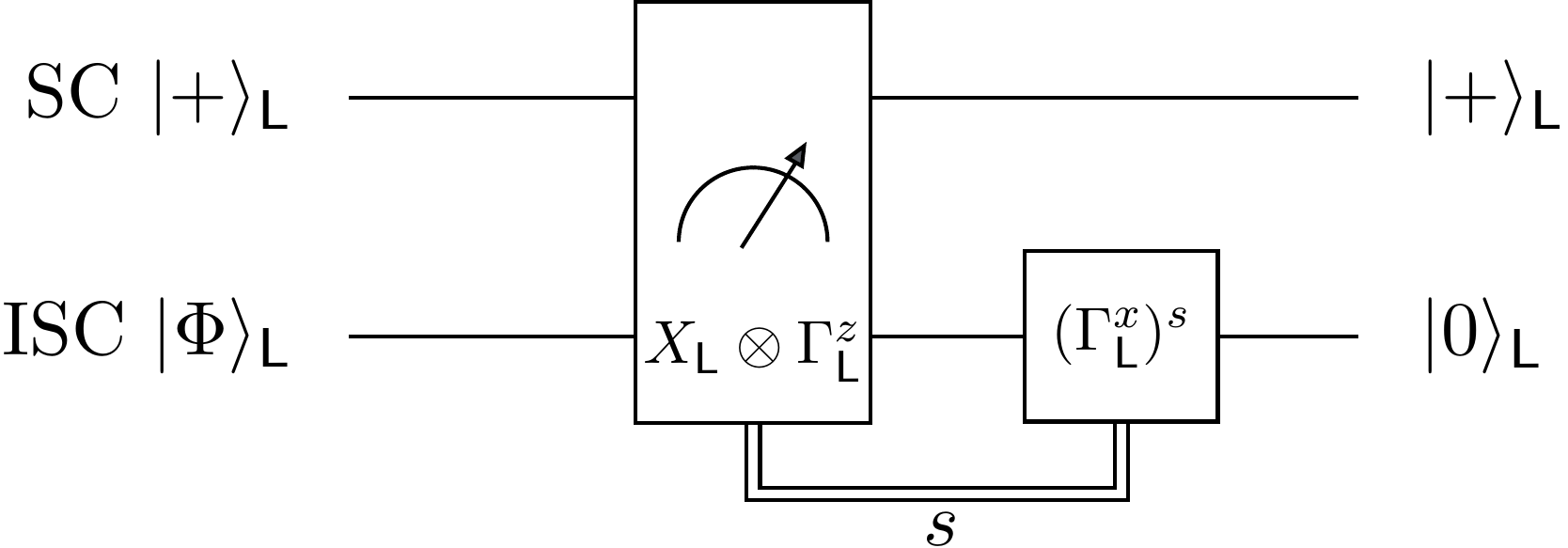}}}\,.
    \end{eqs} 
    This protocol also provides a means of performing a logical $Z$ measurement on the ISC.\\

    \noindent {\bf Preparing $\ket{+}_\LL$:} The procedure is similar to the preparation above, with the roles of $z$ and $x$ interchanged.
    We again prepare an ancillary surface code patch in $\ket{+}_\LL$, and assume the ISC is in an arbitrary code state $\ket{\Phi}_{\LL}$.
    We then measure the joint logical operator $X_\LL\otimes\Gamma^{x}_{\LL}$ using rough-rough surgery, obtaining the outcome $(-1)^{s}$ with $s\in\{0,1\}$.
    Since $X_\LL\ket{+}_\LL=\ket{+}_\LL$, the post-measurement state is, up to normalization,
    \begin{eqs}
        \frac{1+(-1)^{s}\,X_\LL\otimes\Gamma^{x}_{\LL}}{2}\Big(\ket{+}_\LL\otimes\ket{\Phi}_{\LL}\Big)
        \;=\;\ket{+}_\LL\otimes\frac{1+(-1)^{s}\,\Gamma^{x}_{\LL}}{2}\ket{\Phi}_{\LL}\,,
    \end{eqs}
    so the code state of the ISC is projected onto $\ket{+}_{\LL}$ if $s=0$ and onto $\ket{-}_{\LL}$ if $s=1$. In the latter case we apply $\Gamma^{z}_{\LL}$, which maps the code state from $\ket{-}_{\LL}$ to $\ket{+}_{\LL}$.
    The quantum circuit for the $\ket{+}_{\LL}$ state preparation protocol is
    \begin{eqs}\label{eq:+ state prep}
    \vcenter{\hbox{\includegraphics[scale=.25]{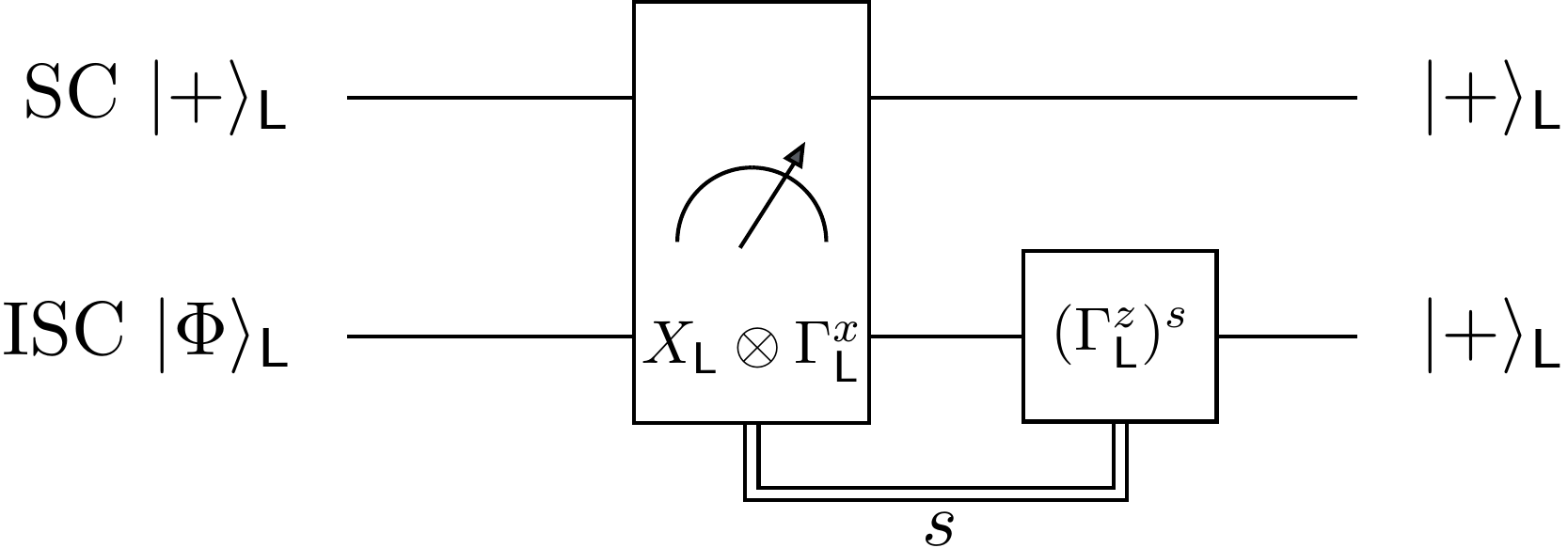}}}.
    \end{eqs} 
    Note that this protocol can be used to perform a logical $X$ measurement of the ISC, more generally.\\

We note that, if the ISC admits a logical Pauli $Z$ operator implemented by $W_\LL$ on the smooth boundary, then the preparation of $\ket{0}_\LL$ can be replaced with:
\begin{eqs}
    \vcenter{\hbox{\includegraphics[scale=.25]{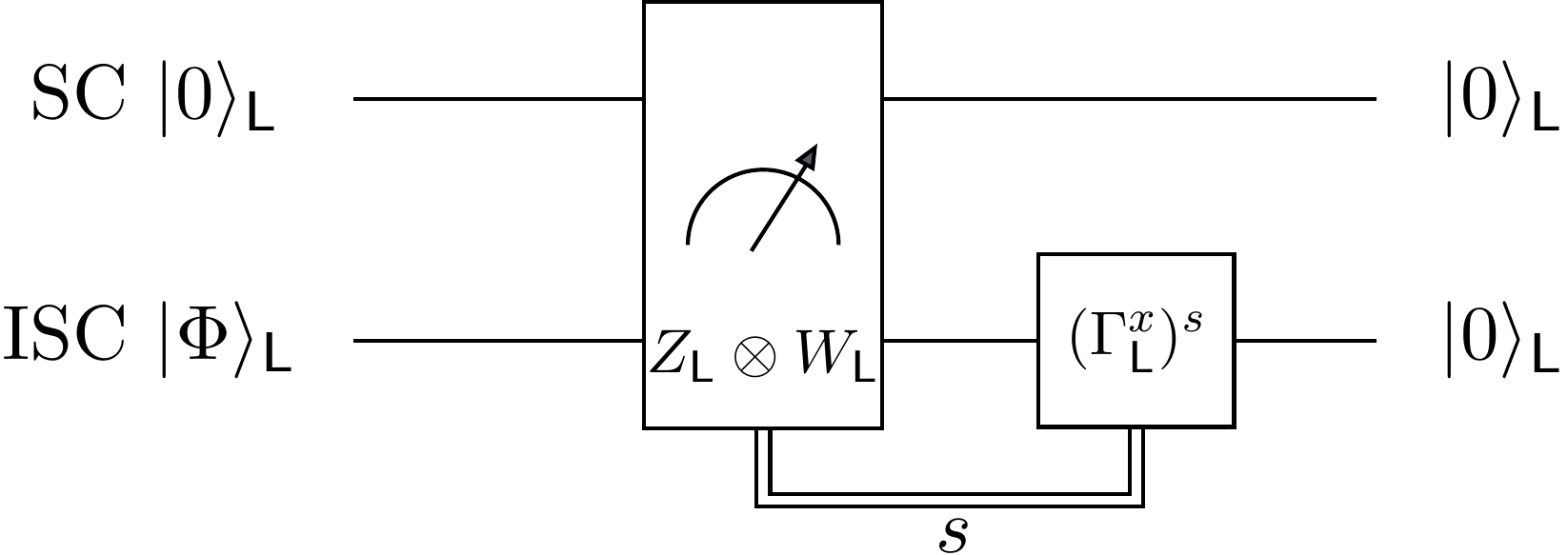}}},
    \label{eq:0 state prep with W}
\end{eqs}
where the surface code is initialized in the $\ket{0}_\LL$ state. For the preparation of the $\ket{+}_\LL$ the feedforward can be implemented by $W_\LL$, giving us the circuit:
\begin{eqs}
    \vcenter{\hbox{\includegraphics[scale=.25]{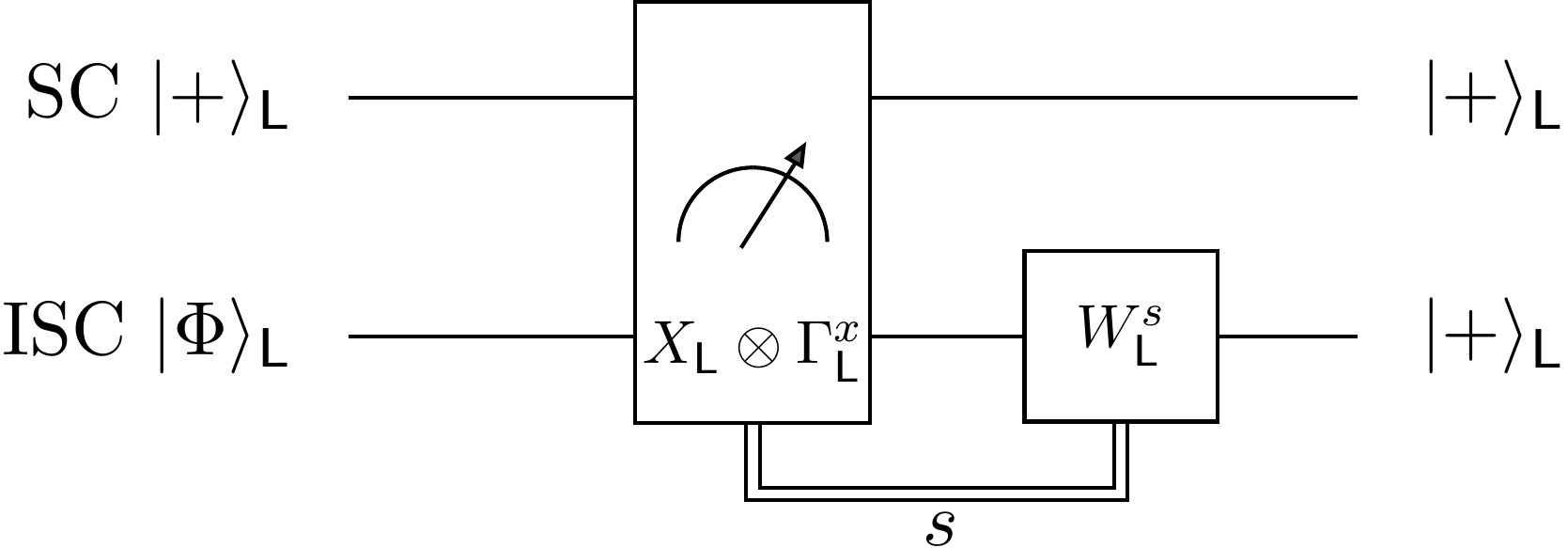}}}.
    \label{eq:+ state prep with W}
\end{eqs}

\subsubsection{Teleporting arbitrary logical states into the ISC}

We now describe how an arbitrary logical state $\ket{\Psi}_{\LL}$, encoded in a surface code, can be teleported into the ISC. The protocol is a one-bit teleportation~\cite{ZhouLeungChuang2000}, in which the joint measurement is implemented by the rough-rough surgery.
It requires the ISC to be initialized in $\ket{0}_{\LL}$, which can be done using the protocol above independently of $\ket{\Psi}_{\LL}$.

Concretely, let the ISC be in code state $\ket{0}_{\LL}$, and prepare a surface code patch in $\ket{\Psi}_{\LL}=\alpha\ket{0}_{\LL}+\beta\ket{1}_{\LL}$.
The teleport-in circuit is
\begin{eqs}    \includegraphics[scale=.25]{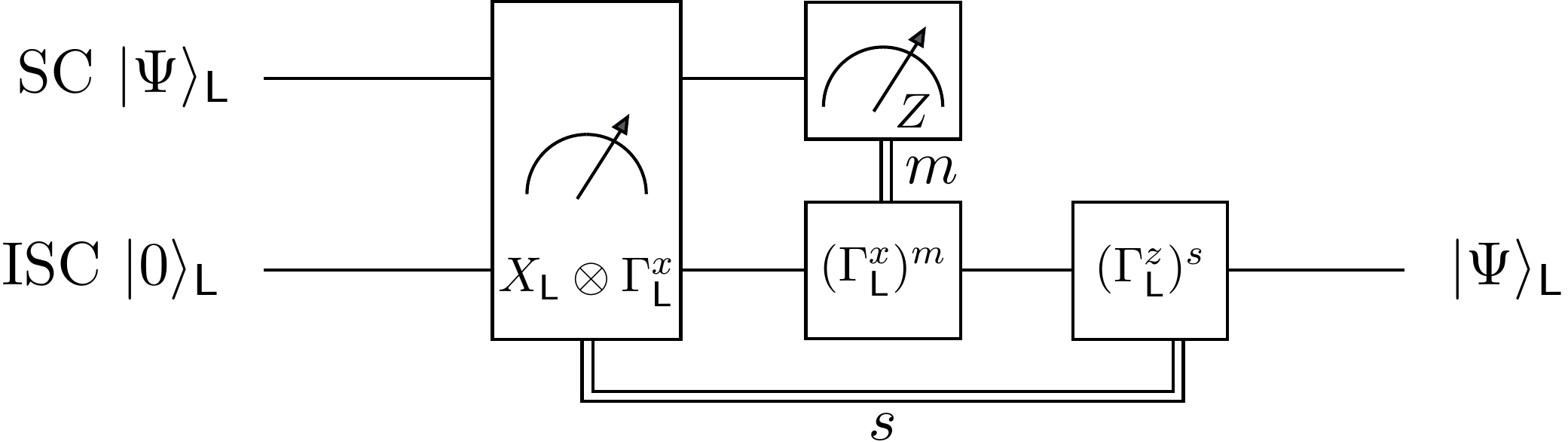}
    \label{eq:teleport in}
\end{eqs}
We first measure the joint logical operator $X_\LL\otimes\Gamma^{x}_{\LL}$ via a rough-rough surgery, obtaining the outcome $(-1)^{s}$ with $s\in\{0,1\}$.
The post-measurement state is, up to normalization,
\begin{eqs}
    \frac{1+(-1)^{s}\,X_{\LL}\otimes\Gamma^{x}_{\LL}}{2}\Big( \ket{\psi}_{\LL}\otimes\ket{0}_{\LL}\Big)
    =\frac{1}{2}\ket{0}_{\LL}\otimes\big(\alpha\ket{0}_{\LL}+(-1)^{s}\beta\ket{1}_{\LL}\big)
    +\frac{1}{2}\ket{1}_{\LL}\otimes\big(\beta\ket{0}_{\LL}+(-1)^{s}\alpha\ket{1}_{\LL}\big)\,.
\end{eqs}
We then perform a logical measurement of $Z_\LL=(-1)^m$ on the surface code. Both $m=0,1$ outcomes occur with probability $1/2$, independently of $\ket{\psi}_{\LL}$.
Since $Z_{\LL}$ anticommutes with $X_{\LL}$, this measurement reveals no information about $\alpha$ and $\beta$.
From the expression above, the code state of the ISC is left in
\begin{eqs}
    \big(\Gamma^{x}_{\LL}\big)^m\big(\Gamma^{z}_{\LL}\big)^{s}\ket{\Psi}_{\LL}\,,
\end{eqs}
up to a global phase.
Finally, we apply $\Gamma^{x}_{\LL}$ if $m=1$ and $\Gamma^{z}_{\LL}$ if $s=1$, in either order, which returns the ISC to $\ket{\Psi}_{\LL}$.
Both feedforward operators are transversal Pauli operators.

The same protocol applies when logical $Z$ is not available on the rough boundary but the ISC admits a $W_{\LL}$ acting as logical $Z$.
One simply replaces $\Gamma^{z}_{\LL}$ by $W_{\LL}$ throughout, initializing ISC in $\ket{0}_{\LL}$ by the smooth-smooth surgery, as in Eq.~\eqref{eq:0 state prep with W}, and applying $W_{\LL}$ in place of $\Gamma^{z}_{\LL}$ in the final feedforward.
The joint measurement and the logical measurement of the surface code are unchanged, since they involve only $\Gamma^{x}_{\LL}$ and the rough boundary.

\subsubsection{Teleporting logical states out of the ISC}

We finally describe the reverse process, in which the logical state of the ISC is transferred to the surface code.
Together with the injection protocol, this allows one to interleave the transversal gates of the ISC with operations that are only available on the surface code.

Let the ISC be in an arbitrary code state $\ket{\Psi}_{\LL}$, our goal is to transfer the state to a surface code.
The protocol is again a one-bit teleportation, now with the ISC as the sender.
Teleporting information out requires two surface code patches.
There is an ancillary surface code $A$ prepared in $\ket{+}_{\LL}$ and a receiver $B$ initialized in $\ket{0}_{\LL}$.
The circuit is 
\begin{eqs}
    \includegraphics[scale=.25]{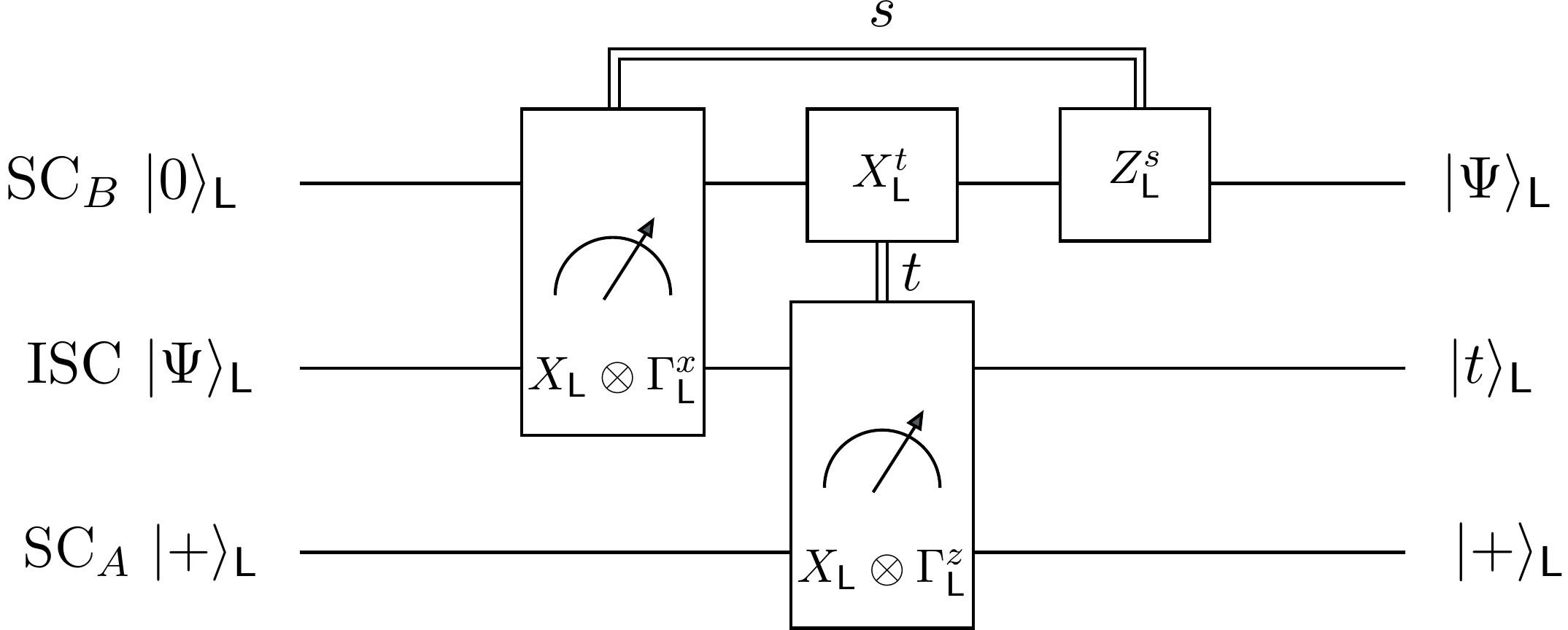}
    \label{eq:teleport out}
\end{eqs}

For definiteness, the circuit assumes that logical $Z$ is available as the rough-boundary logical $\Gamma^{z_{i}}_{\LL}$.
When it is instead available as $W_{\LL}$ on the smooth boundary, the second joint measurement becomes a smooth-smooth surgery measuring $Z^{A}_{\LL}\otimes W_{\LL}$ with $A$ prepared in $\ket{0}_{\LL}$.

\subsubsection{Information transfer when the ISC encodes $k$ qubits}

All of the above protocols can be straightforwardly generalized to the case when the ISC encodes $k$ logical qubits, as long as the Pauli $X$ and $Z$ on the $i$-th logical qubit can be implemented on the rough boundary as $\Gamma^{x_i}_{\LL}$ and $\Gamma^{z_i}_{\LL}$, or the latter is available as $W_{\LL}$ on the smooth boundary. We take the teleport-out protocol as an example here. Suppose that the code state of the ISC is 
\begin{eqs}
    \ket{\Psi}_{\LL}=\ket{0}_{i}\otimes\ket{\Phi_{0}}+\ket{1}_{i}\otimes\ket{\Phi_{1}}\,,
    \label{eq:Psi decomposition}
\end{eqs}
where $\ket{\Phi_{0}}$ and $\ket{\Phi_{1}}$ are unnormalized, not necessarily orthogonal states of the remaining $k-1$ logical qubits.
In particular, the $i$-th qubit may be entangled with the rest of the ISC. Performing the protocol as described above, we can teleport the state to the surface code patch initialized in state $\ket{0}_{B}$:
\begin{eqs}
    \ket{0}_{B}\otimes \left(\ket{0}_{i}\otimes\ket{\Phi_{0}}+\ket{1}_{i}\otimes\ket{\Phi_{1}}\right)
    \;\longmapsto\;
    \ket{t}_{i}\otimes\big(\ket{0}_{B}\otimes\ket{\Phi_{0}}+\ket{1}_{B}\otimes\ket{\Phi_{1}}\big)\,,
    \label{eq:teleport out map}
\end{eqs}
for some known $t\in\{0,1\}$, so that the surface code takes over the role of the $i$-th qubit in the entangled state, while the remaining logical qubits of the ISC are untouched.

The full logical state of the ISC can then be transferred to $k$ surface code patches $A_{1},\dots,A_{k}$ by applying the single-qubit protocol to each $i=1,\dots,k$ logical qubits.
This is justified by Eq.~\eqref{eq:teleport out map}.
The protocol for qubit $i$ acts only on qubit $i$, $A_{i}$ and the ancilla, and notably leaves the states $\ket{\Phi_{0}}$, $\ket{\Phi_{1}}$ unchanged.
Therefore it applies verbatim when these are states of the remaining ISC qubits, after part of the ISC logical state $\ket{\Psi}$ has been teleported out in preceding steps.
At the logical level, the $2k$ measurements can be performed in any order as long as, for each $i$, the measurement of $Z_{i,\LL}$ follows that of $X^{B_{i}}_{\LL}\otimes\Gamma^{x_{i}}_{\LL}$, since operators belonging to different $i$ commute.
They cannot, however, be performed in parallel since every logical operator $\Gamma^{g}_{\LL}$ is supported on all $d$ dangling edges of the rough boundary, so a rough-rough surgery measuring $X^{B}_{\LL}\otimes\Gamma^{x_{i}}_{\LL}$ requires seam stabilizers along the entire boundary, and only one patch can be merged with the ISC at a time.
The $2k$ surgeries are therefore sequential, and the protocol requires $O(kd)$ code cycles, $k$ receiver patches, and a single ancillary patch prepared in the $\ket{+}_A$ state.
Any subset of the logical qubits can also be teleported out in the same way, leaving the remaining qubits in the ISC.

\section{Discussion} \label{sec: discussion}

In this paper, we introduced irrep surface codes (ISCs) as a systematic framework for engineering two-dimensional topological codes around a prescribed set of logical operations. The construction accommodates any irreducible finite group of gates and yields additional transversal operations through projective intertwiners, including gates outside every finite level of the Clifford hierarchy. Representative examples of gates are summarized in Table~\ref{tab: summary}. We further developed lattice-surgery protocols for transferring encoded information between surface codes and ISCs satisfying either of the two conditions in Result~\ref{res: transfer}. This suggests a modular architecture in which ISCs provide selected logical operations within a broader fault-tolerant computation.

A particularly interesting direction is a detailed study of the implementation of the $S_3$ ISC, based on the two-dimensional irreducible representation of $S_3$. This code admits transversal $X$ and $P_{3}$ gates, where notably, the $P_{3}$ gate, a $2\pi/3$ rotation about the $z$ axis, lies outside every finite level of Clifford hierarchy. Although $Z$ does not belong to the input gate group, the code admits a transversal logical $Z$ supported on the smooth boundary, enabling the state-preparation and information-transfer protocols described in Section~\ref{sec: transferring information}. This makes the $S_3$ ISC a useful setting for developing the complete sequence of initialization, logical gate application, and transfer back to the surface code. Explicit local gate decompositions and measurement circuits for this case would also provide a basis for testing the more general construction.

Several questions remain for turning this construction into a practical computational advantage:

\begin{itemize}
    \item Decoding -- There have been various decoders proposed for non-Abelian topological orders, tailored to different anyon models, available measurements, and noise assumptions~\cite{wootton_error_2014,Hutter_2015,wootton_active_2016,Brell_2014,Burton_2017,dauphinais_fault_2017,brown2020fault,schotte_quantum_2022,davydova2025D4,jing_intrinsic_2025,jing2026integerlinearprogrammingdecoder,sajith2026fusion}. An important task is to adapt and extend them to ISCs, taking into account their distinct rough, smooth, and twisted boundaries, as well as the representation-dependent corner. In addition to decoding an isolated ISC, information transfer requires decoding the temporarily merged code during lattice surgery. Developing such high-performance decoders will be essential for assessing the practical fault tolerance of ISCs.
    
    \item Syndrome extraction circuits -- A complementary direction is to develop explicit syndrome-extraction circuits and a corresponding tensor-network description. Combined with a decoder, this would provide a route to Monte Carlo estimates of logical error rates and finite-size scaling of the error threshold~\cite{darmawan2017tensor,piveteau2024tensor}. Beyond numerical simulation, such a spacetime tensor-network description could provide a systematic framework for identifying syndrome constraints, logical observables, and circuit-level fault paths, thereby helping formulate the decoding problem directly~\cite{delfosse2023spacetimecodescliffordcircuits,Bombin2024unifyingflavorsof,rodatz2026faulttoleranceconstruction,magdalena2025xyz,davydova2025D4,xu2026theoryspacetimequantumfault}. It may also facilitate alternative dynamical measurement schedules, including possible Floquet-code realizations~\cite{Bauer2025xplusy,Rodatz2026floquetifying}.
    
    \item Resource estimates -- Our approach with ISCs replaces the conventional logical gate synthesis cost, including for example the magic-state preparation, with the cost of operating a code tailored to the target gate set.~A rough estimate  of the necessary resources follows from the lattice geometry: an ISC with code distance $d$ requires $O(d^2 q_G)$ data qubits with $q_G=\log_2|G|$. The information transfer requires $O(d)$ cycles for fault tolerance, which gives a spacetime estimate of $O(d^3q_G)$.  A more complete estimate must additionally include syndrome qubits, the group-dependent costs of local gates and measurements, repeated syndrome measurement, and decoder runtime. Moreover, although an exactly implemented gate has no synthesis error, approximating continuously variable gates may require $q_G$ to grow with the desired precision. Consequently, a quantitative comparison between our approach and the conventional synthesis method should be performed separately, in the spirit of existing comparisons between distillation and code switching~\cite{Beverland2021comparison, daguerre2025codeswitching, gao2026resourcecomparisonlogicaltstate}.
    
    \item Algorithmic applications -- Several algorithmic settings may benefit from our gate-tailored codes. QAOA repeatedly applies diagonal phase gates generated by the cost Hamiltonian~\cite{farhi2014quantumapproximateoptimizationalgorithm}, while Hamiltonian-simulation algorithms based on product formulas repeatedly apply exponentials of local Hamiltonian terms, often reducible to Pauli rotations and diagonal phase gates by Clifford basis changes~\cite{childs2021theory}. Quantum Fourier transforms and order-finding make extensive use of controlled-phase gates and reversible arithmetic~\cite{shor1997polynomial}, while hidden-subgroup and hidden-shift algorithms provide natural group-theoretic applications~\cite{kuperberg2005subexponential,vandam2006quantum}. These examples motivate tailoring ISCs to frequently repeated diagonal or reversible subroutines. Determining whether this yields a practical advantage will require application-specific resource comparisons that include approximation, initialization, and information-transfer costs.
\end{itemize}

\vspace{0.2in}
\noindent \textit{Acknowledgments} -- ZS is supported by the U.S. National Science Foundation under Grant No.~DRL I-TEST 2148467. VM is grateful for support from the Rasera Family Physics Research Award.  Research at Perimeter Institute is supported in part by the Government of Canada through the Department of Innovation, Science and Economic Development and by the Province of Ontario through the Ministry of Colleges and Universities.
A.T. is funded by Villum Fonden Grant no. VIL60714.

\vspace{0.2in}
\noindent \textbf{AI usage statement.} We used a combination of LLMs in this work (GPT 5.6 Sol, GPT 6 Astra, Claude Opus, Claude Fable), all accessed in October 2026. These were used in Appendix~\ref{app: more examples} for an initial draft and most of the computations, to review various other sections, and to format tables. 

\begin{appendix}
\addtocontents{toc}{\protect\hideappsubsections}

\section{Twisted group algebras} \label{app: irrep review}
In this appendix, we introduce $\cgw$, the twisted group algebra for a finite group $G$, where $\omega$ denotes the 2-cocycle twisting the group action. We closely follow the treatment of Ref.~\cite{serre1977linear} for untwisted group algebras and Ref.~\cite{isaacs1994character} for projective group representations. 

Informally, the elements of the twisted group algebra are linear combinations of elements of $G$ with complex coefficients, and the multiplication is inherited from the (possibly projective) group operation of $G$. More generally, an algebra is a vector space with a bilinear vector operation (a multiplication), so it requires scalars, vectors, and an operation which takes a pair of vectors to a single vector.
For the twisted group algebra, the scalars are complex numbers, the basis vectors are group elements, and the bilinear operation is the projective group multiplication.  
The algebra $\cgw$ has multiple roles in an ISC: it is used to describe local Hilbert spaces, in which case only the vector space structure is relevant, but also describes both local and logical representation actions, for which the bilinear operation is necessary. This dual nature is made clear in section \ref{subapp: action of CG on itself}. The language developed in this appendix is also valuable for understanding the relation between ISCs and projective group surface codes (pGSCs), as described in Appendix~\ref{app: isc from gsc}.

\subsection{The untwisted group algebra $\mathbb{C}[G]$}
We start by considering the untwisted group algebra $\cg$, the special case of $\cgw$ with trivial $\omega$:
\begin{eqs}
    \omega(g,h) = 1, \quad \forall\, g, \, h\, \in \, G.
\end{eqs}
In this section, we describe a natural basis for the algebra, describe its representations, and define the group Fourier transform, through which we obtain a dual basis.

Every element $x \in \cg$ has a unique decomposition:
\begin{eqs}
\label{eq: untiwsted algebra element}
    x = \sum_g c_g g, \quad c_g \in \mathbb{C}, \quad g \in G.
\end{eqs}
Throughout the rest of the appendix, we use $x$ as defined above, and $y = \sum_h c_h h$, when describing operations that require explicit elements of the algebra.
The multiplication of two vectors is given by linearly extending the group operation:
\begin{eqs}
    xy = \sum_g c_g g \sum_h c_h h = \sum_{g,h} c_g c_h gh, \quad gh \in G.
\end{eqs}
The set $\{ g\, |\, g\in G\}$ forms a basis for the algebra, since the group elements are linearly independent and span the space. The former is clear, while the latter is by the definition of Eq.~\eqref{eq: untiwsted algebra element}.
There is another basis for $\mathbb{C}[G]$, which is best understood in terms of representations of $\mathbb{C}[G]$. Therefore, we now describe $\cg$ representations before giving the explicit form of such a basis.

As for the case of the bilinear operation of the algebra, the representation theory of $\cg$ is also inherited from the underlying group $G$, so we first give a brief introduction to group representations. A representation of $G$ is a group homomorphism into the linear group of invertible matrices  of some dimension, over a field like $\mathbb{R}$ or $\mathbb{C}$. We will restrict our attention to complex representations. A complex representation  $\Gamma$ is given by the group homomorphism:
\begin{eqs}
\label{eq: definition of homomorphism}
    \Gamma:\, G \rightarrow \mathrm{GL}(V_\Gamma), \quad g \mapsto \Gamma(g), \quad V_\Gamma \cong \mathbb{C}^{d_\Gamma},
\end{eqs}
where $\mathrm{GL}(V_\Gamma)$ is the group of $d_\Gamma$-dimensional  complex invertible matrices acting on the vector space $V_\Gamma$, and $d_\Gamma$ is the dimension of $\Gamma$. We can always choose a basis in which the $\Gamma(g)$ matrices are unitary, and we assume such a choice has been made throughout the rest of this appendix. Note that, for notational convenience, we write the representation operators as $\Gamma(g)$ in this appendix, as opposed to $\Gamma^g$ used elsewhere in the text.

We now explicitly define representations of $\mathbb{C}[G]$, which are inherited from $\Gamma$ by linearity:
\begin{eqs}
    \Gamma: \cg \rightarrow \Endom (V_\Gamma), \quad  x \mapsto \Gamma(x) = \Gamma\left (\sum_g c_g g \right ) = \sum_g c_g \Gamma(g).
\end{eqs}
Here, $\Gamma$ is a map to $\mathrm{End}(V_\Gamma)$, which is the algebra of operators on $V_\Gamma$, and equivalently the span of $\mathrm{GL}(V_\Gamma)$.
It should be noted that $\mathrm{End}(V_\Gamma)$ contains all complex matrices of dimension $d_\Gamma$, not just invertible ones, thus it corresponds to the full operator algebra on $V_\Gamma$. A convenient basis for $\mathrm{End}(V_\Gamma)$ is given by the matrix units $E_{ij}$:
\begin{eqs}
    E_{ij} = \ket{i}_{\Gamma} \bra{j}_{\Gamma}, \quad \ket{i}_{\Gamma} \in {V_\Gamma}, \quad \bra{j}_{\Gamma} \in {V_\Gamma}^*,
\end{eqs}
where ${V_\Gamma}^*$ is the dual vector space. This notation makes evident the isomorphism $\mathrm{End}({V_\Gamma}) \cong {V_\Gamma} \otimes {V_\Gamma}^*$, which we use moving forward.

We now focus on irreps of $G$ and their extension to irreps of $\mathbb{C}[G]$, to show that any irreducible $\Gamma$ is a surjection from $\cg$ onto ${V_\Gamma} \otimes {V_\Gamma}^*$, and as such, its image spans the operator algebra over ${V_\Gamma}$. Such a property justifies the definition of irreducible group of gates in Def.~\ref{def: irreducible group of gates}, since the center of a matrix algebra is given by the set of scalar multiples of the identity. To give the surjection from $\cg$ onto $V_\Gamma \otimes V_\Gamma^*$, it is sufficient to show that the basis states of ${V_\Gamma} \otimes {V_\Gamma}^*$ can be written as linear combinations of $\Gamma(g)$.
We do this explicitly:
\begin{eqs}
    \Gamma(g) = \sum_{i,j} \Gamma(g)_{ij} \ket{i}_{\Gamma} \bra{j}_{\Gamma}, \quad \ket{i}_{\Gamma} \bra{j}_{\Gamma} = \frac{d_\Gamma}{|G|} \sum_{g} \Gamma(g)^*_{ij} \, \Gamma(g),
\end{eqs}
where the coefficients for for the right expression are derived using Schur orthogonality.
The left expression is just a way to rewrite the $\Gamma(g)$ matrix in terms of matrix units, whereas the right expression shows that any $\ket{i} \bra{j}$ can be expressed as a linear combination of $\Gamma(g)$, hence showing that the matrices for irrep $\Gamma$ span the operator algebra of $V_\Gamma$.

Moreover, we can construct an isomorphism between $\mathbb{C}[G]$ and the direct sum of $V_\Gamma \otimes V_\Gamma^*$ over all irreps $\Gamma$:
\begin{eqs}
\label{eq: CG to direct sum isomorphism}
    \mathbb{C}[G] \cong \bigoplus_{\Gamma \in \mathrm{Irr}(G)} V_\Gamma \otimes V_\Gamma^*.
\end{eqs}
The explicit isomorphism is given by sending each element $g$ to the direct sum of its irrep matrices:
\begin{eqs}
    g \mapsto \bigoplus_{\Gamma \in \mathrm{Irr}(G)} \Gamma(g).
\end{eqs}
In terms of bases, we can write this as:
\begin{eqs}
\label{eq: Fourier transform definition for CG}
    g \mapsto \sum_{\Gamma, 1\leq i,j \leq d_\Gamma} \Gamma(g)_{ij} \ket{i}_{\Gamma} \bra{j}_{\Gamma}, \quad \ket{i}_{\Gamma} \bra{j}_{\Gamma} \mapsto \frac{d_\Gamma}{|G|} \sum_{g} \Gamma(g)^*_{ij} \, g.
\end{eqs}
In matrix form, labeling the irreps from $\Gamma_1$ to $\Gamma_n$, where $n$ is the number of irreps, this is equivalent to:
\begin{eqs}
g \mapsto
\begin{pmatrix}
\Gamma_1(g) & \cdots & 0 \\
\vdots      & \ddots & \vdots \\
0           & \cdots & \Gamma_n(g)
\end{pmatrix}, \quad {
\setlength{\arraycolsep}{0.5pt}
\renewcommand{\arraystretch}{0.5}
\def\topentry#1{%
  \smash{\raisebox{-6pt}{$\vcenter{\hbox{$#1$}}$}}%
}
\begin{pmatrix}
\topentry{0} & & \topentry{\cdots} & & \topentry{0} \\
& \ddots & & & \\
\vdots & & |i\rangle_{{\Gamma_k}}\langle j|_{{\Gamma_k}} & & \vdots \\
& & & \ddots & \\
0 & & \cdots & & 0
\end{pmatrix}
}
\mapsto \frac{d_{\Gamma_k}}{|G|} \sum_{g} {\Gamma_k(g)}^*_{ij} \, g,
\end{eqs}
where every $\Gamma_k(g)$ block is a $d_{\Gamma_k}$-dimensional matrix, and $|i\rangle_{\Gamma_k}\langle j|_{\Gamma_k}$ is a $d_{\Gamma_k}$-dimensional matrix unit.
The basis relations given in Eq.~\eqref{eq: Fourier transform definition for CG} are respectively the $G$ Fourier transform and its inverse. For Abelian groups such a matrix decomposition of $g$ is diagonal, since all irreps are $1$D, whereas for non-Abelian groups there is always at least one block with dimension larger than $1$.

Therefore, $\cg$ has the dual basis $\{|i\rangle_{{\Gamma}}\langle j|_{{\Gamma}}\, |\, 1\leq i,j \leq d_\Gamma, \, \Gamma \in \mathrm{Irr}(G) \}$.
The isomorphism of Eq.~\eqref{eq: CG to direct sum isomorphism} is especially useful in interpreting a patch of group surface code (GSC) as housing a direct sum of pairs of dual charges, one at each of its rough boundaries. We use this analogy in Appendix~\ref{app: isc from gsc} to show how to derive a triangular patch of ISC from a rectangular patch of pGSC.

\subsection{The twisted group algebra $\cgw$}
We now move to the description of the twisted group algebra by introducing projective group representations. A projective representation $\Gamma$ of the group $G$ is defined as:\footnote{It is useful to note that, for any cocycle $\omega$ and twisted representation $\Gamma$ of $G$, there is a finite group $G'$ that has an isomorphic linear representation. Moreover, for any $G$ there is a group, called the Schur cover of $G$, which contains all twisted irreps of $G$, for any cocycle $\omega$, as linear representations.}
\begin{eqs}
    \Gamma: G \rightarrow \mathrm{GL}(V_\Gamma), \quad g \mapsto \Gammomega(g), \quad \Gammomega(g) \Gammomega(h) = \omega(g,h) \Gammomega(gh), \quad \omega: G \times G \rightarrow U(1),
\end{eqs}
where the function $\omega$ must satisfy the cocycle condition of Eq.~\eqref{eq: cocycle condition}, without which the algebra would not be associative.\footnote{We could generally have $\omega$ be a map into any Abelian group, in which case we would get a factor system, but we restrict to the case of the group of phases, $U(1)$.}

The function $\omega$ is also used to define the twisted group algebra $\cgw$, which has the same elements as $\cg$ (hence it is isomorphic to it as a vector space) but a different multiplication. The twisted multiplication, denoted as $\multomega$, reproduces the multiplication of matrices in the projective representation of $G$ with the same cocycle:
\begin{eqs}
    {x} \multomega {y} = \sum_{g,h} c_{g}c_h\, \omega(g,h)\, gh, \quad {x},{y} \in \cgw, \quad {g},{h} \in G.
\end{eqs}
To avoid confusion, we also introduce the twisted inverse operation $\invomega{g}$, which is not simply the inverse of the group element in $G$:
\begin{eqs}
\label{eq: twisted inverse}
     \invomega{h} = {\omega(h,\bar{h})}^{-1}{\bar{h}}, \quad h \multomega \invomega{h} = \frac{{\omega(h,\bar{h})}}{{\omega(h,\bar{h})}} h\bar{h} = 1.
\end{eqs}

Representations of $\cgw$ are obtained by extending projective representations of $G$ with the same $\omega$:
\begin{eqs}
    \Gammomega: \cgw \rightarrow \Endom(V_{\Gammomega}), \quad 
\end{eqs}
which means that we get the same type of isomorphism as for the untwisted case, as shown in Ref.~\cite{cheng2015character}:
\begin{eqs}
\label{eq: decomposition of twisted algebra}
    \cgw \cong \bigoplus_{\Gammomega \in \mathrm{Irr_\omega}(G)} V_{\Gammomega} \otimes V_{\Gammomega}^*.
\end{eqs}
Above, $\mathrm{Irr_\omega}(G)$ is the set of projective irreps with cocycle $\omega$. The Fourier transform and its inverse still provide the explicit isomorphism maps for Eq.~\eqref{eq: decomposition of twisted algebra}:\footnote{To get the inverse Fourier transform we need to use the twisted version of Schur orthogonality, which is given in Ref.~\cite{waldron2020fourier}. This generally contains a cocycle, which can be removed if we consider a matrix inverse $\Gamma^\omega(g)^{-1}$ rather than the matrix for the inverse group element $\Gamma^\omega(\bar{g})$, which generally differ as explained in Eq.~\eqref{eq: twisted inverse}. We use the former to remove cocycles from the inverse Fourier expression, and assume that the representation is unitary so that we end up with the usual complex conjugation of matrix elements, matching the untwisted Fourier relations.}
\begin{eqs}
\label{eq: twisted Fourier transform}
    g \mapsto \sum_{\Gammomega, 1\leq i,j \leq d_{\Gammomega}} {\Gammomega}(g)_{ij} \ket{i}_{{\Gammomega}} \bra{j}_{{\Gammomega}}, \quad \ket{i}_{{\Gammomega}} \bra{j}_{{\Gammomega}} \mapsto \frac{d_{\Gammomega}}{|G|} \sum_{g} {\Gammomega}(g)^*_{ij} \, g.
\end{eqs}

\subsection{Action of $\cgw$ on itself}
\label{subapp: action of CG on itself}
So far, we have discussed the structure of the twisted group algebra, which gives us both the local Hilbert space at each edge and the logical Hilbert space for a rectangular patch of pGSC, as shown in Appendix~\ref{app: isc from gsc}. However, to describe the operations on these Hilbert spaces we must introduce a way to have $\cgw$ act on itself.\footnote{As Hilbert spaces, $\cg$ and $\cgw$ are isomorphic since only the vector space structure is relevant, so we could equivalently take the action of $\cgw$ on $\cg$.} We do so in this section. To avoid confusion, we use ket notation for elements in the copy of $\cgw$ which is being acted on.
Formally, to define an action of $\cgw$ on itself we need to specify a map:
\begin{eqs}
    \tilde{L}: \cgw \times \cgw \rightarrow \cgw, \quad \tilde{L}^x \ket{y} \mapsto \ket{z}, \quad x,y,z \, \in \cgw,
\end{eqs}
satisfying some conditions. Rather than stating the conditions for this $\cgw$ action, we first list the conditions for the action of $G$ onto itself, which we can then extend linearly to $\cg$, and then to $\cgw$ by introducing a twist:
\begin{eqs}
\label{eq: group action conditions}
    L: G \times G \rightarrow G, \quad L^g \ket{h}\mapsto \ket{k}: \quad  L^{g_1} L^{g_2} \ket{h} = L^{g_1 g_2} \ket{h}, \quad L^1 \ket{h} = \ket{h}, \quad \forall \, g_1, g_2, h \, \in G.
\end{eqs}
The above just says that the group action must be compatible with the group multiplication, and that the identity must act trivially. Since group multiplication is associative, we can choose the group action to coincide with it, automatically satisfying both conditions:
\begin{eqs}
    L^g \ket{h} = \ket{gh}.
\end{eqs}
We can define another action from $G$ onto itself, namely right multiplication by the inverse group element, $R^g$:\footnote{The right action must be defined using the group inverse in order to satisfy Eq.~\eqref{eq: group action conditions}.}
\begin{eqs}
    R^g  \ket{h} = \ket{h \bar{g}}.
\end{eqs}
The actions defined above are called the left and right regular representations, are reducible, isomorphic to each other, and have a direct sum decomposition which contains each irrep of $G$ with a multiplicity equal to its dimension:
\begin{eqs}
    L \cong R \cong \bigoplus_{\Gamma \in \mathrm{Irr}} d_\Gamma \Gamma.
\end{eqs}
Under the simultaneous action of $L$ and $R$, the underlying vector space $\cg$ decomposes as the direct sum of irrep spaces given in Eq.~\eqref{eq: CG to direct sum isomorphism}, on each of which the regular representations have action:
\begin{eqs}
\label{eq: action of L and R on irrep spaces}
    L (V_\Gamma \otimes V_\Gamma^*) = (\Gammomega \otimes I_{\Gammomega^*})(V_\Gamma \otimes V_\Gamma^*), \quad R (V_\Gamma \otimes V_\Gamma^*) =  (I_{{\Gammomega}} \otimes \Gamma^*) (V_\Gamma \otimes V_\Gamma^*).
\end{eqs}

Both actions can be extended linearly to provide a $\cg$ action on $\cg$:\footnote{Note that $R^x y$ is not equal to $y \bar{x}$ since elements of $\cg$ are not necessarily invertible. Instead, it is a sum over right actions by its group basis components.}
\begin{eqs}
    L^x \ket{y} = \sum_{g,h} c_g c_h \ket{gh}, \quad R^x \ket{y} = \sum_{h,g} c_h c_g \ket{h\bar{g}}.
\end{eqs}
The action can be defined for the twisted case in a straightforward way, since all we required from the group multiplication was associativity, which also holds for a nontrivial $\omega$. We thus define the projective left and right actions of $G$ on itself:
\begin{eqs}
\label{eq: left and right twisted actions}
    \tilde{L}^g \ket{h} =& \ket{g \multomega h} = \omega(g,h) \ket{g h}\\ 
    \tilde{R}^g \ket{h} =& \ket{h \multomega \invomega{g}} = \omega(h,\bar{g}) \ket{h \invomega{g}} =  \frac{\omega(h,\bar{g})}{\omega(g, \bar{g})} \ket{h \bar{g}} = \omega(h \bar{g}, g)^{-1} \ket{h \bar{g}},
\end{eqs}
where we used Eq.~\eqref{eq: twisted inverse} and the cocycle condition to simplify the $\tilde{R}^g$ action. Again, we can extend this to $\cgw$:
\begin{eqs}
     \tilde{L}^x \ket{y} = \sum_{g,h} c_g c_h \omega(g,h) \ket{gh}, \quad \tilde{R}^x \ket{y} = \sum_{h,g} c_h c_g \omega(h \bar{g}, g)^{-1} \ket{h \bar{g}}.
\end{eqs}

In Appendix~\ref{app: isc from gsc} we use the logical right action $\tilde{R}_\LL$ on a pGSC to form the logical projector from the full Hilbert space $\cgw$ onto an irrep space $V_\Gammomega \otimes V_\Gammomega^*$, in order to obtain an ISC.
Working with the Fourier basis we can intuitively understand the formulae for such projectors, which are given by:
\begin{eqs}
\label{eq: projection onto single irrep}
    \Pi_\Gamma = \frac{d_\Gamma}{|G|}\sum_g \chi_\Gamma(g)\, \tilde{R}^g_\LL,
\end{eqs}
where we have $\chi_\Gammomega$ rather than $\chi_\Gammomega^*$ since $\tilde{R}_\LL^g$ acts via the dual irrep as shown in Eq.~\eqref{eq: action of L and R on irrep spaces}, and $\Gammomega^*$ is the complex conjugate of $\Gammomega$.
By using the Fourier transform of Eq.~\eqref{eq: Fourier transform definition for CG}, we then have:
\begin{eqs}
    \Pi_\Gamma \mapsto & \left ( \frac{d_\Gamma}{|G|}\sum_{g,k} \Gamma(g)_{kk}  \sum_{\Gamma, 1\leq i,j \leq d_\Gamma} \Gamma(g)^*_{ij} \ket{i}_{\Gamma^*} \bra{j}_{\Gamma^*} \right) \\
    =& \frac{d_\Gamma}{|G|} \sum_{i,j,k, \Gamma} \frac{|G|}{d_\Gamma} \delta_{k,i} \delta_{k,j} \delta_{\Gamma, \Gamma} \ket{i}_{\Gamma^*} \bra{j}_{\Gamma^*} \\
    =& \sum_k \ket{k}_{\Gamma^*} \bra{k}_{\Gamma^*} \\
    =& {
\setlength{\arraycolsep}{0.5pt}
\renewcommand{\arraystretch}{0.5}
\def\topentry#1{%
  \smash{\raisebox{-6pt}{$\vcenter{\hbox{$#1$}}$}}%
}
\begin{pmatrix}
\topentry{0} & & \topentry{\cdots} & & \topentry{0} \\
& \ddots & & & \\
\vdots & & I_{\Gamma^*} & & \vdots \\
& & & \ddots & \\
0 & & \cdots & & 0
\end{pmatrix},
}
\end{eqs}
where we used the fact that $\chi$ is a trace and Schur orthogonality for the first equality.
$I_{\Gamma^*}$ is the identity on $V_\Gamma^*$, so that we obtain the projector onto the $V_\Gamma \otimes V_\Gammomega^*$ subspace. 

\section{Irrep surface codes from projective group surface codes} \label{app: isc from gsc}
In this appendix, we introduce pGSCs and describe the connection between pGSCs on a rectangular patch and ISCs on a triangular patch. To do so, we generalize the construction given in Ref.~\cite{manjunath2026GSC} for group surface codes (GSCs) to the projective case, in which the top boundary is twisted by a cocycle $\omega$. Finally, we show that we can deform a rectangular pGSC patch into a triangular patch with the same code space and logical operations as the ISC, hence drawing a direct connection between the two constructions.

We now work with a rectangular lattice with twisted top boundary, smooth bottom boundary, and rough left and right boundaries:
\begin{eqs}
    \vcenter{\hbox{\includegraphics[scale=.25]{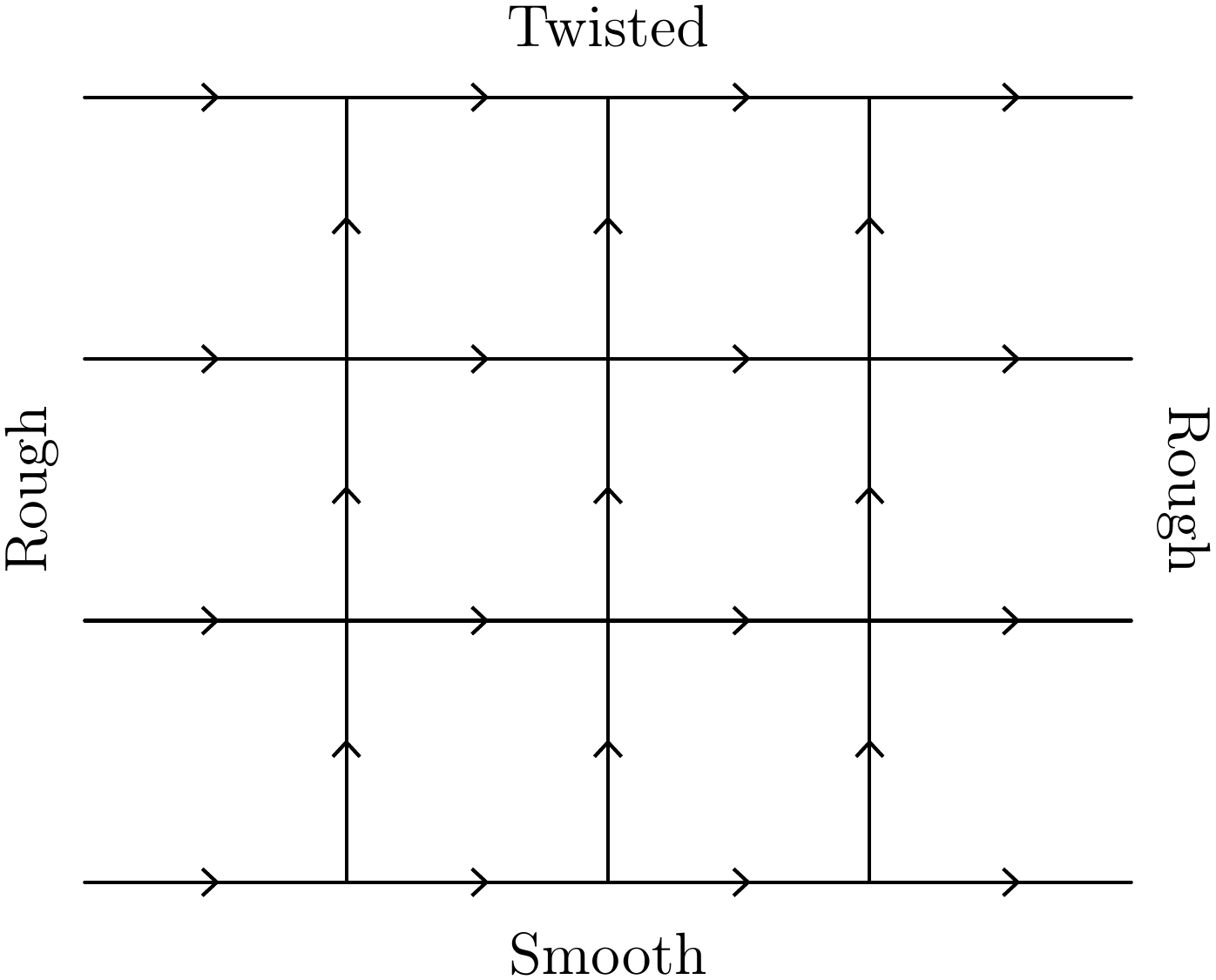}}}.
\end{eqs}
The stabilizers are defined analogously to the ISC, except for the stabilizers on the twisted boundary, where the vertex stabilizers are supported on three edges, and the plaquette stabilizers are supported on four.

\subsection{Logical Hilbert space}
To determine the dimension of the logical Hilbert space, we follow a similar argument to that of Ref.~\cite{manjunath2026GSC} and Section~\ref{sec: code space}, by constructing a set of states that are invariant under all plaquette and vertex stabilizers. We start from a flux free configuration in which the left rough edges have group element $g$, and all other edges have the trivial element $1$:
\begin{eqs}
    \vcenter{\hbox{\includegraphics[scale=.2]{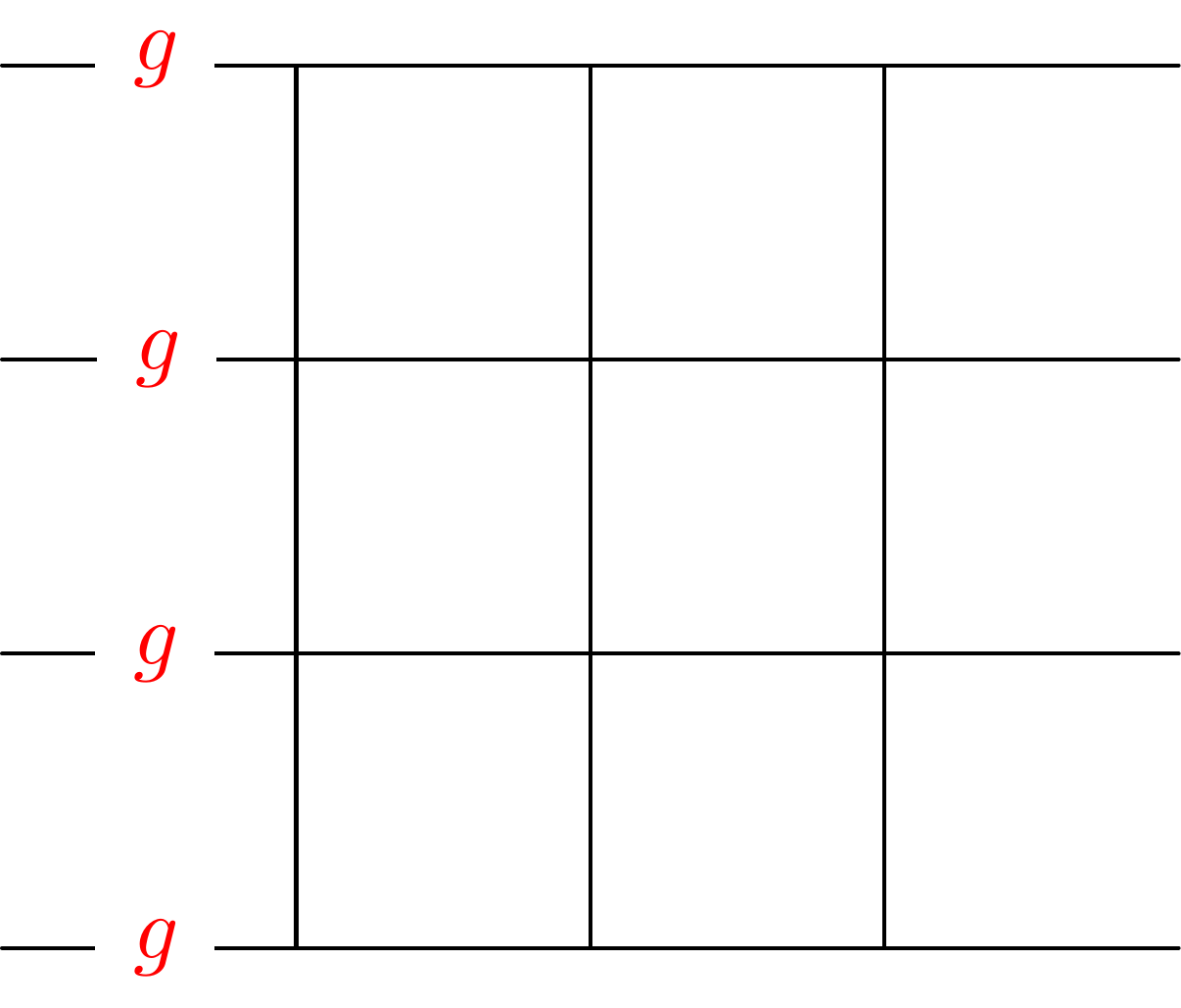}}}.
\end{eqs}
We call this configuration the left gauge, and denote its state by $\ket{\bm{g_\ell}}$. This configuration is flux-free by construction, since the plaquettes on the left rough boundary all have flux $g\bar{g}=1$. We now project onto the common $+1$ eigenspace of all vertex stabilizers by the operator defined in Eq.~\eqref{eq: PiA decomposition}:
\begin{eqs}
    \Pi_A = \prod_v A_v^G.
\end{eqs}
The resulting state $\ket{g}_\mathsf{L}$ is now also invariant under all vertex stabilizers by Eq.~\eqref{eq: absorbing gauge transformations}, and is therefore a code state:
\begin{eqs}
    \ket{g}_\mathsf{L} = \Pi_A \ket{\bm{g_\ell}}, \quad B_p\ket{g}_\mathsf{L} = A_v^h \ket{g}_\mathsf{L} = \ket{g}_\mathsf{L}, \quad \forall \, h \in G.
\end{eqs}
It can be checked that the set $\{\glog{g}\, | \, g \in G\}$, in bijection with set of left-gauge configurations, forms a basis for the logical Hilbert space. The wavefunction for each code state can be written explicitly as:
\begin{eqs}
    \glog{g} = \sum_{\bm{h_e} \in [\bm{g_e}]} \frac{1}{\Phi(\bm{h_e})} \ket{\bm{h_e}}.
\end{eqs}
In the above, $\ket{\bm{h_e}}$ is a configuration which is in the equivalence class $[\bm{g_e}]$, the same as the left gauge configuration $\ket{\bm{g_\ell}}$, meaning that the two configurations are related by the action of vertex stabilizers. Note that any configuration $\ket{\bm{h_e}}$ has a $g$ holonomy on the twisted boundary. The factor $\Phi(\bm{h_e})$, whose explicit form is not essential for the discussion, is the phase accumulated by projectively multiplying all the elements on the twisted boundary, ensuring that $\glog{g}$ is invariant under the vertex stabilizers. This recovers the code state wave functions for the GSC of Ref.~\cite{manjunath2026GSC} by setting $\Phi(\bm{h_e}) = 1$ for all configurations. As a result, the code states are labeled by group elements, and we can identify the logical Hilbert space with the (vector space underlying the) twisted group algebra $\cgw$. 

\subsection{Logical operations}

We now consider two types of transversal logical operations that can be carried out on a pGSC, namely left and right logical multiplication, which match the action of $\cgw$ on itself described in App.~\ref{subapp: action of CG on itself}. On the lattice, these are obtained by left (right) multiplication on each edge of the left (right) rough edge, where the multiplication on the twisted boundary includes a cocycle:
\begin{eqs}
    \tilde{L}^g_\LL = \quad \vcenter{\hbox{\includegraphics[scale=.2]{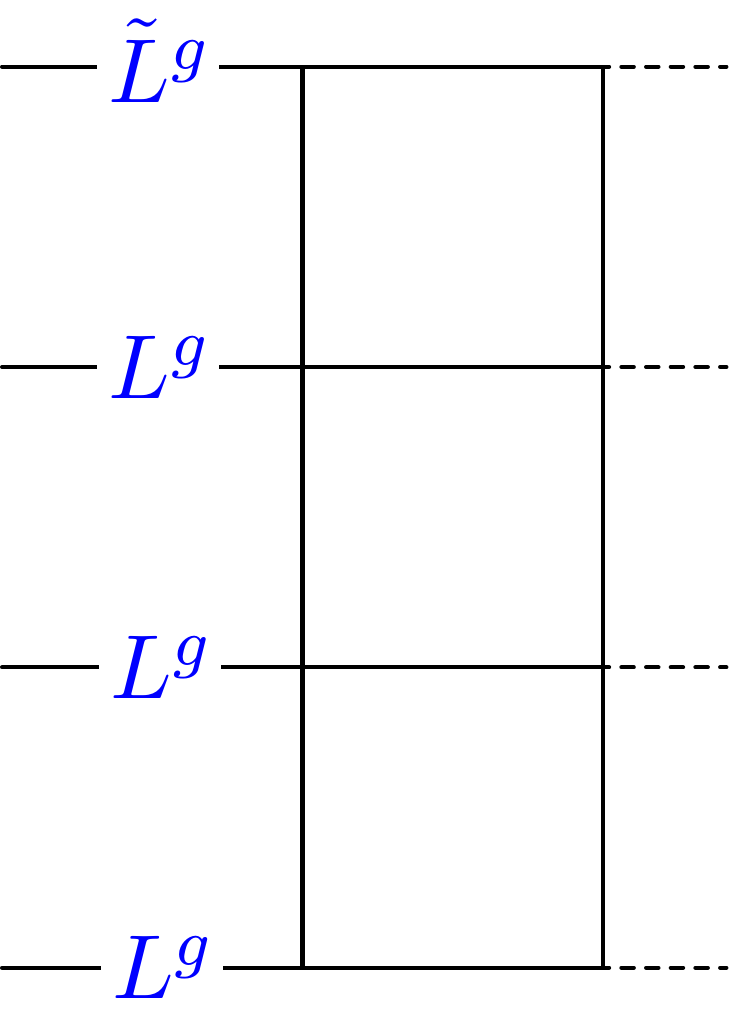}}} \quad, \qquad \tilde{R}^g_\LL = \quad \vcenter{\hbox{\includegraphics[scale=.22]{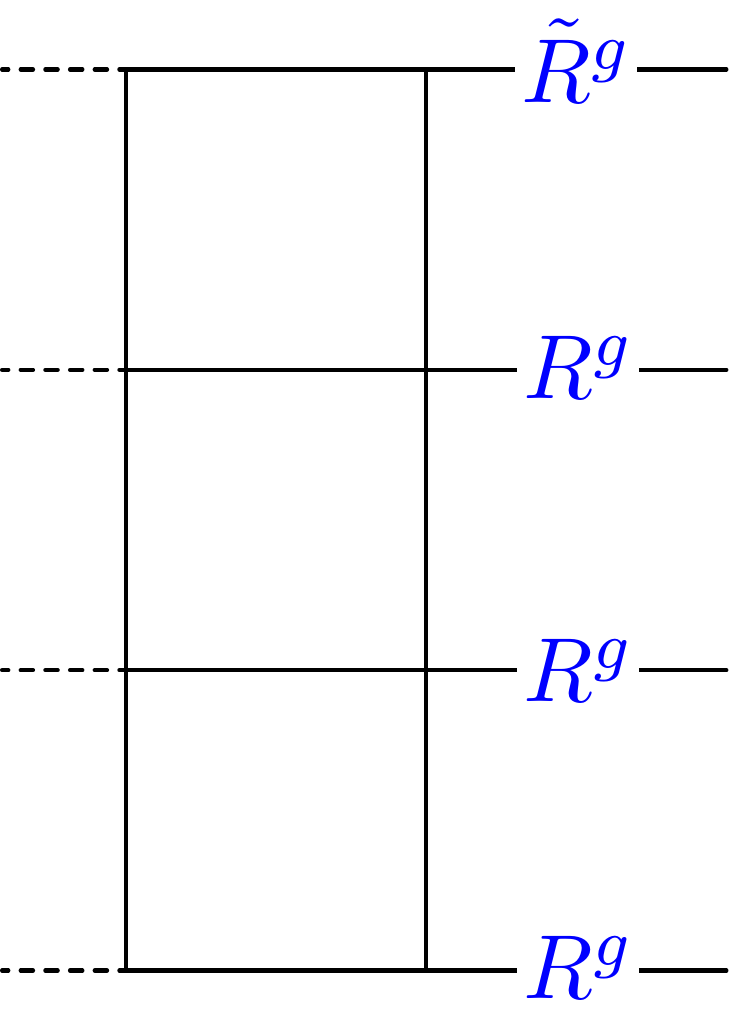}}}\quad .
\end{eqs}
The action of these operators on the code states can be checked explicitly:
\begin{eqs}
    \tilde{L}_\LL^h \glog{g} = \tilde{L}^h_\LL \Pi_A \ket{\bm{g_\ell}} = \Pi_A \tilde{L}^h_\LL \ket{\bm{g_\ell}} = \omega(h,g) \Pi_A \ket{\bm{hg_\ell}} = \omega(h,g)\ket{hg}_\LL,
\end{eqs}
where we used the fact that $\tilde{L}_\LL^h$ commutes with all stabilizers, and the cocycle comes from the twisted multiplication on the top edge.
As a result, $\tilde{L}_\LL^h$ acts as the twisted multiplication by $h$. The right multiplication similarly acts as the right twisted group action:
\begin{eqs}
    \tilde{R}_\LL^h\ket{g}_\LL = \omega(g \bar{h},h)^{-1} \ket{h\bar{g}}_\LL.
\end{eqs}
This shows that the logical action on the Hilbert space is given by $\cgw$ acting on itself from the left and right, as per Eq.~\eqref{eq: left and right twisted actions}.
The twisted logical action implies that the logical Hilbert space can be decomposed according to Eq.~\eqref{eq: decomposition of twisted algebra}:
\begin{eqs}
    \cgw \cong \bigoplus_{\Gammomega \in \mathrm{Irr_\omega}(G)} V_{\Gammomega} \otimes V_{\Gammomega}^*,
\end{eqs}
where $\tilde{L}_\LL$ only acts nontrivially on each $V_{\Gammomega}$ and $\tilde{R}_\LL$ on each $V^*_{\Gammomega}$:
\begin{eqs}\label{eq:left and right}
    \tilde{L}_\LL^g = \bigoplus_{\Gammomega} \Gammomega^g \otimes I_{{\Gammomega^*}}, \quad \tilde{R}_\LL^g = \bigoplus_{\Gammomega} I_{{\Gammomega}} \otimes {\Gammomega^{g}}^*.
\end{eqs}

The above decomposition, alongside the fact that charges are labeled by irreps of $G$, provides the following interpretation for the logical Hilbert space: it comes from the direct sum of all of the (twisted) charge-dual pairs, with the charge living on the left rough boundary (or equivalently, the virtual vertex it houses), and the dual charge living on the right rough boundary.

Consequently, there is a clear difference between the representation actions in the pGSC and ISC: in the pGSC, both left and right multiplication act by the regular representation, whereas in the ISC the left multiplication acts as a single projective irrep. The difference is the number of blocks in the direct sum: in the pGSC, all twisted irreps appear, whereas in the ISC, we have a single one.\footnote{We should note that for linear representations there are always multiple irreps, so that $\cg$ has a direct sum decomposition with multiple blocks. This is not generally true for twisted irreps. For example, the twisted sector of $\ZZ_2^2$ has a single $2$D irrep, so that the underlying logical Hilbert space already has a tensor product decomposition, even in the rectangular geometry.}
The above interpretation of the Hilbert space suggests that, in the ISC, there is a single fixed charge living on the rough boundary, giving the logical Hilbert space, with a corresponding dual charge living on the right corner.

Naively, it would then be sufficient to project down to a single block in the direct sum of the pGSC, to then land in a single irrep subspace. Assuming we want to project onto projective irrep $\Gamma$, we can use Eq.~\eqref{eq: projection onto single irrep}:\footnote{We apply the projector using right multiplication to match the rectangle-to-triangle deformation in the next section. Using left multiplication would yield an equally valid projector.}
\begin{eqs}
    \Pi_\Gamma \, (\cg) \cong \frac{d_\Gamma}{|G|}\sum_g \chi_\Gamma(g)\, \tilde{R}_\LL^g \, (\cg) = V_\Gamma \otimes V_\Gamma^*.
\end{eqs}
In this space, the representation operators act as the $\Gamma$ projective irrep:
\begin{eqs}
     \tilde{L}_\LL^g = \Gammomega^g \otimes I_{{\Gammomega^*}}, \quad \tilde{R}_\LL^g = I_{{\Gammomega}} \otimes {\Gammomega^{g}}^*.
\end{eqs}
However, the support of this projector grows linearly with the system size, so it is not feasible to continuously measure it to remain in the irrep space. In the next section, we circumvent this by deforming the lattice from a rectangle to a triangle. In the latter geometry, the projector becomes a single-edge operator.

\subsection{Deforming the pGSC lattice}
In this section, we show that we can start from a rectangular pGSC lattice and obtain a triangular lattice with a single rough edge on the right boundary by a series of unitary transformations. Since the two configurations are related by unitary transformations, the underlying logical Hilbert space is conserved. Pictorially, we show that there exists a unitary relating the two geometries via an intermediate step, given by a trapezoidal lattice:
\begin{eqs}
    \vcenter{\hbox{\includegraphics[scale=.17]{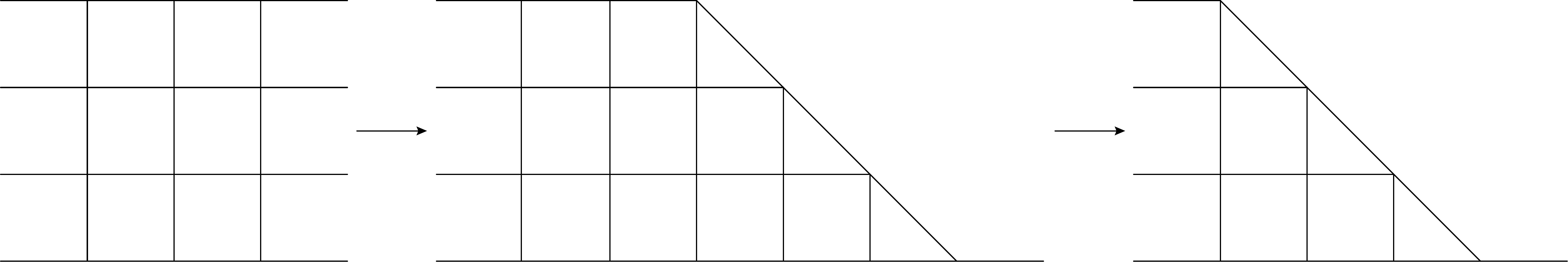}}}.\!\!\!\!\!\!
\end{eqs}
In the triangular configuration, the logical projector onto a given projective irrep sector $V_\Gamma \otimes V_\Gamma^*$ becomes localized to a single edge, and no longer depends on the system size. 
After such a projection, half of the logical Hilbert space can be discarded, since it is fixed by the projector, ending up with a $d_\Gamma$-dimensional logical qudit, matching the ISC. 

We divide the lattice deformation into two steps: \textbf{(i)} rectangle to trapezoid and \textbf{(ii)} trapezoid to triangle. Combining $\textbf{(i)}$ with $\textbf{(ii)}$, we obtain the desired deformation from rectangle to triangle. We introduce five local unitaries as the building blocks for the lattice deformations, namely $U_p'$, $U_v'$, $U_c$, $U_p$ and $U_v$, where the subscripts respectively indicate plaquettes, vertices and corners. We give explicit lattice realizations for all of these local unitaries in terms of controlled group actions. Before doing so, we define pictorial representations of the two body unitaries following the notation of Ref.~\cite{manjunath2026GSC}, where the solid black circle on the left always denotes the source, the target is represented on the right by a dotted or crossed circle, and arrows distinguish between an action controlled on the source or its inverse (if the target is on the twisted boundary, then we use the twisted actions $\tilde{L},\tilde{R}$):
\begin{eqs}
    CL = \vcenter{\hbox{\includegraphics[scale=.3]{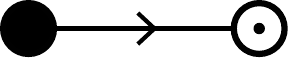}}}, \quad CL\ket{g,h} = \ket{g,gh}, \quad &C\bar{L} = \vcenter{\hbox{\includegraphics[scale=.3]{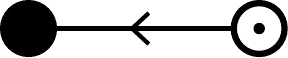}}}, \quad C\bar{L}\ket{g,h} = \ket{g,\bar{g}h} \\
    CR = \vcenter{\hbox{\includegraphics[scale=.3]{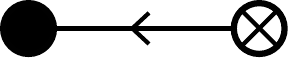}}}, \quad CR\ket{g,h} = \ket{g,h\bar{g}}, \quad &C\bar{R} = \vcenter{\hbox{\includegraphics[scale=.3]{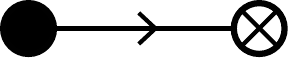}}}, \quad C\bar{R}\ket{g,h} = \ket{g,hg}.
\end{eqs}

We first tackle the deformation from rectangle to trapezoid. For this, we introduce extra qudits to the right of the rough boundary, which we couple to the rest of the lattice using controlled group actions. The first unitary we define is $U_C$:
\begin{eqs}
\label{eq: Uc definition}
    U_C: \,  \vcenter{\hbox{\includegraphics[scale=.3]{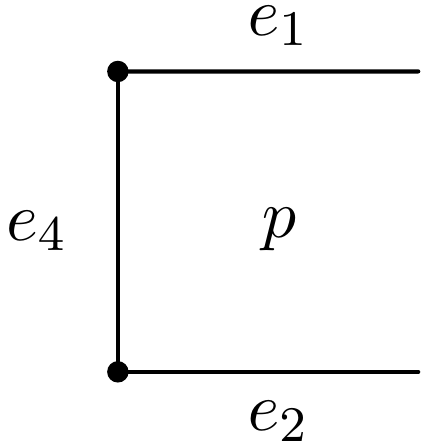}}} \mapsto \vcenter{\hbox{\includegraphics[scale=.3]{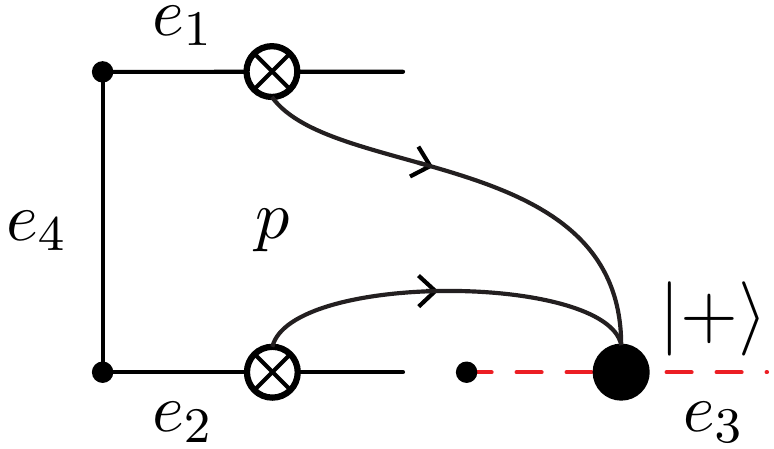}}} = \vcenter{\hbox{\includegraphics[scale=.3]{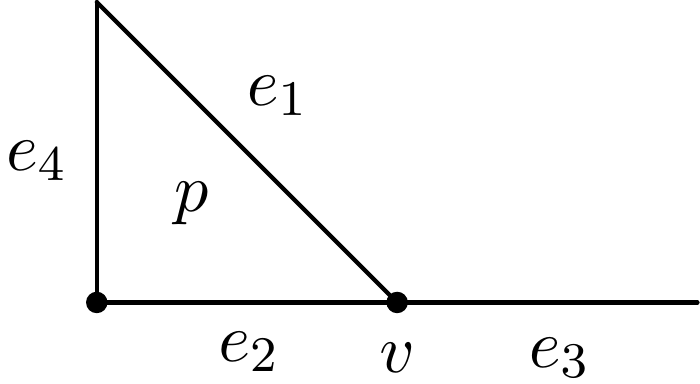}}}
\end{eqs}
Before applying $U_c$, we introduce an extra $G$ qudit in the $\ket{+}$ state on edge $e_3$, which can be thought of as being stabilized by a single-edge vertex stabilizer. The action of $U_c$ couples $e_3$ to both $e_1$ and $e_2$. After the $U_c$ action, there is an extra vertex term at vertex $v$ entangling $e_3$ to the rest of the lattice, while the plaquette stabilizer at plaquette $p$ is invariant.

Next, we define $U_p'$:
\begin{eqs}
\label{eq:Up' definition}
    U_p': \,  \vcenter{\hbox{\includegraphics[scale=.3]{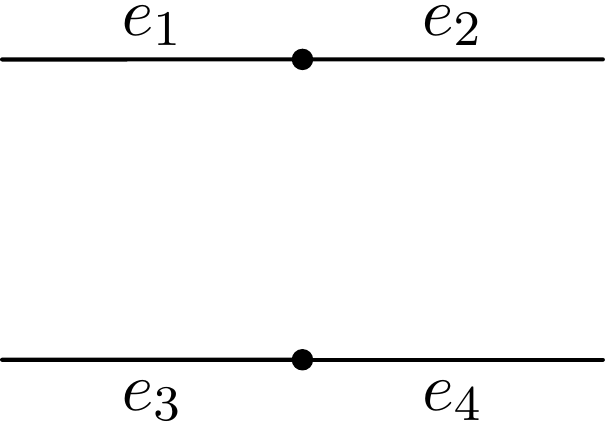}}} \mapsto \vcenter{\hbox{\includegraphics[scale=.3]{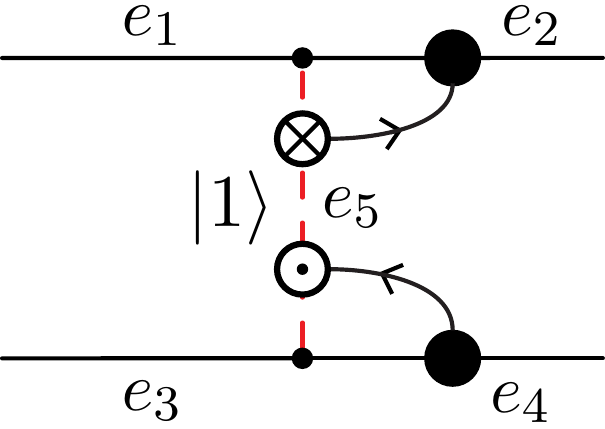}}} = \vcenter{\hbox{\includegraphics[scale=.3]{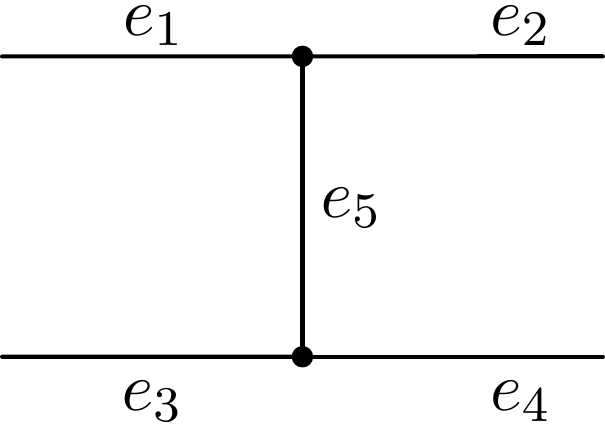}}}
\end{eqs}
This operator takes the extra qudit on edge $e_5$, initially in the $\ket{1}$ state, and entangles it to the $e_2$ and $e_4$ degrees of freedom. After applying $U_p'$, there is now a new plaquette term involving $e_2,e_4,e_5$, as well as vertex operators coupling $e_1,e_2,e_5$ and $e_3,e_4,e_5$.

Finally, we define $U_v'$:
\begin{eqs}
\label{eq:Uv' definition}
    U_v': \,  \raisebox{-2.2mm}{$\vcenter{\hbox{\includegraphics[scale=.3]{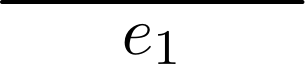}}}$} \mapsto \vcenter{\hbox{\includegraphics[scale=.3]{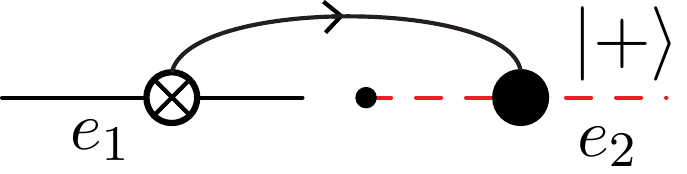}}} = \raisebox{-2.2mm}{$\vcenter{\hbox{\includegraphics[scale=.3]{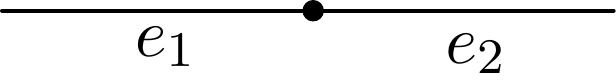}}}$}
\end{eqs}
Similar to $U_c$, we first introduce a $G$ qudit in the $\ket{+}$ state on $e_2$. Afterwards, we entangle it with the qudit on $e_1$, effectively introducing a two-edge vertex term coupling $e_1,e_2$.

\begin{figure*}[t]
        \centering
        \includegraphics[scale=.18]{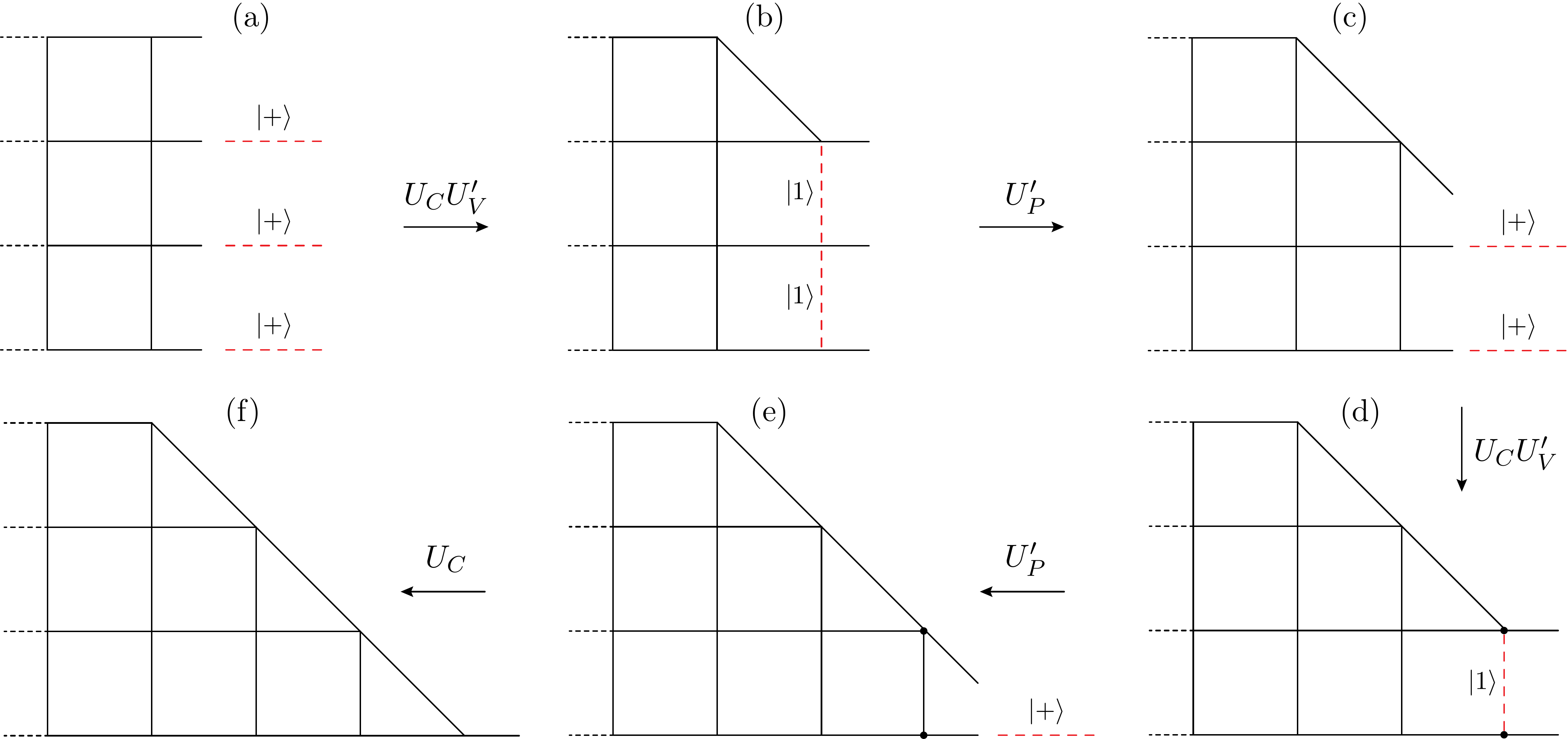}
            \caption{Intermediate states connecting the rectangular and trapezoidal geometries. The red dashed edges denote unentangled qudits which are added to the lattice, whereas the unitaries that connect pairs of consecutive configurations are indicated next to each arrow. The explicit form of such unitaries is given in Fig.~\ref{fig: rectangle to trapezoid unitaries}.} 
    \label{fig: rectangle to trapezoid states}
\end{figure*}
\begin{figure*}[t]
        \centering
        \includegraphics[scale=.18]{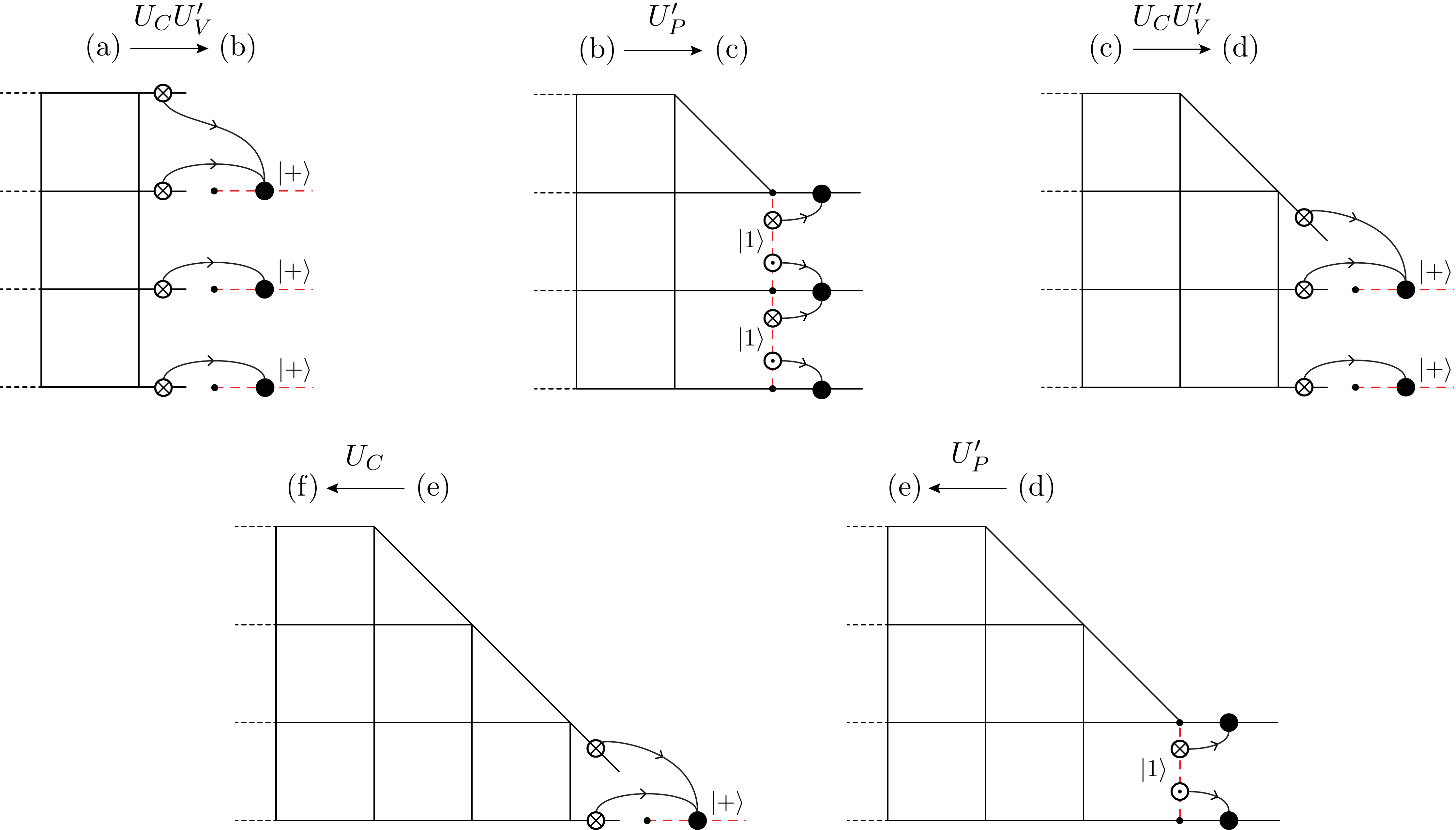}
            \caption{Explicit form of the unitaries connecting the intermediate states in the rectangle to trapezoid deformation shown in Fig.~\ref{fig: rectangle to trapezoid states}. $U_P'$ and $U_V'$ are the products along the rough boundary of the individual unitaries defined in Eqs.~\eqref{eq:Up' definition} and \eqref{eq:Uv' definition}.} 
    \label{fig: rectangle to trapezoid unitaries}
\end{figure*}

Each step of the deformation is obtained by adding $G$ qudits either in the $\ket{+}$ or $\ket{1}$ states, and then acting with $U_c$, $U_v'$ and $U_p'$. More specifically, we define the products $U_P'$ and $U_V'$, which are respectively the product of all of the $U_p'$ and $U_v'$ in a column on the rough boundary, as shown in Fig.~\ref{fig: rectangle to trapezoid unitaries}.
The deformation is then obtained by recursively applying $U_C$, $U_V'$ and $U_P'$, in that order. Fig.~\ref{fig: rectangle to trapezoid states} shows each intermediate state explicitly, whereas Fig.~\ref{fig: rectangle to trapezoid unitaries} shows the unitary transformation between each intermediate state.

The next step is to describe how to shrink the trapezoid into a triangle. This process is, in some sense, dual to what was done so far, since we have to remove edges from the trapezoid by disentangling their degrees of freedom from the rest of the lattice. We do so by introducing two unitaries, $U_v$ and $U_p$, which respectively disentangle horizontal and vertical edges on the left rough boundary. $U_p$ is defined as:
\begin{eqs}
\label{eq:Up definition}
    U_p: \,  \vcenter{\hbox{\includegraphics[scale=.3]{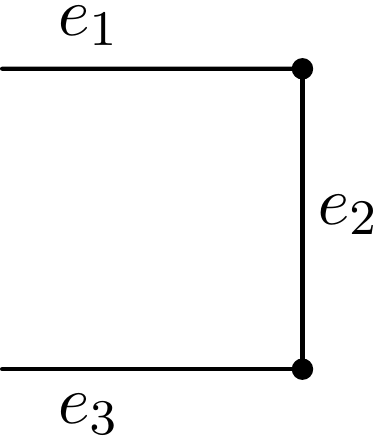}}} \mapsto \vcenter{\hbox{\includegraphics[scale=.3]{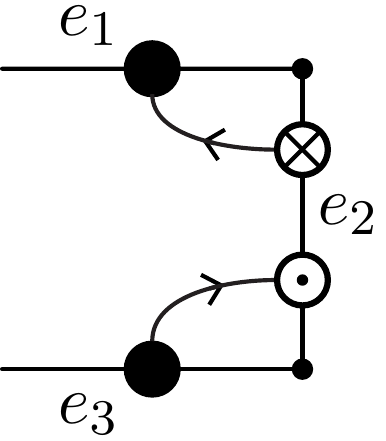}}} = \vcenter{\hbox{\includegraphics[scale=.3]{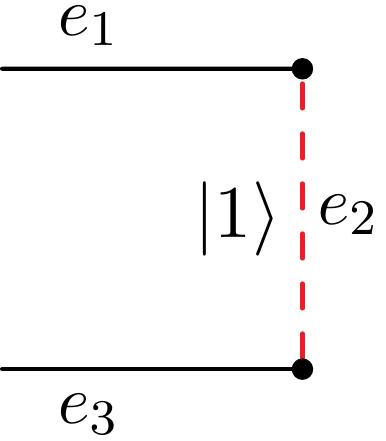}}}.
\end{eqs}
The controlled action of $e_1,e_3$ onto $e_2$ transforms the $e_1,e_2,e_3$ plaquette into a single-edge plaquette constraint on $e_2$, which fixes the corresponding qubit to $\ket{1}$, disentangling it from the rest of the lattice. Moreover, after the $U_p$ action, $e_2$ is no longer in the support of any vertex stabilizer.

$U_v$ is defined as:
\begin{eqs}
\label{eq:Uv definition}
    U_v: \,  \raisebox{-2.4mm}{$\vcenter{\hbox{\includegraphics[scale=.3]{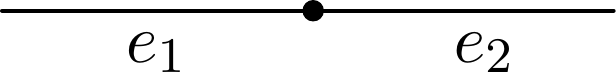}}}$} \mapsto \vcenter{\hbox{\includegraphics[scale=.3]{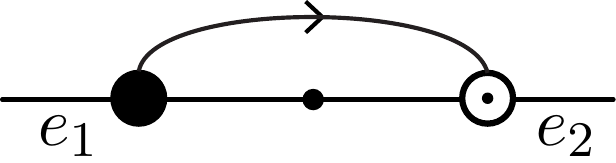}}} = \raisebox{0mm}{$\vcenter{\hbox{\includegraphics[scale=.3]{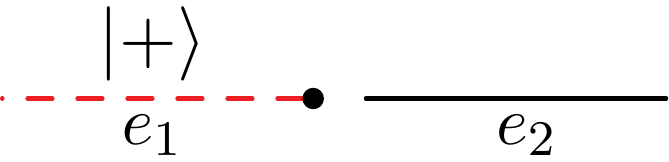}}}$}.
\end{eqs}
It acts on a pair of edges $e_1,e_2$ which are connected by a $2$-vertex stabilizer, disentangling $e_1$ from the rest of the lattice and setting its qudit to the $\ket{+}$ state.

\begin{figure*}[t]
        \centering
        \includegraphics[scale=.18]{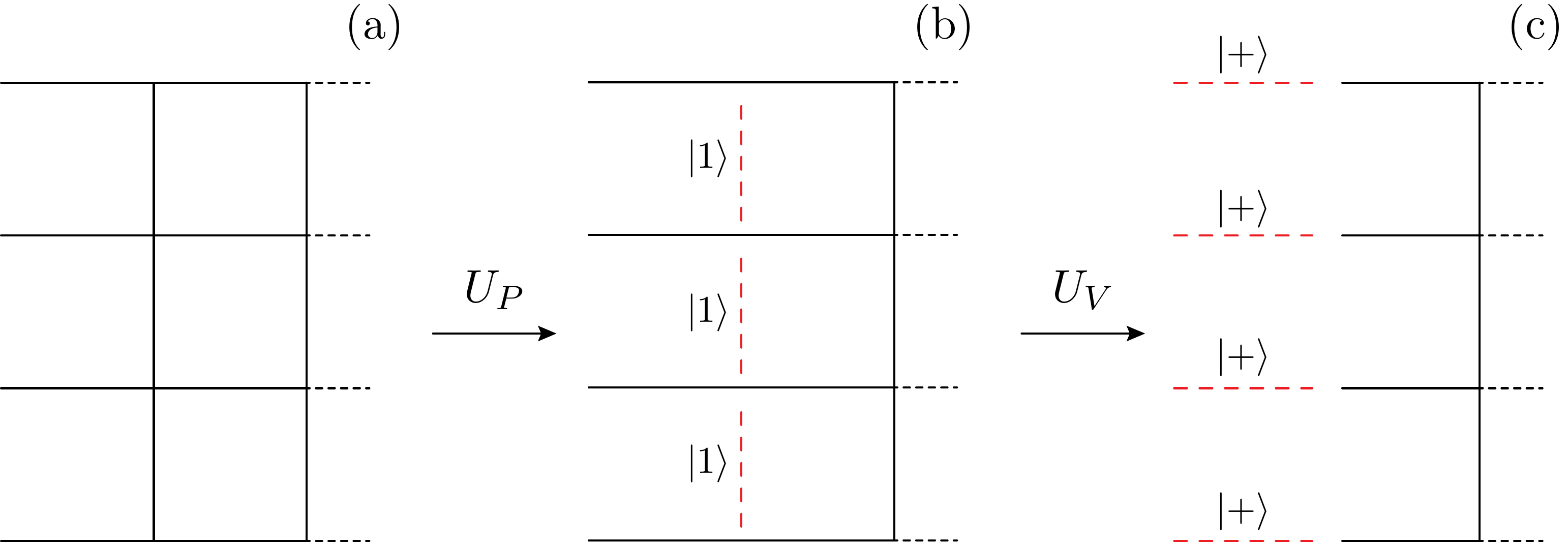}
            \caption{Intermediate states for the trapezoid to triangle deformation. The red dashed edges denote unentangled qudits which are added to the lattice, whereas the unitaries that connect pairs of consecutive configurations are indicated next to each arrow. The explicit form of such unitaries is given in Fig.~\ref{fig: rectangle shrinking unitaries}.} 
    \label{fig: rectangle shrinking states}
\end{figure*}

\begin{figure*}[t]
        \centering
        \includegraphics[scale=.18]{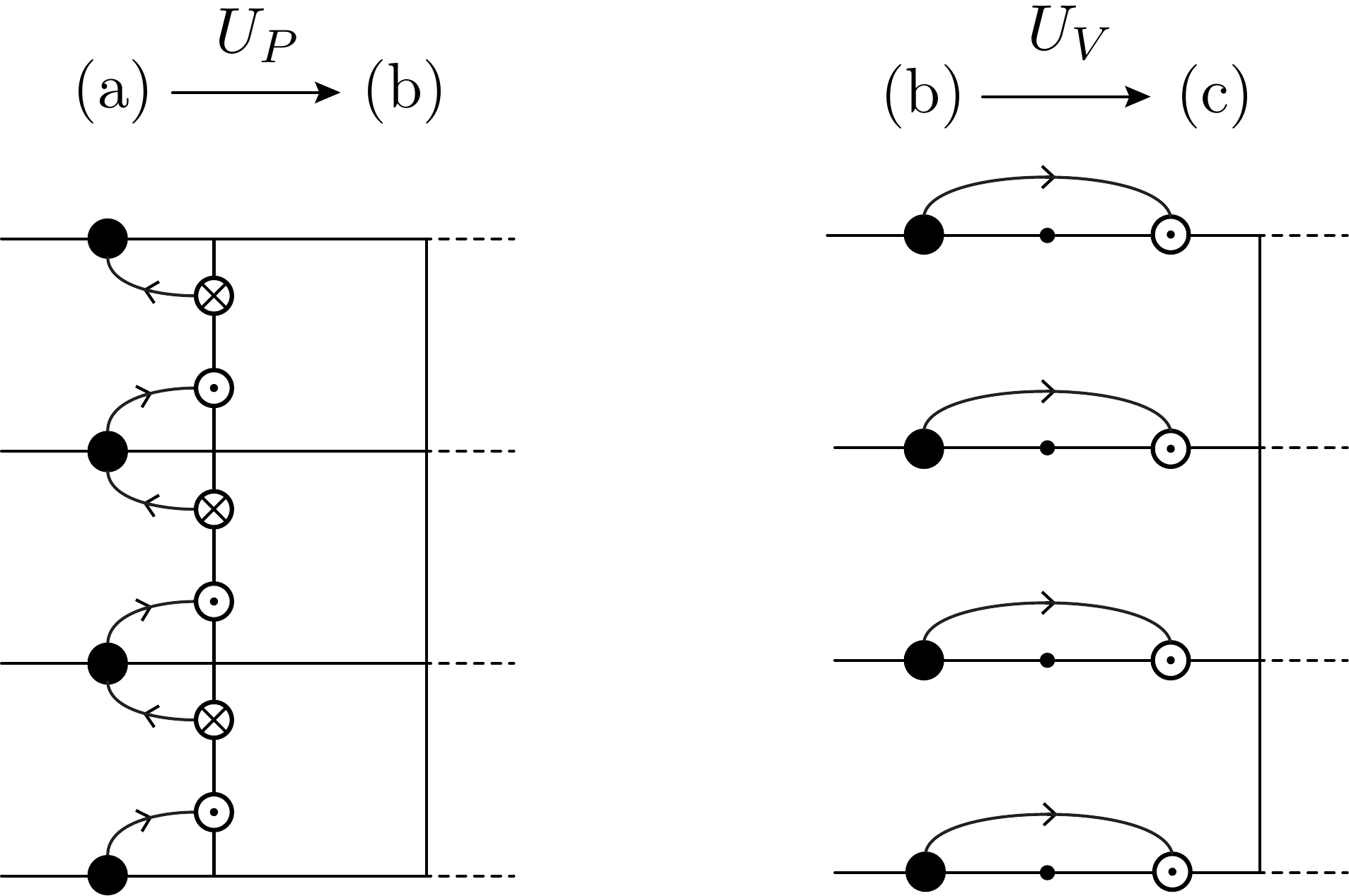}
            \caption{Explicit form of the unitaries connecting the intermediate states in the trapezoid to triangle deformation shown in Fig.~\ref{fig: rectangle shrinking states}. $U_P$ and $U_V$ are the products along the rough boundary of the individual unitaries defined in Eqs.~\eqref{eq:Up definition} and \eqref{eq:Uv definition}.} 
    \label{fig: rectangle shrinking unitaries}
\end{figure*}

In analogy to the previous deformation, we define $U_P$ and $U_V$ to be the product of the individual $U_p$, $U_v$ on the left boundary. This allows us to show the shrinking process from trapezoid to triangle in Fig.~\ref{fig: rectangle shrinking states}. Since shrinking is done by a sequence of identical steps, Fig.~\ref{fig: rectangle shrinking states} shows one of these steps, and Fig.~\ref{fig: rectangle shrinking unitaries} shows the explicit form of the unitaries connecting the intermediate states in this step.

By the deformations described so far, we have shown that the pGSC logical Hilbert space is invariant under the deformation from rectangle to triangle. In the triangular geometry, there is a single edge on the right rough boundary, so that the projector onto a logical irrep subspace of Eq.~\eqref{eq: projection onto single irrep} is now local, with support on a single edge. The action of $\Pi_\Gamma$ projects the rough edge Hilbert space onto the $V_\Gamma \otimes V_\Gamma^*$ subspace, which is now a tensor product space of dimension $d_\Gamma^2$. Since the vertex stabilizer on the bottom right corner acts with left multiplication onto the right rough edge, only the $V_\Gamma$ space is acted upon non-trivially, whereas $V_\Gamma^*$ becomes unentangled with the rest of the system. As a result, the $V_\Gamma^*$ degree of freedom can be removed, leaving only the $V_\Gamma$ entangled to the rest of the lattice. This can be represented by a $d_\Gamma$-dimensional qudit, which is the $\Gamma$-corner of the main text. Pictorially, the above process is given by:
\begin{eqs}
    \Pi_\Gamma: \, \vcenter{\hbox{\includegraphics[scale=.28]{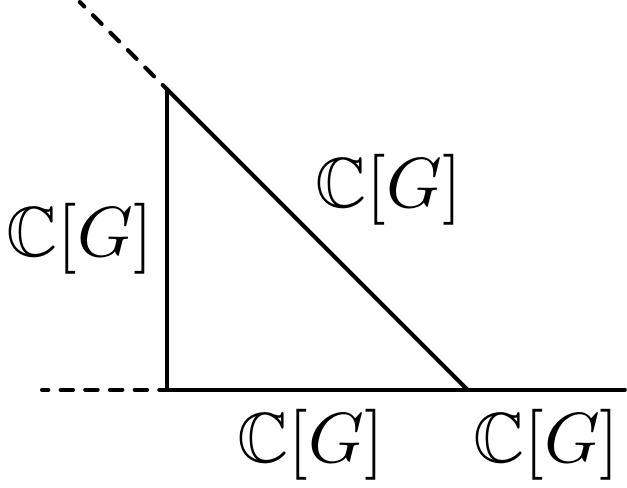}}} \mapsto \vcenter{\hbox{\includegraphics[scale=.28]{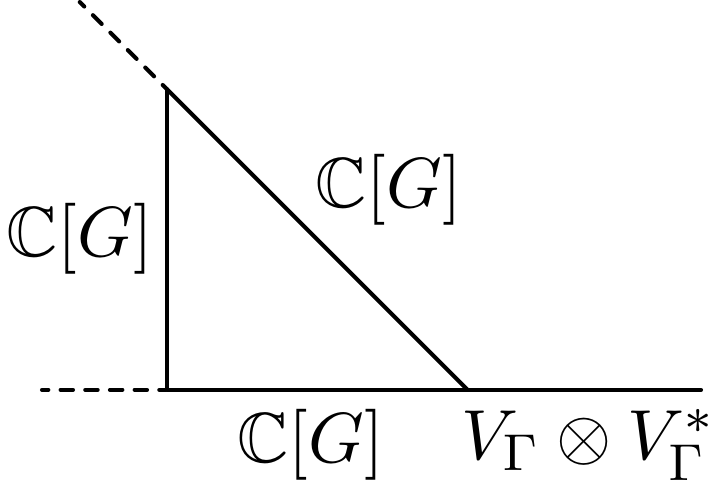}}} = \vcenter{\hbox{\includegraphics[scale=.28]{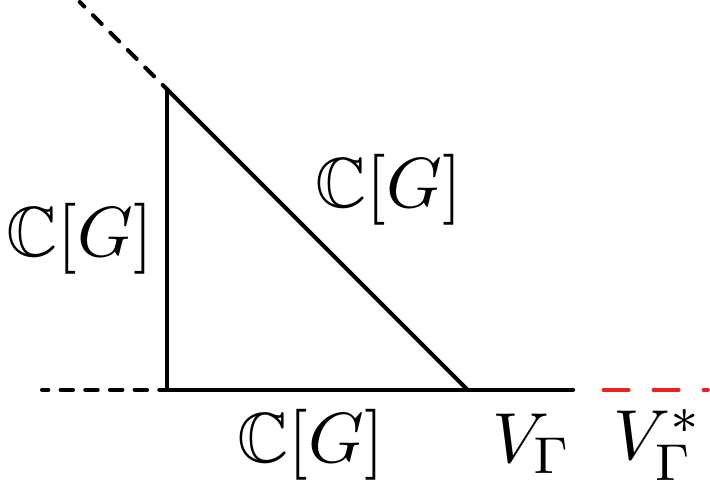}}} = \vcenter{\hbox{\includegraphics[scale=.28]{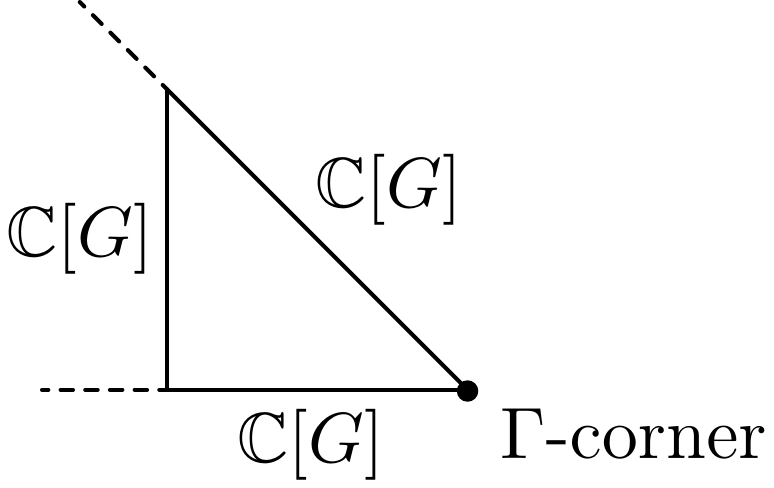}}}
\end{eqs}

After the $\Pi_\Gamma$ projection and introduction of the $\Gamma$-corner, we are now in the ISC, with the logical space being given by $V_\Gamma$. This procedure can be interpreted as fixing the $\Gamma^*$ charge in the corner, so that the rough boundary on the left now hosts a $\Gamma$ charge. The $V_\Gamma^*$ component of the logical space is then removed alongside the local $V_\Gamma^*$ component of the right rough edge, since there is no right action on the remaining $\Gamma$-corner. As a result, this shows the ISC has logical Hilbert space $V_\Gamma$, and the representation operators act as the matrices for the projective irrep $\Gamma$, demonstrating the connection between pGSCs and ISCs.

\section{Minimal irrep surface codes} \label{app: minimal isc}

In this appendix, we use a unitary circuit to sequentially disentangle the ISC into a minimal ISC, which is composed of one $|G|$-dimensional qudit and a $d_\Gamma$ qudit as follows:
\begin{align}
    \vcenter{\hbox{\includegraphics[scale=.1]{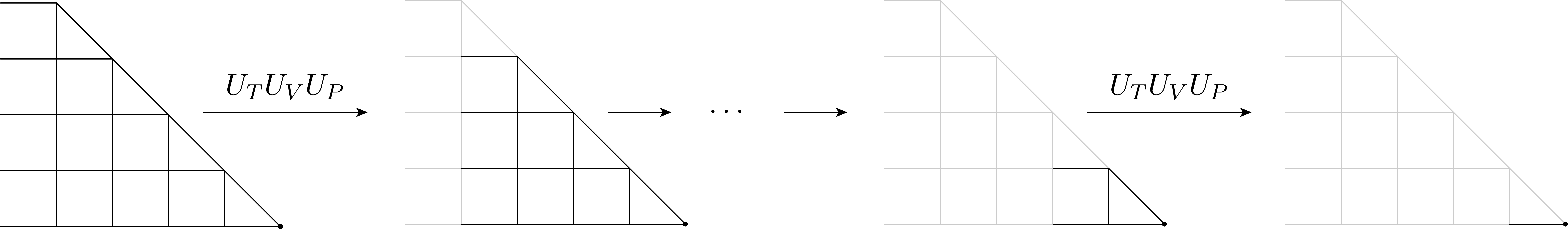}}}
\end{align}
From this minimal model, it can be immediately seen that the dimension of the code space is $d_\Gamma$, and the logical operator $\Gamma_\LL^g$ acts as
\begin{align}
    \Gamma_\LL^g \ket{j}_\LL = \sum_i \Gamma^g_{ij}|i\rangle_\LL.
\end{align}

\subsection{Disentangling circuit}
Before getting into the details of the disentangling circuit, we recall that the ISC stabilizers along the rough boundary include the vertex terms on the top vertex and other vertices respectively as
\begin{align}
    A_v^g =\vcenter{\hbox{\ \includegraphics[scale=.25]{Figures/Avroughtwisted.pdf} }}, \qquad 
    \vcenter{\hbox{\ \includegraphics[scale=.25]{Figures/Avbulk.pdf} }},
\end{align}
and the plaquette term $B_p$ for each open plaquette $p$ acting as
\begin{align}
    B_p \Bigggvert\,
    \vcenter{\hbox{\includegraphics[scale=.25]{Figures/Bprough.pdf}}}\,
    \Bigggrangle = \delta_{g_1 g_2 \bar{g}_3,1}\Bigggvert\,
    \vcenter{\hbox{\includegraphics[scale=.25]{Figures/Bprough.pdf}}}\,
    \Bigggrangle.
\end{align}

Similar to Appendix~\ref{app: isc from gsc}, the disentangling circuit acts on one column after another from left to right. Here, however, we maintain the triangular shape of the code patch. It is composed of four types of local unitary gates. The first type of gate is defined in Eq.~\eqref{eq:Up definition} as $U_p$, along with the product of $U_p$ over all the open plaquettes on the left, denoted as $U_P$. Conjugating the vertex stabilizers along the rough boundary gives:
\begin{align}
    U_{P} A_v^g U_{P}^\dagger=\vcenter{\hbox{\ \includegraphics[scale=.25]{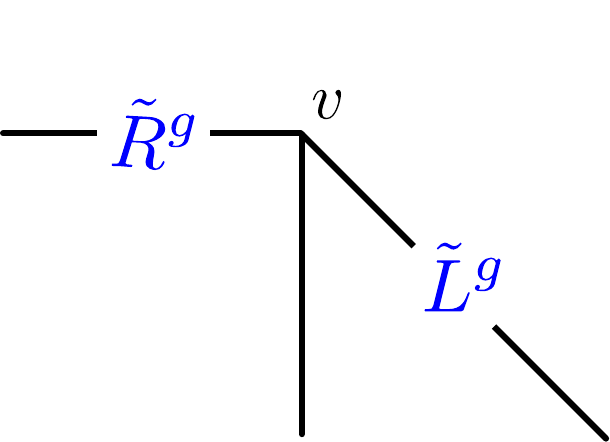} }}, \qquad 
    \vcenter{\hbox{\ \includegraphics[scale=.25]{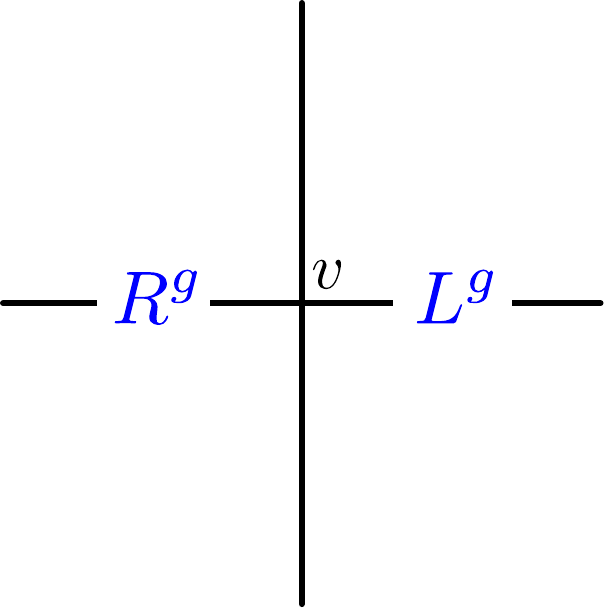} }},
\end{align}
and $B_p$ become a projection operator on a single edge:
\begin{align}
    U_{P} B_p U_{P}^\dagger \Bigggvert\,
    \vcenter{\hbox{\includegraphics[scale=.25]{Figures/Bprough.pdf}}}\,
    \Bigggrangle = \delta_{g_2,1}\Bigggvert\,
    \vcenter{\hbox{\includegraphics[scale=.25]{Figures/Bprough.pdf}}}\,
    \Bigggrangle.
\end{align}

The second type of gate is a two-qudit unitary supported on the two horizontal edges of vertex $v$, and is defined in Eq.~\eqref{eq:Uv definition} as $U_v$. Similarly, the third type of gate $\tilde{U}_v$ is a two-qudit unitary supported on the two non-vertical edges of vertex $v$ on the first column of the twisted boundary:
\begin{align}
    \tilde{U}_v \Bigggvert\,
    \vcenter{\hbox{\includegraphics[scale=.25]{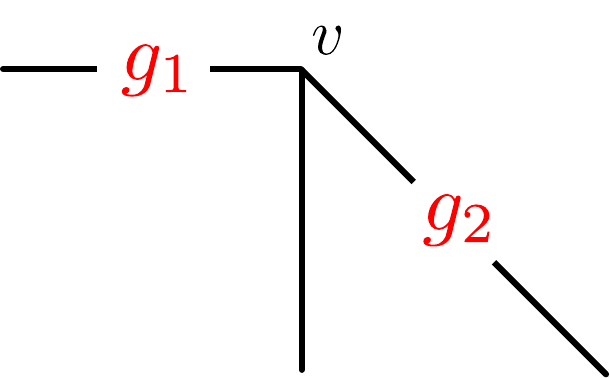}}}\,
    \Bigggrangle = \omega(g_1,g_2)\Bigggvert\,
    \vcenter{\hbox{\includegraphics[scale=.25]{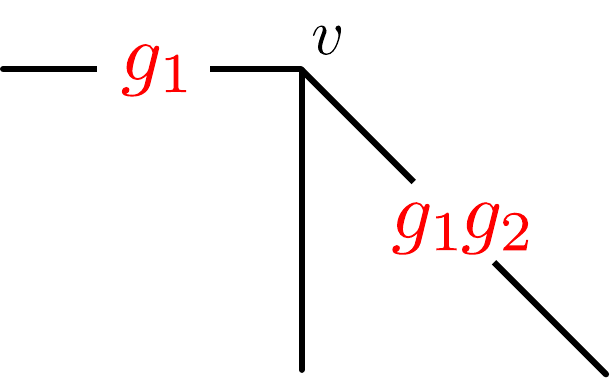}}}\,
    \Bigggrangle.
\end{align}
Denoting as $U_{V}$ the product of $\tilde{U}_v$ at the top vertex and $U_v$ at all other vertices in the first column, after its conjugation, the vertex stabilizers are further reduced to:
\begin{equation}
    \Big(U_{V}U_{P}\Big) A_v^g \Big(U_{V}U_{P}\Big)^\dagger=\vcenter{\hbox{\ \includegraphics[scale=.25]{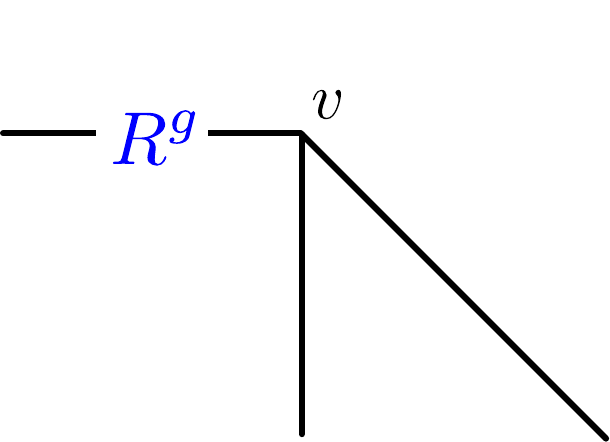} }}, \qquad 
    \vcenter{\hbox{\ \includegraphics[scale=.25]{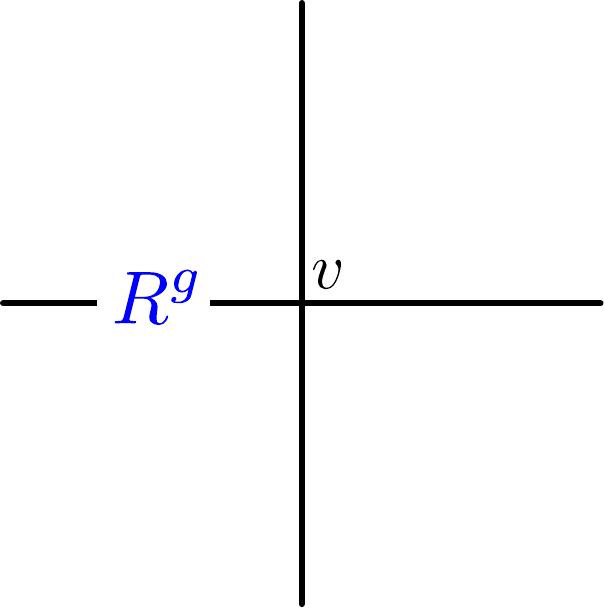} }}.
\end{equation}

As a result, after the conjugation of $U_{V}U_{P}$, all the qudits on the first column are disentangled. In the code space, the vertical edges on the first column are at state $\ket{1}$ while the horizontal edges are at state $\ket{+}=\frac{1}{\sqrt{|G|}}\sum_g \ket{g}$. Furthermore, the plaquette stabilizers on the second column are conjugated to
\begin{align}
    \Big(U_{V}U_{P}\Big) B_p \Big(U_{V}U_{P}\Big)^\dagger \Bigggvert\,
    \vcenter{\hbox{\includegraphics[scale=.25]{Figures/Bpbulk.pdf}}}\,
    \Bigggrangle = \delta_{g_2 g_3 \bar{g}_4,1}\Bigggvert\,
    \vcenter{\hbox{\includegraphics[scale=.25]{Figures/Bpbulk.pdf}}}\,
    \Bigggrangle
\end{align}
in the bulk and 
\begin{align}
    \Big(U_{V}U_{P}\Big) B_p \Big(U_{V}U_{P}\Big)^\dagger \Bigggvert\,
    \vcenter{\hbox{\includegraphics[scale=.25]{Figures/Bptwisted.pdf}}}\,
    \Bigggrangle = \delta_{g_2 \bar{g}_3,1}\Bigggvert\,
    \vcenter{\hbox{\includegraphics[scale=.25]{Figures/Bptwisted.pdf}}}\,
    \Bigggrangle
\end{align}
on the twisted boundary. The last type of gate $U_T$ acts on these two edges as
\begin{eqs}
    U_T: \,  \vcenter{\hbox{\includegraphics[scale=.3]{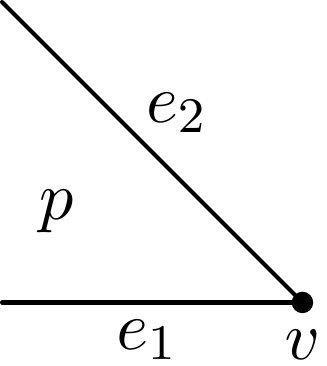}}} \mapsto \vcenter{\hbox{\includegraphics[scale=.3]{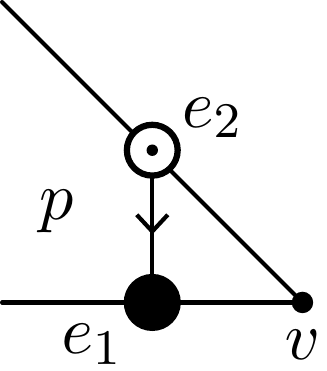}}} = \vcenter{\hbox{\includegraphics[scale=.3]{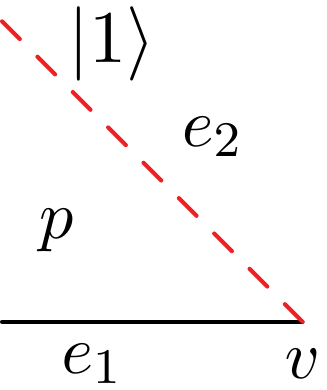}}}.
\end{eqs}
It conjugates the stabilizer on the triangle plaquette of the second column to
\begin{align}
    \Big(U_T U_{V}U_{P}\Big) B_p \Big(U_T U_{V}U_{P}\Big)^\dagger \Bigggvert\,
    \vcenter{\hbox{\includegraphics[scale=.25]{Figures/Bptwisted.pdf}}}\,
    \Bigggrangle = \delta_{g_3,1}\Bigggvert\,
    \vcenter{\hbox{\includegraphics[scale=.25]{Figures/Bptwisted.pdf}}}\,
    \Bigggrangle.
\end{align}

\begin{figure*}[t]
        \centering
        \includegraphics[scale=.18]{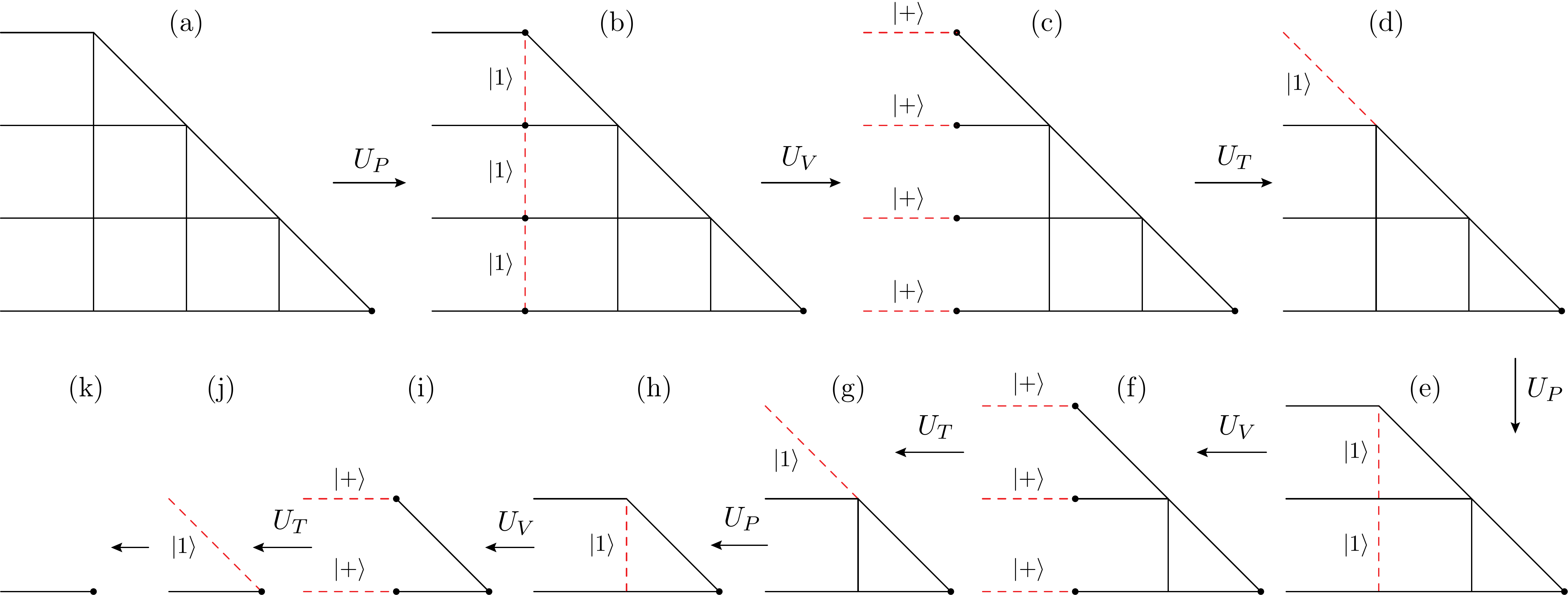}
            \caption{Under the sequential actions of disentangler gates, an ISC is eventually shrunk to the minimal model consisting of one qudit on a horizontal edge and a $d_\Gamma$-dimensional qudit at the corner.} 
    \label{fig: disentangler}
\end{figure*}

\begin{figure*}[t]
        \centering
        \includegraphics[scale=.18]{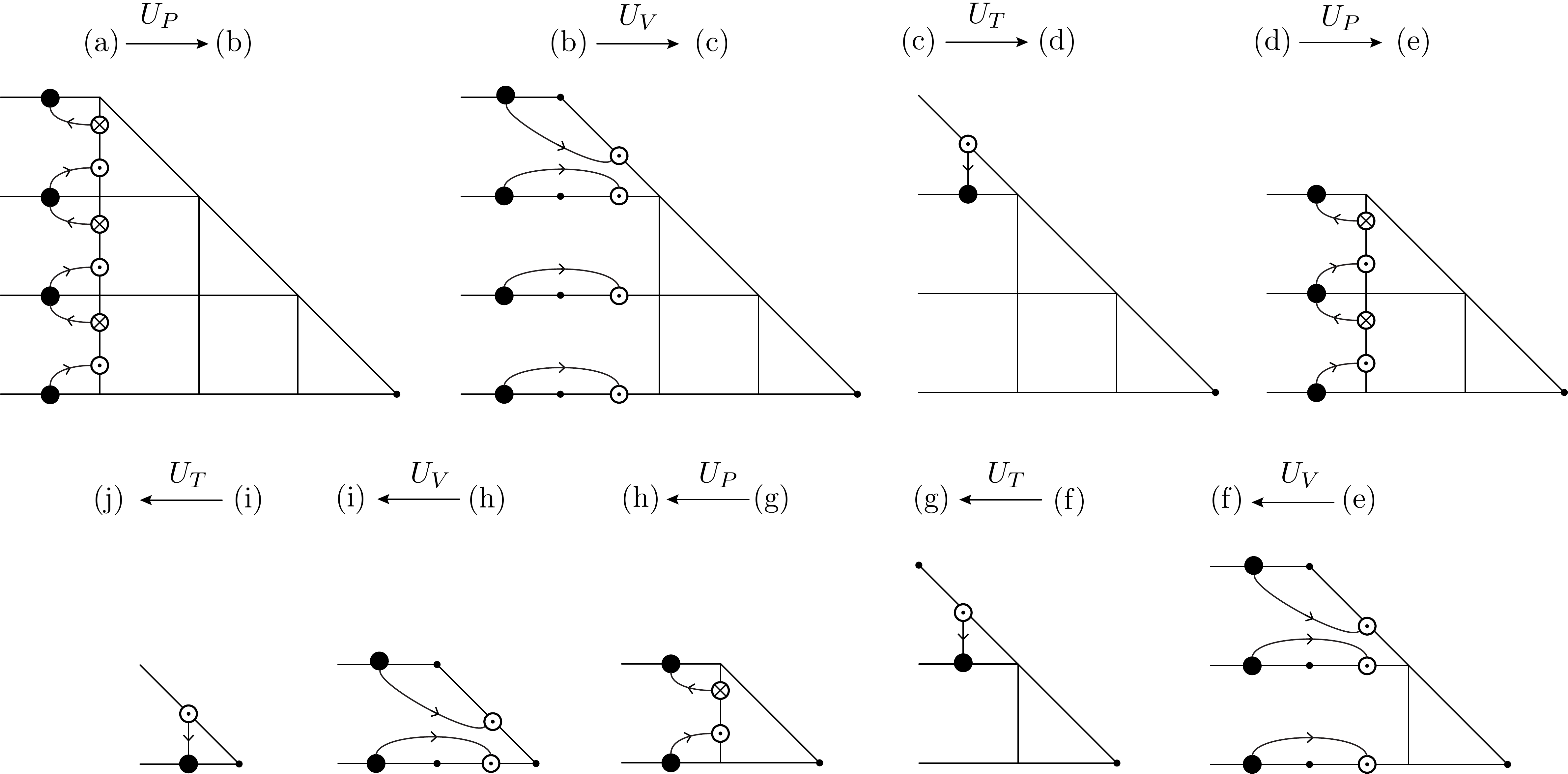}
            \caption{The explicit disentangler gates $U_P$, $U_V$, and $U_T$. } 
    \label{fig: disentangler2}
\end{figure*}

To conclude, under the conjugation of $U_T U_V U_P$ circuit, we obtain exactly an ISC with one less column on the left,  after discarding the disentangled qudits. Iterating this process column by column, we eventually obtain a minimal ISC with one qudit on a horizontal edge and a $d_\Gamma$-dimensional qudit at the corner as shown in Fig.~\ref{fig: disentangler} and Fig.~\ref{fig: disentangler2}. The only stabilizers in the minimal model are given by
\begin{align}
    A^g_{\text{min}}=\vcenter{\hbox{\includegraphics[scale=.25]{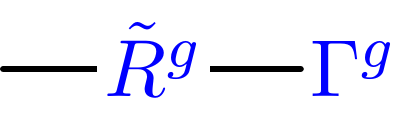}}}
\end{align}
for arbitrary $g\in G$. 

\subsection{Code space and logical operators}
The code space dimensions in the ISC and the minimal model should be identical as the two models are related by a unitary circuit up to fixed disentangled qudits. The Hilbert space of the minimal model is the tensor product of a twisted group algebra $\cgw$ and $V_\Gamma$. The twisted group algebra has a decomposition in terms of projective irreps according to Appendix~\ref{app: irrep review}. As a result,
\begin{equation}
    \mathcal{H}_{\text{min}} = \mathbb{C}_\omega[G]\otimes V_\Gamma = \Bigg(\bigoplus_{\rho\in \text{Irr}_{\omega}(G)} V_\rho \otimes V^*_\rho\Bigg) \otimes V_\Gamma.
\end{equation}
The projector onto the mutual $+1$ eigenspace of the vertex stabilizers is given by 
\begin{equation}
    \Pi_A= \frac{1}{|G|}\sum_{g\in G}A^g_{\text{min}}=\frac{1}{|G|}\sum_{g\in G}\tilde{R}^g\otimes \Gamma^g.
\end{equation}
It is a rank-$1$ projector into $V_1$ on $V^*_\rho \otimes V_\Gamma$ for $\rho\cong\Gamma$, and is zero on $V^*_\rho \otimes V_\Gamma$ for $\rho\ncong\Gamma$. Therefore, the code space $\Pi_A\cdot \mathcal{H}_{\text{min}} = V_{\Gamma}\otimes V_1$ is of dimension $d_\Gamma$. This agrees with the corner gauge result in Section~\ref{sec: code space}.

The logical operator $\Gamma_\LL^g$ in ISC is mapped by the disentangler to $\tilde{L}^g$ on the remaining edge times $L^g$ on the disentangled horizontal edges. In the decomposition of $\cgw$, the action of $\tilde{L}^g$ is given by
\begin{equation}
    \tilde{L}^g = \bigoplus_{\rho\in \text{Irr}_{\omega}(G)} \rho^g \otimes I_{\rho^*}.
\end{equation}
Therefore, this logical operator acts on the code space $V_{\Gamma}\otimes V_1$ as
\begin{equation}
    \Gamma_\LL^g = \Pi_A \tilde{L}^g \Pi_A = \Gamma^g\otimes I.
\end{equation}

Omitting the one-dimensional space $V_1$, we conclude that the code space of the ISC (or the minimal model) is $V_\Gamma$, while the logical operator $\Gamma_\LL^g$ acts exactly as $\Gamma^g$. Due to the irreducibility of the projective representation $\Gamma$, the set of operators $\Gamma_\LL^g$ generates the full algebra of logical operations, as argued in Appendix~\ref{app: irrep review}. 

\section{Deforming logical operators} \label{app: line-like logical}

In this appendix, we discuss the deformations of the logical operators of the ISC. The discussion consists of three parts. In the first part, we present the general bulk representation of $\Gamma_\LL^g$ deformed from the initial representation on the rough boundary. In the second part, we discuss, in particular, the representation of $\Gamma_\LL^g$ deformed to the smooth boundary, and show that when $g$ is in the center of group $G$, this representation is transversal. In the third part, we present a reduction of the operators discussed in Section~\ref{sec: logical projective intertwiners} for trivial automorphism. These logical operators are only supported on a path connecting the rough boundary and the $\Gamma$-corner. The bulk representations of the former logical operators are used in the derivation of code distance in Section~\ref{sec: logical representation operators}. The representations of both logical operators on the smooth boundary are used in the discussion of smooth-smooth surgery in Section~\ref{sec: smoothsmooth surgery}.

\subsection{Deformations of $\Gamma_\LL^g$}

In this section, we present the deformation of $\Gamma_\LL^g$ into the bulk. We show that the logical operator $\Gamma_\LL^g$ has a representation that is supported on an arbitrary column and arbitrary row, as illustrated below: 
\begin{align} \label{eq: deformation support}
    \vcenter{\hbox{\includegraphics[scale=.18]{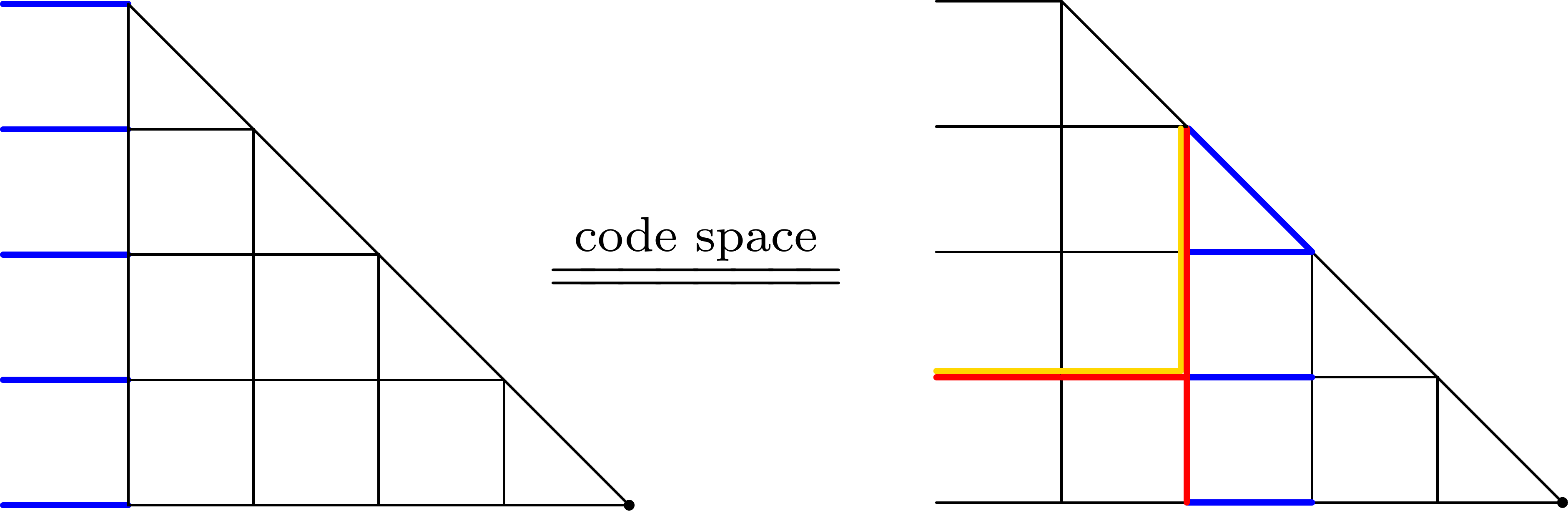}}}.
\end{align}
The thickened edges denote the support of the logical operator, and the meaning of color is clarified later in the section. 

Let us start by making the representation of $\Gamma_\LL^g$ in the bulk explicit. To this end, we introduce notation for the holonomy. For an edge $i$ and a given group-basis configuration $\bm{h_e}$, we denote by $\Omega_{i}(\bm{h_e})$ the product of group elements on the horizontal edges connecting the rough boundary to the leftmost or topmost vertex of $i$. For a flux-free configuration, $\Omega_{i}$ can be calculated using any path from the rough boundary to the left or top vertex of $i$. For example, if we label a small region around the top left corner as:
\begin{align} \label{eq: roughtwisted}
    \vcenter{\hbox{\includegraphics[scale=.25]{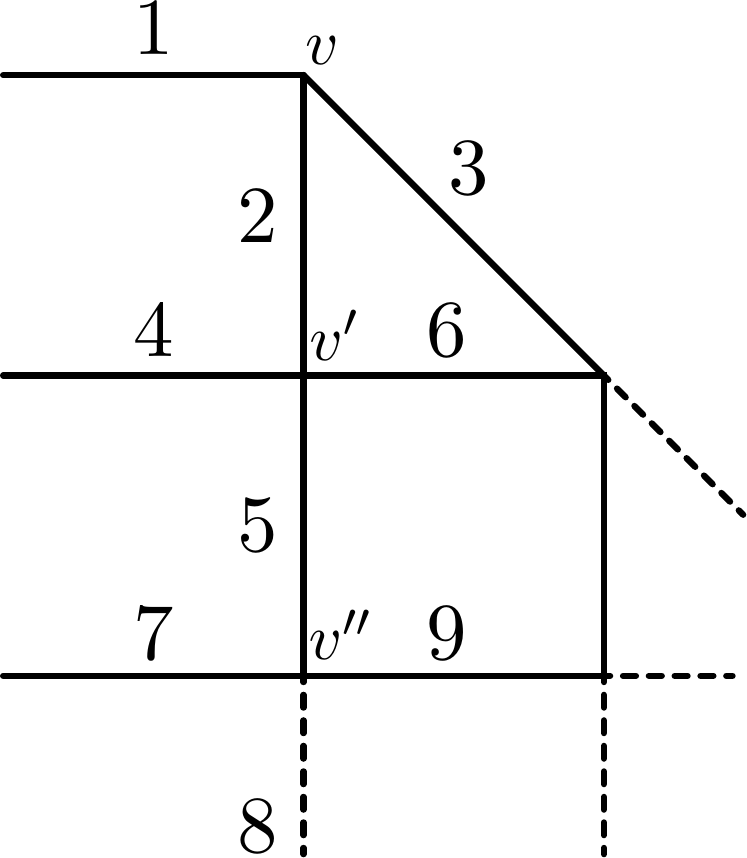}}},
\end{align}
for a flux-free configuration $\bm{h_e}$ we have
\begin{equation}
    \Omega_3(\bm{h_e}) = h_1=h_4 h_2=\cdots.
D\end{equation}
Furthermore, we define the (twisted) controlled-group multiplication operators as
\begin{equation}
\begin{aligned}
    L^{\Omega_i}_j = \sum_{\bm{h_e}}L_j^{\Omega_i(\bm{h_e})}\ket{\bm{h_e}}\bra{\bm{h_e}},\qquad R^{\Omega_i}_j = \sum_{\bm{h_e}}R_j^{\Omega_i(\bm{h_e})}\ket{\bm{h_e}}\bra{\bm{h_e}},\\
    \tilde{L}^{\Omega_i}_j = \sum_{\bm{h_e}}\tilde{L}_j^{\Omega_i(\bm{h_e})}\ket{\bm{h_e}}\bra{\bm{h_e}},\qquad \tilde{R}^{\Omega_i}_j = \sum_{\bm{h_e}}\tilde{R}_j^{\Omega_i(\bm{h_e})}\ket{\bm{h_e}}\bra{\bm{h_e}}.
\end{aligned}
\end{equation}

Now we are ready to give the representation of $\Gamma_\LL^g$ on the $m$-th column and $n$-th row. Let us denote the set of edges on the $m$-th column except for the top edge as $\bar{\gamma}$, and denote the top edge as $f$,
\begin{equation} \label{eq: Gamma column row}
    \Gamma_\LL^g = \Bigg[\prod_{e\in \bar{\gamma}} (L^{\Omega_e}_e)^\dagger L^g_e L^{\Omega_e}_e\Bigg]\Bigg( (\tilde{L}^{\Omega_f}_f)^\dagger \tilde{L}^g_f \tilde{L}^{\Omega_f}_f\Bigg),
\end{equation}
where the holonomy of each edge is taken on a path along the $n$-th row and the $m$-th column, as shown in red in Eq.~\eqref{eq: deformation support}. As a sanity check, when $m=1$ the holonomies of all these edges are equal to identity, and the expression reduces to the initial representation of $\Gamma_\LL^g$. It can also be checked straightforwardly that the above operator commutes with all the ISC stabilizers.

Now we discuss the derivation of the deformed logical operator. Using the labeling of the edges shown in Eq.~\eqref{eq: roughtwisted}, the logical operator $\Gamma_\LL^g$ is defined as
\begin{equation}
    \Gamma_\LL^g = \cdots L^g_7 L^g_4 \tilde{L}^g_1.
\end{equation}
The stabilizers at the top vertex sum to a projector onto the code space:
\begin{equation}
    \frac{1}{|G|}\sum_{g\in G}A^g_v=\frac{1}{|G|}\sum_{g\in G}\tilde{R}^g_1 R^g_2 \tilde{L}^g_3.
\end{equation}
This projects the space of three qudits ($\mathbb{C}[G]^{\otimes 3}$) into a $\mathbb{C}[G]^{\otimes 2}$ subspace. We denote the basis of this subspace as 
\begin{equation}
    \overline{\ket{h,h'}}=\sum_{h_1\in G}\omega(h_1, \bar{h}_1 h)^{-1}\ket{h_1}\otimes \ket{h' h_1}\otimes \ket{\bar{h}_1 h},
\end{equation}
where the three states are at edge $1$, $2$, and $3$ respectively. As a consistency check, we note that
\begin{equation}
\begin{aligned}
    A^g_v \overline{\ket{h,h'}} &= A^g_v \sum_{h_1}\omega(h_1, \bar{h}_1 h)^{-1}\ket{h_1}\otimes \ket{h' h_1}\otimes \ket{\bar{h}_1 h}\\
    &=\sum_{h_1}\frac{\omega(g,\bar{h}_1 h)}{\omega(h_1, \bar{h}_1 h)\omega(h_1\bar{g},g)}\ket{h_1\bar{g}}\otimes \ket{h' h_1\bar{g}}\otimes \ket{g\bar{h}_1 h}\\
    &=\sum_{h_1}\omega(h_1\bar{g}, g\bar{h}_1 h)^{-1}\ket{h_1\bar{g}}\otimes \ket{h' h_1\bar{g}}\otimes \ket{g\bar{h}_1 h}\\
    &=\sum_{h'_1}\omega(h'_1, \bar{h}'_1 h)^{-1}\ket{h'_1}\otimes \ket{h' h'_1}\otimes \ket{\bar{h}'_1 h}\\
    &=\overline{\ket{h,h '}}.
\end{aligned}
\end{equation}
The cocycle condition is used to go from the second to the third line. Furthermore, the operator $\tilde{L}^g_1$ acts on the states as
\begin{equation}
    \begin{aligned}
    \tilde{L}^g_1 \overline{\ket{h,h'}}&= \tilde{L}^g_1 \sum_{h_1\in G}\omega(h_1, \bar{h}_1 h)^{-1}\ket{h_1}\otimes \ket{h' h_1}\otimes \ket{\bar{h}_1 h}\\
    &=\sum_{h_1\in G}\frac{\omega(g,h_1)}{\omega(h_1, \bar{h}_1 h)}\ket{gh_1}\otimes \ket{h' h_1}\otimes \ket{\bar{h}_1 h}\\
    &=\sum_{h'_1\in G}\frac{\omega(g,\bar{g}h'_1)}{\omega(\bar{g}h'_1, \bar{h}'_1 g h)}\ket{h'_1}\otimes \ket{h'\bar{g}h'_1}\otimes \ket{\bar{h}'_1 gh} \\
    &=\sum_{h'_1\in G}\frac{\omega(g,h)}{\omega(h'_1, \bar{h}'_1 g h)}\ket{h'_1}\otimes \ket{h'\bar{g}h'_1}\otimes \ket{\bar{h}'_1 gh} \\
    &= \omega(g,h)\overline{\ket{gh,h'\bar{g}}},
\end{aligned}
\end{equation}
where the cocycle condition is used from the third to the fourth line. On the other hand, a controlled twisted multiplication operator 
\begin{equation} \label{eq: operator replacement}
    \sum_{h_1\in G}R^{\bar{h}_1gh_1}_2(\tilde{L}^{h_1}_3)^\dagger \tilde{L}^{g}_3 \tilde{L}^{h_1}_3 \ket{h_1}\bra{h_1}
\end{equation}
can be shown to have exactly the same action on $\overline{\ket{h,h'}}$ as $\tilde{L}^g_1$. As a result, in the ISC code space, $\tilde{L}^g_1$ in the expression for $\Gamma_\LL^g$ can be replaced by the operator in Eq.~\eqref{eq: operator replacement}:
\begin{equation}
    \Gamma_\LL^g = \cdots L^g_7 L^g_4 \cdot\Bigg(\sum_{h_1\in G}R^{\bar{h}_1gh_1}_2(\tilde{L}^{h_1}_3)^\dagger \tilde{L}^{g}_3 \tilde{L}^{h_1}_3 \ket{h_1}\bra{h_1}\Bigg).
\end{equation}
We can entirely remove the support of the logical operator from the edge $1$ by noting that the code space satisfies the flux-free constraint $h_1=h_4 h_2$ due to the plaquette stabilizer:
\begin{equation}
    \Gamma_\LL^g = \cdots L^g_7 L^g_4 \cdot\Bigg(\sum_{h_1, h_4}L^{\bar{h}_4 \bar{g}h_4}_2(\tilde{L}^{h_1}_3)^\dagger \tilde{L}^{g}_3 \tilde{L}^{h_1}_3 \ket{h_1, h_4}\bra{h_1,h_4}\Bigg).
\end{equation}

Following a similar argument, in the subspace where $A^g_{v'} = R^g_4 L^g_2 R^g_5 L^g_6= 1$, we have relation
\begin{equation}
    \sum_{h_4} L^g_4 L^{\bar{h}_4 \bar{g}h_4}_2 \ket{h_4}\bra{h_4} = \sum_{h_4} R^{\bar{h}_4 gh_4}_5 L^{\bar{h}_4 gh_4}_6 \ket{h_4}\bra{h_4}.
\end{equation}
Combining with the flux-free constraint $h_4=h_7 h_5$, it leads to a further deformation:
\begin{equation}
    \Gamma_\LL^g = \cdots L^g_7 \Bigg(\sum_{h_1, h_4}L^{\bar{h}_7 \bar{g}h_7}_5 L^{\bar{h}_4 gh_4}_6(\tilde{L}^{h_1}_3)^\dagger \tilde{L}^{g}_3 \tilde{L}^{h_1}_3 \ket{h_1, h_4}\bra{h_1,h_4}\Bigg).
\end{equation}

As such, we have deformed the support of the logical operator as follows:
\begin{align}
    \vcenter{\hbox{\includegraphics[scale=.25]{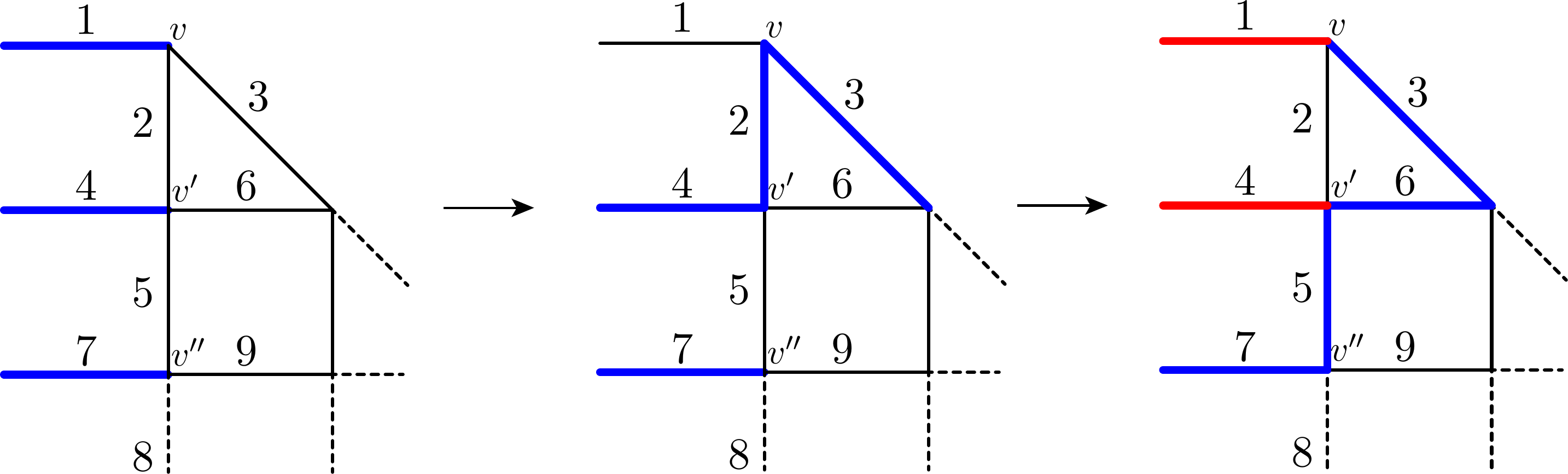}}}
\end{align}
The red color means that the operator is diagonal on the corresponding edge. The controlled left multiplications on edge $5$ and $7$ can be further replaced using stabilizer $A^g_{v''}$, and so on.

Continuing the deformations described above, we eventually obtain a logical operator composed of controlled left (twisted) multiplications on the second column:
\begin{align}
    \Gamma_\LL^g &= \cdots \Bigg((L^{\Omega_9}_9)^\dagger L^{g}_9 L^{\Omega_9}_9\Bigg) \Bigg((L^{\Omega_6}_6)^\dagger L^{g}_6 L^{\Omega_6}_6\Bigg)\Bigg((\tilde{L}^{\Omega_3}_3)^\dagger \tilde{L}^{g}_3 \tilde{L}^{\Omega_3}_3\Bigg).
\end{align}

To demonstrate the deformation on the subsequent columns, we label the adjacent edges as follows:
\begin{align}
    \vcenter{\hbox{\includegraphics[scale=.25]{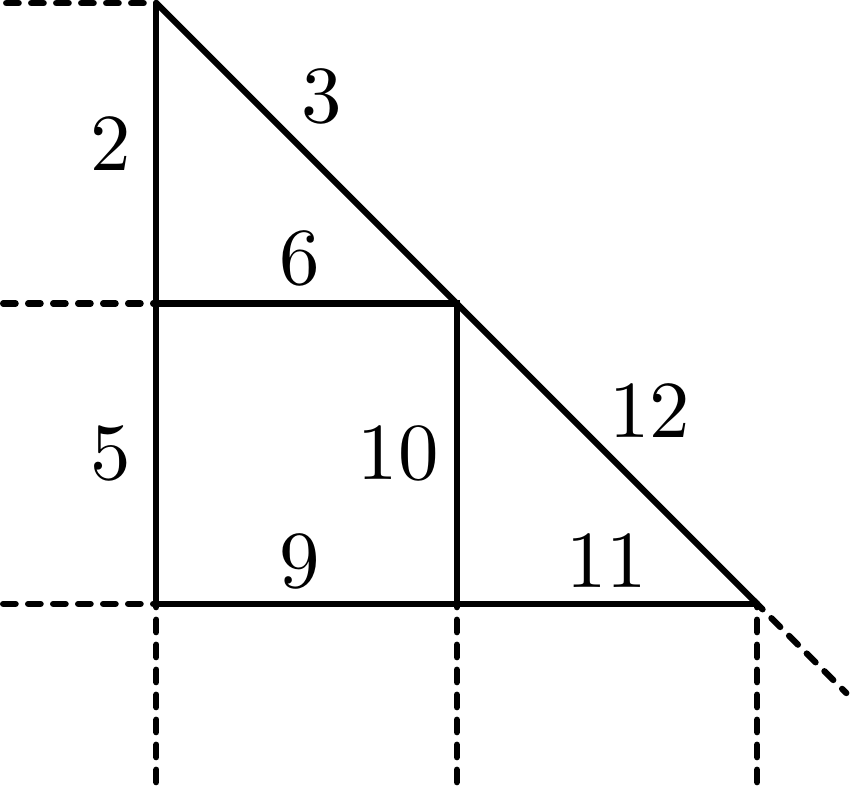}}}.
\end{align}
The following identity is then particularly useful:
\begin{equation}
    (\tilde{L}^{h})^\dagger \tilde{L}^{g} \tilde{L}^{h} = \theta_g(h)\tilde{L}^{\bar{h}gh},
\end{equation}
where $\theta_g(h)= \frac{\omega(g,h)}{\omega(h,\bar{h}gh)}$ is called the slant product of the cocycle $\omega$. Using this notation, we are able to rewrite the logical operator as
\begin{equation}
    \Gamma_\LL^g = \cdots  L^{\overline{\Omega}_9g\Omega_9}_9 L^{\overline{\Omega}_6g\Omega_6}_6 \Bigg(V^{\theta_g}_{\Omega_3} \tilde{L}^{\overline{\Omega}_3g\Omega_3}_3\Bigg),
\end{equation}
where 
\begin{equation}
    V^{\theta_g}_{\Omega_3}\ket{\bm{h_e}} = \theta_g(\Omega_3)\ket{\bm{h_e}}.
\end{equation}
Similar to the procedure below Eq.~\eqref{eq: operator replacement}, we have deformations:
\begin{equation}
\begin{aligned}
    \Gamma_\LL^g&= \cdots L^{\overline{\Omega}_9g\Omega_9}_9 L^{\overline{\Omega}_6g\Omega_6}_6\Bigg(V^{\theta_g}_{\Omega_3} \tilde{L}^{\overline{\Omega}_3g\Omega_3}_3\Bigg)\\
    &=\cdots L^{\overline{\Omega}_9g\Omega_9}_9 L^{\overline{\Omega}_6g\Omega_6}_6\Bigg(V^{\theta_g}_{\Omega_3} V^{\theta_{\overline{\Omega}_3g\Omega_3}}_{h_3} L^{\overline{\Omega}_6\bar{g}\Omega_6}_6 L^{\overline{\Omega}_{10}\bar{g}\Omega_{10}}_{10}\tilde{L}^{\overline{\Omega}_{12}g\Omega_{12}}_{12}\Bigg)\\
    &=\cdots L^{\overline{\Omega}_9g\Omega_9}_9 L^{\overline{\Omega}_{10}\bar{g}\Omega_{10}}_{10} \Bigg(V^{\theta_g}_{\Omega_{12}}\tilde{L}^{\overline{\Omega}_{12}g\Omega_{12}}_{12}\Bigg)\\
    &=\cdots L^{\overline{\Omega}_{11}g\Omega_{11}}_{11} \Bigg(V^{\theta_g}_{\Omega_{12}}\tilde{L}^{\overline{\Omega}_{12}g\Omega_{12}}_{12}\Bigg)\\
    &=\cdots \Bigg((L^{\Omega_{11}}_{11})^\dagger L^{g}_{11}L^{\Omega_{11}}_{11}\Bigg)\Bigg((\tilde{L}^{\Omega_{12}}_{12})^\dagger\tilde{L}^{g}_{12}\tilde{L}^{\Omega_{12}}_{12}\Bigg),
\end{aligned}
\end{equation}
where in the third line we used the identity
\begin{equation} \label{eq: slant product}
    \theta_{\bar{h}gh}(h')\theta_g(h) = \theta_g(hh'),
\end{equation}
and that $\Omega_{12}=\Omega_3\cdot h_3$. Continuing the procedure, we can deform $\Gamma_\LL^g$ to the third column. Furthermore, we recall that the path used to calculate the holonomy can be arbitrarily deformed for flux-free configurations. Thus, we can obtain the representation of $\Gamma_\LL^g$ on any given column and row as in Eq.~\eqref{eq: Gamma column row}, by choosing the paths to be along that particular column and row.

More generally, we can deform $\Gamma_\LL^g$ to any given path $\tilde{\gamma}$ on the dual lattice connecting the smooth boundary to the twisted boundary. We assume that $\tilde{\gamma}$ is oriented from the smooth boundary to the twisted boundary. Denoting as $f$ the edge on the twisted boundary at which $\tilde{\gamma}$ ends, and denoting as $\bar{\gamma}$ the set of bulk edges that $\tilde{\gamma}$ passes through, we can write the logical operator as
\begin{equation}
\begin{aligned}
    \Gamma_\LL^g &= \Bigg[\prod_{e\in \bar{\gamma}} \Big(L^{\overline{\Omega}_{e}g\Omega_{e}}_e\Big)^{s_{\bar{\gamma}}(e)}\Bigg] \Bigg(V^{\theta_g}_{\Omega_f}\tilde{L}^{\overline{\Omega}_{f}g\Omega_{f}}_f\Bigg)
    =\Bigg[\prod_{e\in \bar{\gamma}} \Big((L^{\Omega_e}_e)^\dagger L^{g}_e L^{\Omega_e}_e\Big)^{s_{\bar{\gamma}}(e)}\Bigg]\Bigg((\tilde{L}^{\Omega_f}_f)^\dagger\tilde{L}^g_f \tilde{L}^{\Omega_f}_f\Bigg),
\end{aligned}
\end{equation}
where $s_{\bar{\gamma}}(e)=-1$ when the orientation of $e$ points to the left of $\tilde{\gamma}$, and $s_{\bar{\gamma}}(e)=1$ otherwise.

We would like to make a further remark on the physical meaning of the above deformed logical operator. According to Eq.~\eqref{eq: slant product}, the slant product $\theta_g(\cdot)$ gives a representation of the centralizer group 
\begin{equation}
    C_G(g)=\{h\in G| hg=gh \}.
\end{equation}
The conjugacy class of $g$ together with this representation of centralizer $\theta_g(\cdot)$ is associated with a dyon in the quantum double, that condenses on the twisted boundary. In the continuum, the deformed logical operator can be understood as a flux originating from the smooth boundary that fuses with a charge from the rough boundary into a dyon that condenses at the twisted boundary:
\begin{align} 
    \vcenter{\hbox{\includegraphics[scale=.18]{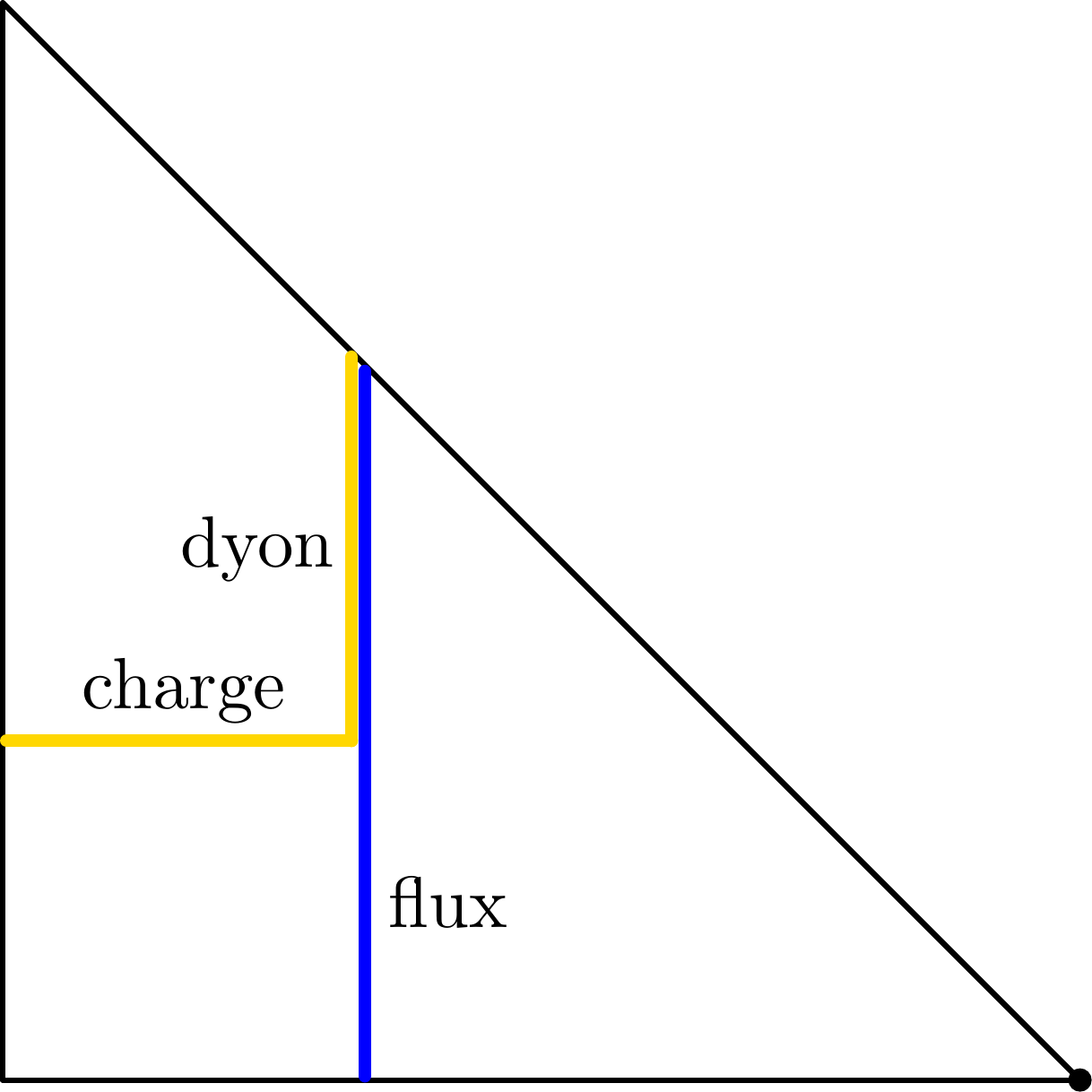}}}.
\end{align}

\subsection{Representation of $\Gamma_\LL^g$ on the smooth boundary} \label{app: smooth-boundary-representation}

Here, we consider a deformation of $\Gamma_\LL^g$ to the smooth boundary. We label the edges near the $\Gamma$-corner as follows:
\begin{align} 
    \vcenter{\hbox{\includegraphics[scale=.18]{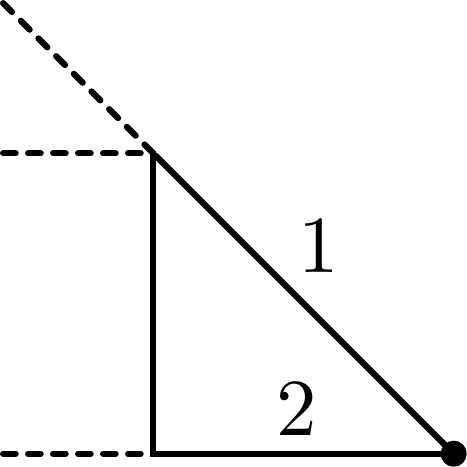}}}.
\end{align}
According to our previous result, the bulk representation of $\Gamma_\LL^g$ near this corner is given by
\begin{equation} \label{eq: logical at corner}
    \Gamma_\LL^g = L^{\overline{\Omega}_{2}g\Omega_{2}}_2\Bigg(V^{\theta_g}_{\Omega_1}\tilde{L}^{\overline{\Omega}_{1}g\Omega_{1}}_1\Bigg).
\end{equation}

Recall that the corner vertex stabilizer is
\begin{align}
    A_c^g =\vphantom{\rule[-2em]{1pt}{2em}}\vcenter{\hbox{\includegraphics[scale=.25]{Figures/Avcorner.pdf}}},
\end{align}
and $A^G_c = \frac{1}{|G|}\sum_g A_c^g$ projects the local Hilbert space from $\mathbb{C}[G]\otimes \mathbb{C}[G]\otimes V_\Gamma$ to $\mathbb{C}[G]\otimes V_\Gamma$. If we use $i=1,\cdots,d_\Gamma$ to label the basis in $V_\Gamma$, then the basis states in the subspace where $A_c^g=1$ can be given by
\begin{equation}
    \overline{\ket{h,i}}= \sum_{h_1\in G, j=1,\cdots,d_\Gamma}\omega(h_1,\bar{h}_1)^{-1}\ket{h_1}\otimes\ket{h h_1}\otimes \Big(\Gamma^{\bar{h}_1}_{ji}\ket{j}\Big).
\end{equation}
To verify this, we note that $\overline{\ket{h,i}}$ for any $h\in G$ and $i$ is invariant under the vertex stabilizer:
\begin{equation}
\begin{aligned}
    A_c^g \overline{\ket{h,i}} &= \sum_{h_1\in G, j,k=1,\cdots,d_\Gamma}\frac{1}{\omega(h_1,\bar{h}_1)\omega(h_1 \bar{g},g)} \ket{h_1\bar{g}}\otimes\ket{h h_1\bar{g}}\otimes \Big(\Gamma^{\bar{h}_1}_{ji}\Gamma^g_{kj}\ket{k}\Big)\\
    &= \sum_{h_1\in G, k=1,\cdots,d_\Gamma}\frac{\omega(g,\bar{h}_1)}{\omega(h_1,\bar{h}_1)\omega(h_1\bar{g},g)} \ket{h_1\bar{g}}\otimes\ket{h h_1\bar{g}}\otimes \Big(\Gamma^{g\bar{h}_1}_{ki}\ket{k}\Big)\\
    &= \sum_{h_1\in G, k=1,\cdots,d_\Gamma}\omega(h_1\bar{g},g\bar{h}_1)^{-1}\ket{h_1\bar{g}}\otimes\ket{h h_1\bar{g}}\otimes \Big(\Gamma^{g\bar{h}_1}_{ki}\ket{k}\Big)\\
    &=\overline{\ket{h,i}},
\end{aligned}
\end{equation}
where we have used the cocycle condition in the third line. To simplify the expression, suppose we are in a space of flux-free configurations where the holonomy of edge $1$ and $2$ are given by $\Omega_1,\Omega_2\in G$. In this space, the action of the logical operator as in Eq.~\eqref{eq: logical at corner} on the basis state is:
\begin{equation} \label{eq: Omegac def}
\begin{aligned}
    \Gamma_\LL^g \overline{\ket{h,i}} &= \sum_{h_1, j}\frac{ \omega(\overline{\Omega}_{1}g\Omega_{1},h_1)\theta_g(\Omega_1)}{\omega(h_1,\bar{h}_1)}\ket{\overline{\Omega}_{1}g\Omega_{1}\cdot h_1}\otimes\ket{\overline{\Omega}_{2}g\Omega_{2}\cdot h h_1}\otimes \Big(\Gamma^{\bar{h}_1}_{ji}\ket{j}\Big)\\
    &= \sum_{h'_1, j}\frac{\omega(\overline{\Omega}_{1}g\Omega_{1},\overline{\Omega}_{1}\bar{g}\Omega_{1}\cdot h'_1)\theta_g(\Omega_1)}{\omega(\overline{\Omega}_{1}\bar{g}\Omega_{1}\cdot h'_1,\bar{h}'_1\cdot\overline{\Omega}_{1}g\Omega_{1})}\ket{h'_1}\otimes\ket{h h'_1}\otimes \Big(\Gamma^{\bar{h}'_1\cdot\overline{\Omega}_{1}g\Omega_{1}}_{ji}\ket{j}\Big)\\
    &=\Gamma^{\overline{\Omega}_{c}g\Omega_{c}}\sum_{h'_1,k}\frac{\omega(\overline{\Omega}_1 g\Omega_1,\overline{\Omega}_1\bar{g} \Omega_c)\theta_g(\Omega_1)}{\omega(\overline{\Omega}_1 \bar{g} \Omega_c,\overline{\Omega}_c g \Omega_1)\omega(\overline{\Omega}_c g \Omega_c,\bar{h}'_1)}\ket{h'_1}\otimes\ket{h h'_1}\otimes \Big(\Gamma^{\bar{h}'_1}_{ki}\ket{k}\Big)\\
    &= \Gamma^{\overline{\Omega}_{c}g\Omega_{c}}\sum_{h'_1,k}\frac{\theta_{\overline{\Omega}_{1}\bar{g}\Omega_{1}}(h'_1)\theta_g(\Omega_1)}{\omega(h'_1,\bar{h'_1})}\ket{h'_1}\otimes\ket{h h'_1}\otimes \Big(\Gamma^{\bar{h}'_1}_{ki}\ket{k}\Big)\\
    &=\theta_g(\Omega_c) \Gamma^{\overline{\Omega}_{c}g\Omega_{c}}\overline{\ket{h,i}}.
\end{aligned}
\end{equation}
In the second line, we changed the variable $h'_1= \overline{\Omega}_{1}g\Omega_{1}\cdot h_1$. In the third line, we defined $\Omega_c= \Omega_1\cdot h'_1$. In the fourth line, we used the cocycle conditions together with the definition of the slant product. In the last line, we used the identity in Eq.~\eqref{eq: slant product}. Therefore, after deforming the logical operator entirely to the smooth boundary, we conclude that in the code space
\begin{equation}
    \Gamma_\LL^g = V^{\theta_g}_{\Omega_c} \Gamma^{\overline{\Omega}_{c}g\Omega_{c}},
\end{equation}
where $\Omega_c$ is the holonomy of the $\Gamma$-corner and the diagonal gate $V^{\theta_g}_{\Omega_c}$ is defined as
\begin{equation}
    V^{\theta_g}_{\Omega_c}\ket{\bm{h_e}} = \theta_g(\Omega_c(\bm{h_e}))\ket{\bm{h_e}}.
\end{equation}

This representation is particularly useful for an element $z$ in the center of $G$. The identity in Eq.~\eqref{eq: slant product} then reduces to
\begin{equation}
    \theta_z(h)\theta_z(h')=\theta_z(hh').
\end{equation}
As a result, $\theta_z(\cdot)$ is a one-dimensional representation of $G$. The logical operator $\Gamma_\LL^z$ then has a transversal representation on the smooth boundary:
\begin{equation}
    \Gamma_\LL^z = \Bigg[\prod_{e\in E_\text{smooth}} V^{\theta_z}_e\Bigg]\Gamma^z,
\end{equation}
where $E_{\text{smooth}}$ denotes the set of edges along the smooth boundary, and the operator $V^{\theta_z}_e$ is supported on edge $e$ acting as 
\begin{equation}
    V_e^{\theta_z}\ket{h}=\theta_z(h)\ket{h}.
\end{equation}
The above representation is used in Section~\ref{sec: smoothsmooth surgery} when discussing smooth-smooth surgery.

\subsection{$W_\LL$ for trivial automorphism}

In Section~\ref{sec: logical projective intertwiners}, we introduced the class of logical operators arising from projective intertwiners satisfying
\begin{align}
    W\Gamma^g W^\dagger  = \beta(g)\Gamma^{\phi(g)},
\end{align}
where $\phi$ is an automorphism and the map $\beta:\ G\rightarrow U(1)$ satisfies
\begin{align} \label{eq: beta constraint}
    \omega(\phi(g),\phi(h)) = \frac{\beta(gh)}{\beta(g)\beta(h)}\omega(g,h).
\end{align}
The corresponding logical operator is realized transversally as in Eq.~\eqref{eq: W logical operator}:
\begin{align}
    W_\LL = \Bigg [ \prod_{e } U^\phi_e \Bigg ] \Bigg [ \prod_{e \in {E}_\text{twist}} V^\beta_e \Bigg ] W_c.
\end{align}

Here, we focus on the case where $\phi$ is a trivial automorphism and the corresponding logical operator becomes line-like. Since $\phi(g)=g$ for all elements, the gate $U^{\phi}_e$ becomes the identity. Moreover, the above constraint in Eq.~\eqref{eq: beta constraint} reduces to
\begin{equation}
    \beta(gh)=\beta(g)\beta(h).
\end{equation}
In other words, $\beta$ is a one-dimensional representation (associated to an Abelian charge). As such, the logical operator reduces to
\begin{align}
    W_\LL = \Bigg [ \prod_{e \in {E}_\text{twist}} V^\beta_e \Bigg ] W_c = V^\beta_{\Omega_c} W_c,
\end{align}
where $\Omega_c$ is the holonomy of the $\Gamma$-corner defined in Eq.~\eqref{eq: Omegac def}. Due to the deformability of the holonomy operator in the code space, $W_\LL$ can be supported on any path from the rough boundary to the $\Gamma$-corner. In particular, we can choose a representation with support on the smooth boundary:
\begin{equation}
    W_\LL = \Bigg [ \prod_{e \in {E}_\text{smooth}} V^\beta_{e} \Bigg ] W_c,
\end{equation}
where $E_{\text{smooth}}$ denotes the set of edges along the smooth boundary, including the dangling edge at its left end.

\section{More on examples}\label{app: more examples}

In this appendix, we fill in the details of Table~\ref{tab: summary}. We first consider two projective irreps of $A_5$: the projective irrep in Example~\ref{example3}, defined on a qubit, and the irrep appearing in Table~\ref{tab: summary}, defined on a pair of qubits. We also show that the projective intertwiners in Example~\ref{example4} for $G=D_m$ are exhaustive. We then derive the group structure for a projective irrep with a controlled Hadamard, as well as its projective intertwiners. Finally, we consider irreducible finite groups of gates containing an arbitrary finite group of diagonal gates and take a group containing $\mathrm{CCZ}$ as a concrete example.

\subsection{ISCs with $G = A_5$}\label{app: A5-2qubits}

\subsubsection{Single qubit: gates from intertwiners}

Our discussion of the projective irrep of $A_5$ on a qubit in Example~\ref{example3} focused on the transversal logical operations arising from the representation operators. Here, we discuss the possibility of additional gates from projective intertwiners.

The automorphism group of $A_5$ is $S_5 \cong A_5 \rtimes \Z_2$. The subgroup $A_5$ consists of the inner automorphisms, which can already be implemented through the operators $\Gamma^g$, with $g \in A_5$. The $\Z_2$ outer automorphism is generated by the odd permutations in $S_5$. In particular, it sends a five-cycle $a$ to $a^2$ up to conjugacy. According to Section~\ref{sec: logical projective intertwiners}, a projective intertwiner $W$ must satisfy 
\begin{equation}
    W\Gamma^a W^\dagger = \beta(a)\Gamma^{a^2},
\end{equation}
for some phase $\beta(a)$. On the other hand, using the explicit form of the projective irrep in Example~\ref{example3} we have
\begin{equation}
\operatorname{Tr}\Gamma^a=\varphi\quad \neq\quad \operatorname{Tr}\Gamma^{a^2}=\varphi-1.
\end{equation}
Thus, the existence of $W$ associated with the non-trivial outer automorphism is ruled out.

A projective intertwiner $W$ associated with the trivial automorphism of $A_5$ acts by transforming $\Gamma^g$ to $\beta(g) \Gamma^g$, where $\beta: A_5 \rightarrow \text{U}(1)$ is a group homomorphism. However, one can show that $A_5$ does not have any non-trivial homomorphisms~\cite{Kanemitsu2023Icosahedral}, meaning that the only intertwiner associated to the trivial automorphism is the identity. 
Putting these facts together, we conclude that there is no additional transversal logical gate from projective intertwiner for this projective irrep of $A_5$ other than $\Gamma^g$.

\subsubsection{Two qubits}

Next, we consider an ISC with $G=A_5$ that encodes two logical
qubits. Let $\Gamma$ denote the four-dimensional linear irrep
of $A_5$, which has the standard presentation
\begin{equation}
    A_5=\langle a,b\mid a^5=b^2=(ab)^3=e\rangle.
\end{equation}
A useful feature of this irrep is that it admits a basis in
which the gate group contains Pauli $X$ operators on both qubits.

To see this, consider the subgroup
\begin{equation}
    \langle b,c\rangle\cong\Z_2\times\Z_2,
\end{equation}
where $c$ is a second order-two element commuting with $b$.
The restriction of $\Gamma$ to this subgroup is the regular representation
of $\Z_2\times\Z_2$~\cite{conway1985atlas}. We may therefore
choose a two-qubit basis in which
\begin{equation}
    \Gamma^b=X_1=X\otimes I,
    \qquad
    \Gamma^c=X_2=I\otimes X.
\end{equation}

In the same basis, the order-five generator $a$ can be represented by
\begin{equation}
    \Gamma^a
    =
    \frac{1}{4}
    \begin{pmatrix}
    -\varphi & -\varphi & -\varphi & 2-3\varphi\\
    \varphi-2 & \varphi-2 & \varphi+2 & -\varphi\\
    \varphi+2 & \varphi-2 & \varphi-2 & -\varphi\\
    \varphi-2 & \varphi+2 & \varphi-2 & -\varphi
    \end{pmatrix},
    \label{eq:A5-linear-4d}
\end{equation}
where $\varphi=(1+\sqrt{5})/2$ is the golden ratio.
These matrices satisfy
\begin{equation}
    (\Gamma^a)^5
    =
    (\Gamma^b)^2
    =
    (\Gamma^a\Gamma^b)^3
    =
    I,
\end{equation}
and generate the four-dimensional linear irrep of $A_5$.
In particular, the second Pauli generator is obtained by taking
$c=a^2ba^{-2}ba^2$, giving
\begin{equation}
    X_2
    =
    (\Gamma^a)^2 X_1
    (\Gamma^a)^{-2} X_1
    (\Gamma^a)^2.
\end{equation}

The four-dimensional irrep of $A_5$ is obtained by restricting
the standard representation of $S_5$ to $A_5$. Consequently,
it admits intertwiners implementing the nontrivial outer
automorphism of $A_5$, since
$\operatorname{Out}(A_5)\cong\Z_2$. In the basis above, one
particularly simple intertwiner is
\begin{equation}
    W_{\mathrm{out}}
    =
    \operatorname{CX}_{21}
    =
    \begin{pmatrix}
    1&0&0&0\\
    0&0&0&1\\
    0&0&1&0\\
    0&1&0&0
    \end{pmatrix}.
\end{equation}
Its action on the Pauli generators is
\begin{equation}
    W_{\mathrm{out}}X_1W_{\mathrm{out}}^\dagger
    =
    X_1,
    \qquad
    W_{\mathrm{out}}X_2W_{\mathrm{out}}^\dagger
    =
    X_1X_2.
\end{equation}
Conjugation of $\Gamma^a$ by $W_{\mathrm{out}}$ also produces
an element of $\Gamma(A_5)$. Explicitly,
\begin{equation}
    W_{\mathrm{out}}
    \Gamma^a
    W_{\mathrm{out}}^\dagger
    =
    \Gamma^a X_1
    (\Gamma^a)^2 X_1
    (\Gamma^a)^{-1}.
\end{equation}
These relations establish that
\begin{equation}
    W_{\mathrm{out}}
    \Gamma(A_5)
    W_{\mathrm{out}}^\dagger
    =
    \Gamma(A_5),
\end{equation}
so $W_{\mathrm{out}}$ implements the nontrivial outer
automorphism as an intertwiner. Other representatives of
the same outer-automorphism class give additional intertwiners,
including $\operatorname{CX}_{12}$ and
$\operatorname{SWAP}_{12}$.

\begin{graybox}
\noindent \textbf{Remark:} Although this irrep of $A_5$ includes Pauli $X$ operators, it does not include Pauli $Z$ operators. It also does not have a projective intertwiner that implements a logical Pauli $Z$ operator. Therefore, it falls outside of the conditions in Result~\ref{res: transfer} and Section~\ref{sec: transferring information} for transferring information between the ISC and a surface code. We leave the problem of transferring information in and out of this $A_5$ ISC as an open question.
\end{graybox}

\subsection{Gates from intertwiners with $G=D_m$}

In Example~\ref{example4}, we discussed the logical gates arising from projective intertwiners for $G=D_m$. Here, we show that these intertwiners are exhaustive by considering the automorphisms of $D_m$.

For $m>2$, the automorphism group of $D_m$ is
\begin{equation}
    \operatorname{Aut}(D_m)
    \cong \Z_m \rtimes \Z_m^\times,
\end{equation}
where $\Z_m^\times$ denotes the multiplicative group of integers modulo $m$ that are coprime with $m$. Every automorphism is specified by the transformations
\begin{equation}
    \phi(s)= sr^a,
    \quad
    \phi(r)= r^b,
\end{equation}
where $a\in\Z_m$ and $b\in\Z_m^\times$. We classify the corresponding projective intertwiners by considering (i) trivial automorphisms, (2) automorphisms that fix $r$, and (iii) automorphisms that fix $s$.

\subsubsection{Trivial automorphism}

For the trivial automorphism, the intertwiner condition becomes
\begin{equation}
    W\Gamma^g W^\dagger
    =
    \beta(g)\Gamma^g.
\end{equation}
In this case, $\beta$ must be a one-dimensional representation of $D_m$, as follows from Eq.~\eqref{eq: beta_omega_phi}. The one-dimensional representations are classified by $\Z_2$ when $m$ is odd and by $\Z_2\times\Z_2$ when $m$ is even.
For any $m$, the one-dimensional representation
\begin{equation}
    \beta(s)=-1,
    \qquad
    \beta(r)=1
\end{equation}
corresponds the intertwiner $W=Z$, since
\begin{equation}
    ZXZ=-X,
    \qquad
    ZP_mZ=P_m.
\end{equation}

When $m$ is even, there is an additional representation with $\beta(s)=1$ and $\beta(r)=-1$. However, for $m>2$, no corresponding intertwiner exists: unitary conjugation preserves the trace, whereas $\Tr P_m\neq 0$. The same obstruction rules out the product of this representation with the one above. Thus, up to an overall phase, the only intertwiners associated with the trivial automorphism are $I$ and $Z$.

\subsubsection{Automorphisms fixing $r$}

The automorphisms fixing $r$ form a group isomorphic to $\Z_m$, generated by
\begin{equation}
    \phi_r(s)=sr,
    \qquad
    \phi_r(r)=r.
\end{equation}
A corresponding projective intertwiner is
\begin{equation}
    W_m=
    \begin{pmatrix}
    1&0\\
    0&e^{-i\pi/m}
    \end{pmatrix},
\end{equation}
which satisfies
\begin{equation}
    W_m X W_m^\dagger
    =
    e^{-i\pi/m}XP_m,
    \qquad
    W_m P_m W_m^\dagger
    =
    P_m.
\end{equation}
Thus, $W_m$ implements $\phi_r$ up to phases, and its powers implement all automorphisms fixing $r$. 

\subsubsection{Automorphisms fixing $s$}

The automorphisms fixing $s$ are specified by
\begin{equation}
    s\mapsto s,
    \qquad
    r\mapsto r^b,
\end{equation}
where $b\in\Z_m^\times$. We show that projective intertwiners exist only for $b=\pm1$.
To see this, an intertwiner must satisfy
\begin{equation}
    WP_mW^\dagger
    =
    \beta(r)P_m^b.
\end{equation}
Since unitary conjugation preserves the eigenvalues, the eigenvalue ratios of $P_m$ and $P_m^b$ must agree up to inversion. This requires $b=\pm1\pmod m$.

The case $b=1$ corresponds to the trivial automorphism discussed above. The remaining automorphism is
\begin{equation}
    \phi_s(s)=s,
    \qquad
    \phi_s(r)=r^{-1}.
\end{equation}
A corresponding projective intertwiner is $X$, which satisfies
\begin{equation}
    XXX=X,
    \qquad
    XP_mX=e^{2\pi i/m}P_m^{-1}.
\end{equation}
Thus, $X$ implements $\phi_s$ up to phases.

Every automorphism of $D_m$ can be expressed as a composition of an automorphism fixing $r$ and an automorphism fixing $s$. The preceding results show that the automorphisms admitting projective intertwiners are generated by $\phi_r$ and $\phi_s$.
Furthermore, any two intertwiners implementing the same automorphism differ by an intertwiner associated with the trivial automorphism. Since
\begin{equation}
    W_m^m=Z,
\end{equation}
the latter are already generated by $W_m$. Consequently, every projective intertwiner for $m>2$ can be expressed, up to an overall phase, as
\begin{equation}
    W=W_m^aX^b,
    \qquad
    a\in\Z_{2m},
    \quad b\in\{0,1\}.
\end{equation}
Equivalently, the group of projective intertwiners is generated by $W_m$ and $X$.

\subsection{Controlled Hadamard on two qubits}\label{app: CH}

In Table~\ref{tab: summary}, we included a two-qubit group of gates generated by a controlled Hadamard and Pauli $X$ operators. Here, we examine this example in greater detail, establishing the irreducibility of the group of gates and showing that the projective intertwiners are exhaustive.

The group of gates is:
\begin{equation}
    \cG=\langle X_1,X_2,\mathrm{CH}_{12}\rangle,
\end{equation}
where $X_1=X\otimes I$, $X_2=I\otimes X$, and $\mathrm{CH}_{12}$ is defined as:
\begin{equation}
    \mathrm{CH}_{12}
    =
    |0\rangle\langle0|\otimes I
    +
    |1\rangle\langle1|\otimes H.
\end{equation}
The group contains the operators
\begin{equation}
    (X_1\mathrm{CH}_{12})^2=H_2,
    \qquad
    (X_2\mathrm{CH}_{12})^4=Z_1.
\end{equation}
Since $H_2X_2H_2=Z_2$, the group contains all
two-qubit Pauli operators up to phases and is
therefore irreducible.

We first observe that $S_1$ and $\mathrm{CX}_{12}$
are projective intertwiners. Indeed, $S_1$ satisfies
\begin{equation}
    S_1X_1S_1^\dagger=iX_1Z_1
\end{equation}
and commutes with $X_2$ and $\mathrm{CH}_{12}$.
Similarly, $\mathrm{CX}_{12}$ satisfies
\begin{equation}
\begin{aligned}
    \mathrm{CX}_{12}X_1\mathrm{CX}_{12}
    &=X_1X_2,\\
    \mathrm{CX}_{12}X_2\mathrm{CX}_{12}
    &=X_2,\\
    \mathrm{CX}_{12}\mathrm{CH}_{12}\mathrm{CX}_{12}
    &=X_2\mathrm{CH}_{12}X_2.
\end{aligned}
\end{equation}
Thus, both operators preserve $\cG$ up to phases.

We now show that all projective intertwiners are
generated by $\cG$, $S_1$, and $\mathrm{CX}_{12}$.
To this end, define
\begin{equation}
    r_1=X_2H_2,
    \qquad
    r_2=\mathrm{CH}_{12}r_1\mathrm{CH}_{12}.
\end{equation}
These operators commute and satisfy
$r_1^4=r_2^4=-I$. Every element of $\cG$ can be
written as
\begin{equation}
    g=\pm r_1^k(H_2X_1)^e
    r_2^\ell X_1^f\mathrm{CH}_{12}^s,
\end{equation}
where $k,\ell\in\Z_4$ and $e,f,s\in\Z_2$.
A direct calculation shows that the only group
elements with trace magnitude $2\sqrt{2}$ are
$\pm r_1^{\pm1}$ and $\pm r_2^{\pm1}$.

Since the magnitude of the trace is preserved
under unitary conjugation and multiplication by
phases, any projective intertwiner must permute
$r_1$ and $r_2$, possibly taking their inverses.
Composing with elements of $\cG$, we may therefore
assume that an arbitrary projective intertwiner $W$
commutes with both $r_1$ and $r_2$.

The four joint eigenspaces of $r_1$ and $r_2$
are one-dimensional, with eigenvalues
$\omega^{\pm1}$, where $\omega=e^{i\pi/4}$.
In the joint eigenbasis, ordered by the eigenvalues
$(++,+-,-+,--)$, we may write
\begin{equation}
    W=\operatorname{diag}(1,u,v,w),
\end{equation}
up to an overall phase.

The diagonal elements of $\cG$ in this basis,
up to phases, are precisely
\begin{equation}
    \operatorname{diag}(1,i^a,i^b,i^{a+b}),
    \qquad a,b\in\Z_4.
\end{equation}
Requiring conjugation by $W$ to preserve
$H_2X_1$, $X_1$, and $\mathrm{CH}_{12}$
up to elements of $\cG$ and overall phases
gives
\begin{equation}
    u,v\in\langle\omega\rangle,
    \qquad
    u/v\in\langle i\rangle,
    \qquad
    w=\pm uv.
\end{equation}
Modulo the diagonal elements of $\cG$,
these conditions leave only two independent
binary choices: the common parity of the
exponents of $u$ and $v$, and the sign of
$w/(uv)$.

In the same basis, the two intertwiners
identified above satisfy
\begin{equation}
\begin{aligned}
    S_1
    &=\operatorname{diag}(1,i,i,1),\\
    \mathrm{CH}_{12}\mathrm{CX}_{12}
    &=\operatorname{diag}(1,\omega^{-1},\omega,1).
\end{aligned}
\end{equation}
The first changes the sign of $w/(uv)$,
while the second changes the common parity
of the exponents of $u$ and $v$.
Consequently, every allowed $W$ can be
reduced to an element of $\cG$ by composing
with these intertwiners.

We conclude that all projective intertwiners
are generated, up to overall phases, by
$\cG$, $S_1$, and $\mathrm{CX}_{12}$.

\subsection{Arbitrary finite group of diagonal gates on $m$ qubits}\label{app: arbitrary diagonal gates}

\subsubsection{Group structure}

Here, we study the group of gates generated by Pauli $X$ operators and a finite group of diagonal gates. For simplicity, we consider a system of $m$ qubits, but the discussion can be generalized to qudits straightforwardly. Explicitly, we let $\langle X_1,\ldots,X_m\rangle\cong \mathbb{Z}_2^m$ be the group generated by the single-qubit Pauli $X$ operators and $\mathcal{D}$ be an arbitrary finite group of diagonal gates on the $m$ qubits. Our goal is to characterize the quotient group $G$ generated by $\mathcal{D}$ and $\Z_2^m$ modulo phases. We note that the group of diagonal gates can always be extended to include the Pauli $Z$ operators. This ensures that the group of gates is irreducible.

To characterize the group, we first define the $X$-closure of $\mathcal{D}$ by
\begin{equation}
\mathcal{D}_X
=
\frac{\left\langle
xdx^{}
:\;
x\in \Z_2^m,\ d\in \mathcal{D}
\right\rangle}{\{x d x \sim e^{i \theta} x d x,~ e^{i \theta} \in \text{U}(1)\}}
~.
\end{equation}
Since conjugation by an $m$-qubit Pauli $X$ just permutes the computational basis states,
$x d x^{}$ is diagonal whenever $d$ is diagonal. By construction, $\mathcal{D}_X$ is closed
under conjugation by $\Z_2^m$. It is the full diagonal subgroup generated by the original group
$\mathcal{D}$ together with all of its conjugates under the Pauli $X$ operators.

The full group generated by $\mathcal{D}$ and the Pauli $X$ operators is
\begin{equation}
G\cong \mathcal{D}_X\rtimes\mathbb{Z}_2^m.
\end{equation}
One can verify that $G$ must have this semidirect product structure by checking the following properties: (i) $\mathcal{D}_X$ and $\Z_2^m$ are both subgroups of $G$, (ii) $\mathcal{D}_X \cap \Z_2^m = \{1\}$, (iii) $\mathcal{D}_X$ is a normal subgroup of $G$. In particular, $m$ and $|\mathcal{D}|$ are both finite, and so is $|\mathcal{D}_X| \le |\mathcal{D}|^{2^m}$, the number of diagonal operators obtained by permuting the computational basis without taking the quotient. Therefore $|G| = 2^m \times |\mathcal{D}_X|$ is also finite. 

\subsubsection{Examples}

Let us now work out two examples. To begin, let $m=1$ and consider $\mathcal{D} = \Z_q$, generated by the operator $P_q$ defined in Example~\ref{example2}. In this case, we can follow Example~\ref{example2} to recover the result $\mathcal{D} \cong \mathcal{D}_X$, and the well-known result $G \cong \Z_q \rtimes \Z_2 = D_q$.

For our second example, let us take $\mathcal{D} = \langle \text{C}^{m-1} Z \rangle \cong \Z_2$, for an arbitrary $m$. In this case, the only nontrivial element of $\mathcal{D}$ is a multi-controlled $Z$ gate controlled on the first $m-1$ qubits, with target on the last qubit. In the $2^m$-dimensional Hilbert space spanned by the computational basis states $\ket{i_1 i_2 \dots i_m}$ where $i_k = 0$ or 1, we see that $\text{C}^{m-1} Z = \operatorname{diag}(1,1, \dots , 1, -1)$: it acts by a minus sign only on the all-1 basis state, and trivially on the other basis states. By conjugating with the Pauli $X$ operators, we can permute the basis states and obtain all other diagonal gates that have a single minus sign on their diagonal. One can now check that these $2^m$ operators generate a group $\Z_2^{2^m}$ consisting of {all} real diagonal gates on $m$ qubits. Modding out by the scaler subgroup $\{I,-I\}$, we obtain the desired closure $\mathcal{D}_X = \Z_2^{2^m-1}$, which is {much} bigger than $\mathcal{D}$, in contrast to the previous example.\footnote{The group generated by Pauli $X$ and $\text{C}^{m-1}X$ was recently studied in Ref.~\cite{manjunath2026GSC}, within the related framework of {group} surface codes, and the corresponding group of gates has the same order as $G$, but a more complicated iterated semidirect product structure.}

Let us fix $m=3$ in the above example, so that $\mathcal{D} = \langle \text{CCZ}_{123}\rangle$. We now derive that the projective intertwiners in this case are of the form $\{S_i\}$ and  $\{\text{CX}_{ij}\}$ where $i, j \in \{1,2,3\}$ and $i \neq j$.

Note that the subgroup $\mathcal{D}_X$ is the distinguished maximal normal Abelian
subgroup of $G$, and hence any intertwiner must satisfy (up to global phases)
\begin{equation}
W\mathcal{D}_XW^\dagger=\mathcal{D}_X.
\end{equation}
To see the consequence of this condition, temporarily restore the global
sign and consider the group of all diagonal sign matrices.  Its simultaneous
one-dimensional eigenspaces are precisely the computational basis states $|x\rangle$ with
$x\in\mathbb{Z}_2^3.$ Indeed, any two distinct basis states can be distinguished by a diagonal
sign gate. Therefore a unitary normalizing $\mathcal{D}_X$ must permute
these one-dimensional eigenspaces:
\begin{equation}
W|x\rangle
=
e^{i\theta_x}|\pi(x)\rangle
\end{equation}
for some permutation $\pi$ of $\mathbb{Z}_2^3$.  Equivalently every intertwiner is of the form $W=DP$, where $D$ is diagonal and $P$ is a permutation matrix.

We first constrain the permutation part.  Write $X^a
=
X_1^{a_1}X_2^{a_2}X_3^{a_3},~
a\in\mathbb{Z}_2^3.$ The operator $X^a$ acts as the translation
\begin{equation}
|x\rangle\longmapsto|x+a\rangle.
\end{equation}
Now $P X^a P^\dagger$ is within the group, and its restriction to $\Z_2^3$ is another translation $X^{L(a)}$ for some invertible linear map $L\in GL(3,2)$. Therefore,
\begin{equation}
\pi(x+a)
=
\pi(x)+L(a) \implies \pi(x)=Lx+b,
\quad
L\in GL(3,2),~
b\in\mathbb{Z}_2^3.
\end{equation}

The translation $b$ is already implemented by the Pauli operators $X^b$.
The remaining linear transformations $L\in GL(3,2)$ are generated by
CNOT gates of the form
\begin{equation}
\text{CX}_{ij}:
x_j\longmapsto x_j+x_i~.
\end{equation}
Since a CNOT only permutes computational-basis states, it also maps an
arbitrary element of $\mathcal{D}_X$ to another element of
$\mathcal{D}_X$. Hence every $\text{CX}_{ij}, i \neq j$ is an intertwiner.

It remains to constrain the diagonal part. Let
\begin{equation}
D
=
\sum_x
e^{i\theta_x}|x\rangle\langle x|.
\end{equation}
Then
\begin{equation}
D X^a D^\dagger
=
\left[
\sum_x
e^{i(\theta_{x+a}-\theta_x)}
|x+a\rangle\langle x+a|
\right]
X^a.
\end{equation}
For this to belong projectively to $G$, the diagonal factor must be a real
sign gate up to an overall phase.  Therefore, for each $a$,
\begin{equation}
\theta_{x+a}-\theta_x
=
c_a
\pmod{\pi},
\end{equation}
where $c_a$ is independent of $x$.  Taking $a$ to be the standard basis vector $e_i$ and using
$X_i^2=I$, implies
\begin{equation}
c_{e_i}
=
0
\quad\text{or}\quad
\frac{\pi}{2}
\pmod{\pi}.
\end{equation}
Consequently, up to an overall phase,
\begin{equation}
\theta_x
=
\frac{\pi}{2}
\left(
u_1x_1+u_2x_2+u_3x_3
\right)
\mod \pi~,
\quad
u_i\in\mathbb{Z}_2~.
\end{equation}
Hence up to an overall phase,
\begin{equation}
D
=
D_f
S_1^{u_1}
S_2^{u_2}
S_3^{u_3},
\end{equation}
where $D_f$ is some diagonal sign gate belonging to $\mathcal{D}_X$. The $S_i$ gates are indeed intertwiners: they commute with every diagonal gate,
while mapping $X_i$ to $Y_i\sim X_iZ_i$ and $X_j$ to $X_j$ when $j \neq i$. Since $Z_i\in\mathcal{D}_X$, these images again belong to $G$.

\end{appendix}

\bibliography{bib}

\end{document}